\documentclass[10pt,journal,compsoc]{IEEEtran}
\usepackage{times}
\usepackage{epsfig}
\usepackage{graphicx}
\usepackage{epstopdf}
\usepackage{amsmath}
\usepackage{amssymb}
\usepackage{mathtools}
\usepackage{multirow}
\usepackage{enumerate}
\usepackage{enumitem}
\usepackage{makecell}
\usepackage[caption=false,font=footnotesize]{subfig}
\usepackage{url}
\usepackage[noend]{algpseudocode}
\usepackage[ruled,vlined]{algorithm2e}
\usepackage{verbatim}
\usepackage{flushend}
\usepackage[utf8]{inputenc}
\usepackage[english]{babel}
\graphicspath{ {Figures/} }
\usepackage{placeins}
\usepackage{breqn}
\usepackage{color}
\usepackage{soul}
\usepackage{tipa}

\ifCLASSOPTIONcompsoc
  \usepackage[nocompress]{cite}
\else
  \usepackage{cite}
\fi

\begin{document}
\setlength{\abovedisplayskip}{3pt}
\setlength{\belowdisplayskip}{3pt}
	\title{DeepVOX: Discovering Features from Raw Audio for Speaker Recognition in 
		Non-ideal Audio Signals}
	
	\author{Anurag~Chowdhury,~\IEEEmembership{Member,~IEEE,}
		Arun~Ross,~\IEEEmembership{Senior Member,~IEEE}
		\thanks{Both authors are with the Department
			of Computer Science Engineering, Michigan State University, East Lansing,
			MI, 48823, USA, 
			e-mail: \{chowdh51, rossarun\}@msu.edu.}
	}

	\maketitle
	
	\begin{abstract}
		Automatic speaker recognition algorithms typically use pre-defined filterbanks, such as Mel-Frequency and Gammatone filterbanks, for characterizing speech audio. However, it has been observed that the features extracted using these filterbanks are not resilient to diverse audio degradations. In this work, we propose a deep learning-based technique to deduce the filterbank design from vast amounts of speech audio. The purpose of such a filterbank is to extract features robust to non-ideal audio conditions, such as degraded, short duration, and multi-lingual speech. To this effect, a 1D convolutional neural network is designed to learn a time-domain filterbank called DeepVOX directly from raw speech audio. Secondly, an adaptive triplet mining technique is developed to efficiently mine the data samples best suited to train the filterbank. Thirdly, a detailed ablation study of the DeepVOX filterbanks reveals the presence of both vocal source and vocal tract characteristics in the extracted features. Experimental results on VOXCeleb2, NIST SRE 2008, 2010 and 2018, and Fisher speech datasets demonstrate the efficacy of the DeepVOX features across a variety of degraded, short duration, and multi-lingual speech. The DeepVOX features also shown to improve the performance of existing speaker recognition algorithms, such as the xVector-PLDA and the iVector-PLDA.
		
	\end{abstract}

	\begin{IEEEkeywords}
		Speaker Recognition, Degraded Audio, Deep Learning, Feature Extraction, Filterbanks
	\end{IEEEkeywords}
	
	\IEEEpeerreviewmaketitle
	
	\vspace{-0.2cm}
	\section{Introduction}
	
	\IEEEPARstart{A}{utomatic} speaker recognition entails recognizing an individual from their voice. One of the key applications of speaker recognition is securing devices with voice-controlled user interfaces (VUI) such as digital voice assistants~\cite{googlevoice} and telephone banking systems~\cite{Wellsfargo}. VUIs are gaining popularity due to the ease-of-access provided by their hands-free operation. However, in practice, the voice input to the speaker recognition systems often exhibits non-ideal speech audio characteristics, such as degraded~\cite{chowdhury2020fusing}, multi-lingual~\cite{lu2009effect}, and short-duration~\cite{kanagasundaram2011vector} speech. The unfavorable nature of these non-ideal inputs propagates through different components of the speaker recognition system and lowers its performance~\cite{chowdhury2020fusing,ross2019some}. Therefore, it is important to develop speaker recognition techniques that are robust to a wide variety of non-ideal audio conditions, thereby providing generalizable speaker recognition performance.
    
    Some of the current speaker recognition enabled consumer devices address the issues of non-ideal audio conditions at the sensor-level by employing specialized hardware, such as far-field microphone arrays~\cite{googlevoice}. However, the use of specialized hardware interfaces limits their backward compatibility with existing speaker recognition systems. On the other hand, some of the latest speaker recognition techniques address these issues at the software-level by designing robust matchers~\cite{snyder2018x,chowdhury2020fusing,chowdh2017extract}. However, these techniques rely on traditional handcrafted speech features such as Mel-Frequency Cepstral Coefficients (MFCC) and Linear Predictive Coding (LPC), whose representation capability varies with the quality of input audio, thus limiting the effectiveness of the subsequent matcher~\cite{guo2017robust}. While some of the recent work in end-to-end speaker recognition can perform speaker recognition directly from raw input audio, their robustness to non-ideal audio conditions is yet to be determined~\cite{ravanelli2018speaker, jung2020improved}.

    We position our work with the existing literature by approaching the issue of non-ideal audio conditions at the feature-level. We design \textbf{a raw audio-based speech feature extractor, called DeepVOX, that is robust to non-ideal audio conditions and compatible with existing speaker recognition algorithms}. Our method delivers generalizable noise-robust speaker recognition performance without any specialized hardware interface or relying on any handcrafted feature extraction techniques. The main contributions in this work are as follows:
    \vspace{-0.06cm}
    \begin{enumerate}[leftmargin=0cm,itemindent=.5cm,labelwidth=\itemindent,labelsep=0cm,align=left]
	\setlength\itemsep{0em}
        \item We propose a Convolutional Neural Network (CNN) based approach for learning a robust speech filterbank, referred to as DeepVOX, directly from raw speech audio. 
        \item We propose an adaptive triplet mining technique for efficiently training the DeepVOX filterbank in conjunction with 1D-Triplet-CNN~\cite{chowdhury2020fusing}, a CNN based speech feature embedding technique, to perform speaker verification.
        \item We experimentally demonstrate the compatibility and the associated performance benefits of the DeepVOX features with some existing speaker recognition algorithms such as the xVector-PLDA~\cite{snyder2018x} and the iVector-PLDA~\cite{dehak2011front}.
        \item We further study the impact of a large variety of audio degradations, multi-lingual speech data, and varying length speech audio on the representation capability of DeepVOX features.
        \item Finally, we perform a detailed ablation study to identify the type of speech features extracted by DeepVOX and characterize their frequency-response to a various degraded speech audio.
    \end{enumerate}
    

	\section{Related work}~\label{sec:related_work}
	\textit{Speech} recognition\textemdash i.e., recognition and translation of spoken language into machine-readable format\textemdash has been one of the most popular tasks in the speech processing community for decades. Therefore, most of the initial speech feature representations were developed from the \textit{speech} recognition perspective. The widely popular Mel-Frequency Cepstral Coefficients (MFCC) was initially proposed for performing monosyllabic word recognition and was later observed to be efficient for performing \textit{speaker} recognition as well~\cite{reynolds1994experimental}. However, MFCC features are not robust to audio degradations~\cite{guo2017robust} and are, therefore, not very suitable for speaker recognition tasks in presence of noisy speech data. This lack of generalizability in the performance of the handcrafted features, such as MFCC, primarily stems from the fact that they are derived from auditory experiments of limited scale. Although these auditory experiments have been later revised multiple times~\cite{o1987speech,umesh1999fitting} to improve their robustness to a wider variety of audio conditions, the scope for improvement remains. This motivated the development of robust speech features for performing speaker recognition in varied non-ideal audio conditions. These speech features are also adept at encoding various physical and acoustic properties of human voice and can be accordingly partitioned into several feature categories, as follows~\cite{kinnunen2010overview}: 
	\begin{itemize}[leftmargin=0cm,itemindent=.5cm,labelwidth=\itemindent,labelsep=0cm,align=left]
	\setlength\itemsep{0em}
		\item Short-term spectral features encode the vocal tract shape.
		\item Vocal source features characterize the glottal excitation signal.
		\item Prosodic features model the speaking style of a speaker.
		\item High-Level Features model the lexicon of a speaker.
	\end{itemize}
	
	According to the source-filter model of speech~\cite{milner2002speech}, human vocal tract can be assumed to behave like a time-varying digital filter due to the articulatory movements. Therefore, in order to model the vocal tract, short-term audio frames (usually 25 to 50ms) are used for extracting the stable voice characteristics in the form of short-term spectral features. Majority of the popular techniques for extracting stable voice characteristics are based on either MFCC or Linear Predictive Coding (LPC)~\cite{kim2016power}. The MFCC feature extraction process uses triangular-filters placed on the Mel-scale for modeling the human auditory perception system~\cite{voicerecoMFCC}. The LPC, on the other hand, estimates an all-pole model of filter design for modeling the vocal tract~\cite{milner2002speech}.
	
	Humans are noted to be efficient in performing speaker recognition in the presence of unknown audio degradations. However, the MFCC feature, which is based on human auditory processing, is unable to cope well in such scenarios~\cite{zhao2013analyzing}. Motivated by this, the authors in~\cite{zhao2012casa} propose the Gammatone Filterbank-based Gammatone Frequency Cepstral Coefficients (GFCC) features as a noise-robust alternative to MFCC. Compared to the Mel-filterbank, the Gammatone Filterbank has finer resolution at lower frequencies, which is claimed to better represent the human auditory model~\cite{fedila2015consolidating} and potentially improve speaker recognition performance. Another drawback of the MFCC feature extraction process is its disregard of phase information in the speech data, as the features are extracted only from the amplitude spectrum. The initial motivation behind this was based on human auditory system experiments~\cite{fedila2015consolidating}, where short-term phase spectrum did not provide enough performance benefits to justify the associated computational expenses. However, recent studies have reported comparable and complementary speaker recognition performance of both magnitude- and phase-based features~\cite{murty2006combining, sadjadi2011hilbert}.

    \begin{figure*}[t]		
		\centering
		\includegraphics[scale = 0.46]{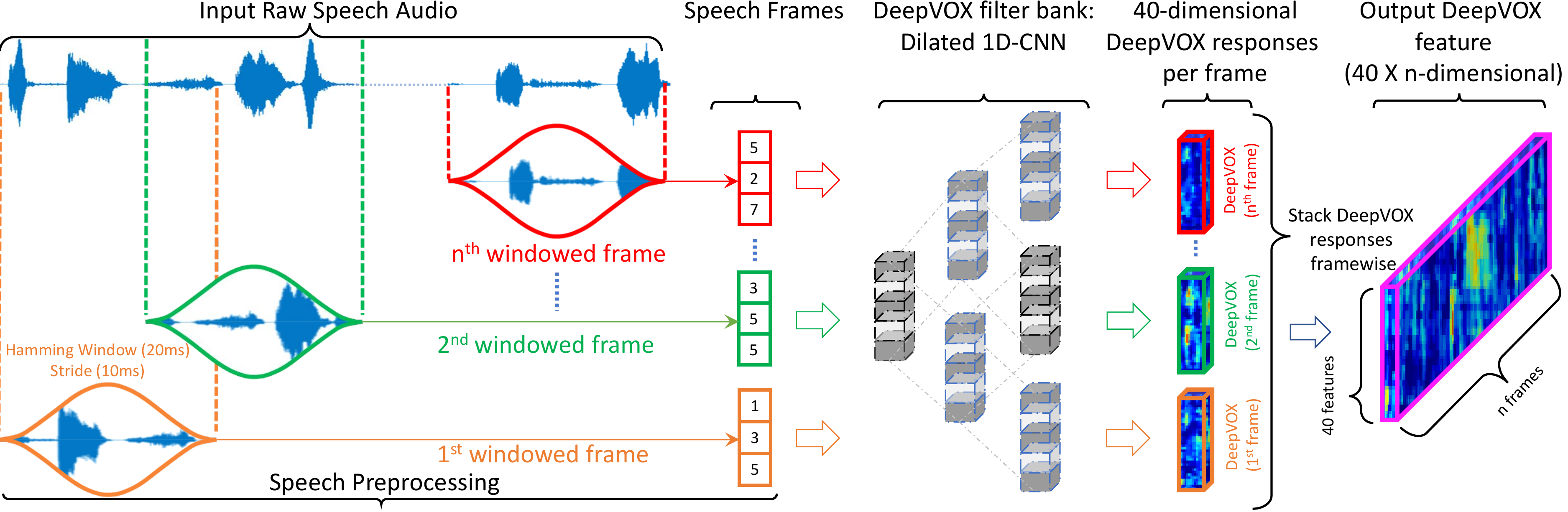}
		\caption{A visual representation of the proposed Dilated 1D-CNN based DeepVOX feature extraction process. The input raw speech audio is framed and windowed in the speech pre-processing step to extract short duration speech frames. The speech frames are then fed to the DeepVOX filterbank to extract corresponding frame-level DeepVOX features.
		}
		\label{fig:speech_patch}
		\vspace{-0.4cm}
	\end{figure*}

	LPC-based methods~\cite{wong2001comparison} in comparison, attempt to characterize the speech production model using an all-pole filter model. Linear Prediction Cepstral Coefficients (LPCC) are the cepstral representation of LPC features and are often considered more reliable than the regular LPC features~\cite{wong2001comparison}. One of the major disadvantages of the LPC and LPCC-based techniques is that they provide a linear approximation of speech at \textit{all} frequencies, whereas the spectral resolution of human hearing is known to reduce with frequency beyond 800Hz. This issue was addressed by Hermansky et al.~\cite{hermansky1990perceptual} in their work on Perceptual Linear Prediction (PLP) Coefficients. For extracting the PLP features on a non-linear scale, that resembles the human auditory system, several spectral transformations~\cite{hermansky1990perceptual} were applied	to the power spectrum of the speech audio prior to the all-pole model approximation by the autoregressive	model. It is important to note that both LPC and MFCC based features are usually augmented with `spectro-temporal features' in the form of delta and delta-delta coefficients, which are the first and second order time-derivatives of the short-term spectral features, respectively. Spectro-temporal features are one way of adding temporal features, such as formant transitions and energy modulations, to the short-term spectral features.

	The human vocal tract contributes to a majority of the speaker dependent features in the human voice. Short-term spectral features, such as LPC, that attempt to model the human vocal tract are particularly effective in performing speaker recognition. However, vocal tract modeling is not the only way of approaching speaker recognition. Vocal source features~\cite{kinnunen2010overview} can also be used for the task. Vocal source features refer to the characteristics of the source of human voice originating in the form of glottal excitation pulses. Features such as glottal pulse shape, rate of vocal fold vibration, and fundamental frequency can potentially be extracted~\cite{espy2006new} to perform speaker recognition. One such work in~\cite{zheng2007integration} used an LPC-based inverse vocal tract filter and wavelet transform for extracting vocal source features called Wavelet Octave Coefficients Of Residues (WOCOR). It also combined MFCC and WOCOR features for improving overall speaker verification performance.

	In the past decade, deep learning based methods have been successfully designed and implemented for solving many speech processing tasks, including speaker recognition~\cite{li2017deep,chowdh2017extract,chowdhury2020fusing}. A majority of such speaker recognition methods use some type of hand-crafted features, e.g. MFCC, LPCC, as input to their network for solving the problem. For example, authors in~\cite{li2017deep} developed an end-to-end Neural Speaker Embedding System called Deep Speaker that learns speaker-specific embeddings from $64$-dimensional log Mel-filterbank coefficients using ResCNN and GRU architectures.
	However, some of the recent works~\cite{muckenhirn2018towards, ravanelli2018speaker} have proposed to feed the raw speech waveform directly as input to deep neural networks for performing a variety of tasks such as speaker recognition and detection of voice presentation attacks. The authors in~\cite{ravanelli2018speaker}, for example, propose to learn the cut-off frequency of pre-defined band-pass filters for performing speaker recognition on clean (un-degraded) speech data.
	
	In this paper, we propose a new approach for extracting robust short-term speech features from raw audio data using 1D-Convolutional Neural Networks (1D-CNN). We draw design cues from our previous work on 1D-CNN~\cite{chowdh2017extract} and 1D-Triplet-CNN~\cite{chowdhury2020fusing} based architectures for performing speaker identification and verification respectively from degraded audio signals. However, both these architectures use MFCC and LPC-based feature representation as input and are, therefore, limited by the representation power of MFCC and LPC features. We, instead, propose a 1D-CNN based feature extraction module, termed as \textit{DeepVOX}, to learn and extract speech feature representation directly from raw audio data, in the time-domain itself. The DeepVOX learns filterbanks directly from a large quantity of degraded raw speech audio samples, thereby laying its emphasis on learning robust and highly discriminative speech audio features.
	
	Note that, unlike the work in~\cite{ravanelli2018speaker}, we learn the proposed DeepVOX filterbank without imposing any constraints on the design of the constituent filters. Also, unlike any of the current raw-waveform based speaker recognition methods~\cite{muckenhirn2018towards, ravanelli2018speaker, jung2020improved}, we demonstrate the compatibility of the proposed DeepVOX features with state-of-the-art deep learning-based speaker recognition methods such as xVectors~\cite{snyder2018x} and 1D-Triplet-CNN~\cite{chowdhury2020fusing} and even on classical methods such as the iVector-PLDA~\cite{dehak2011front}. 
	
	\begin{figure*}[t]		
		\centering
		\includegraphics[scale = 0.36]{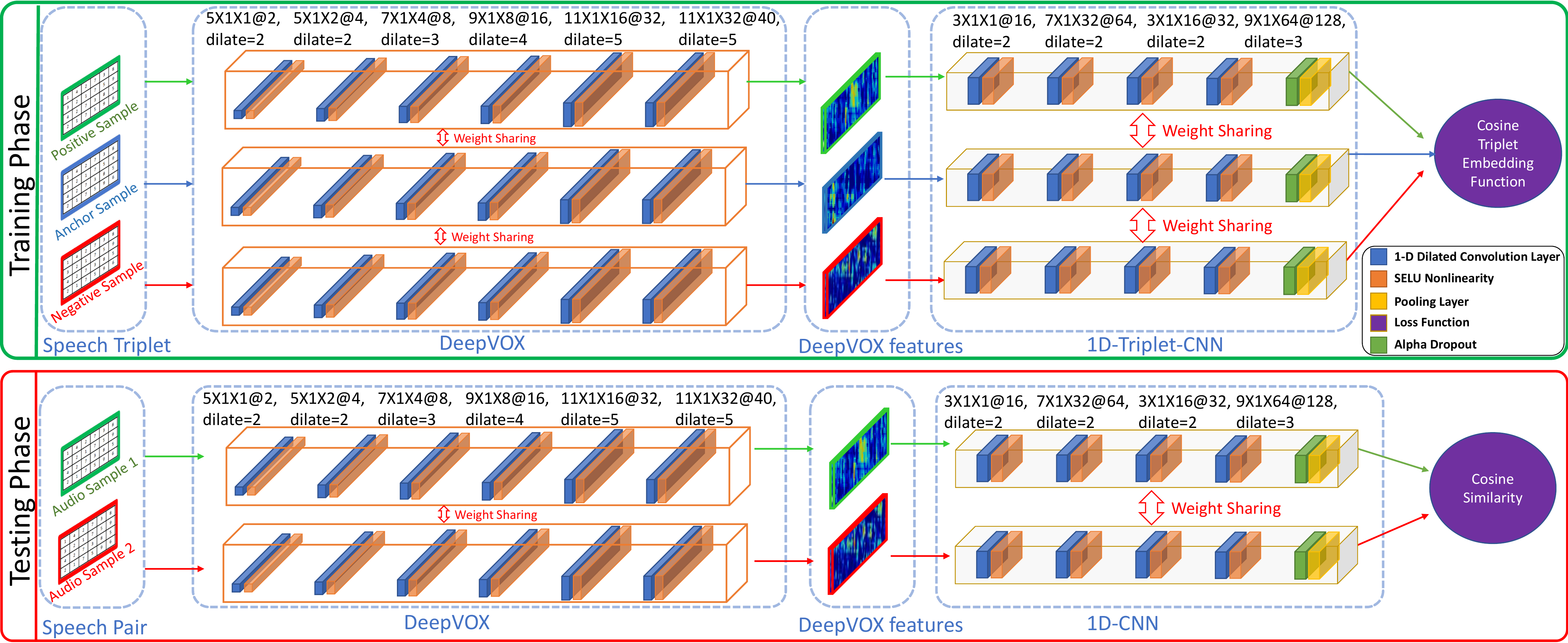}
		\caption{A visual representation of the training and testing phases of the proposed DeepVOX architecture. A 1D-Triplet-CNN is used to train the DeepVOX on speech triplets. A siamese 1D-CNN is used to evaluate the trained DeepVOX on pairs of speech audio. }
		\label{fig:arch}
		\vspace{-0.4cm}
	\end{figure*}
	
	\section{Proposed Algorithm}
	In the previous section, we discussed some popular speech feature extraction techniques. Depending upon the type of the features being extracted, the algorithms were further categorized into four different feature categories. As discussed, human vocal tract significantly contributes to the majority of speaker dependent features in the human voice. Short-term spectral features are, therefore, well-suited for speaker recognition due to their ability to model the human vocal tract. In this work, we propose a method for learning a new type of short-term speech features, referred to as \textit{DeepVOX features}, using 1D-Convolutional Neural Networks (1D-CNN). It is important to note that, unlike short-term spectral feature extraction algorithms like MFCC, where the extracted speech features are not specifically geared towards speaker recognition, our proposed algorithm learns to extract features directly from raw speech data, specifically suited for the task of speaker recognition. 
		
    \subsection{Speech Feature Extraction Using DeepVOX} \label{algo}
    
    In this work, we use the proposed DeepVOX feature extractor jointly with a 1D-Triplet-CNN~\cite{chowdhury2020fusing}-based feature embedding network for performing speaker recognition. The 1D-Triplet-CNN~\cite{chowdhury2020fusing} was initially developed for performing speaker verification in degraded audio signals by combining the MFCC and LPC features into a joint-embedding space. However, here the 1D-Triplet-CNN network is used jointly with the DeepVOX to map the DeepVOX features to a highly discriminative speaker embedding space. The proposed joint architecture (see Figure~\ref{fig:arch}), also referred to as 1D-Triplet-CNN(DeepVOX), consists of four separate units described below:
    
    \subsubsection{Speech Preprocessing} \label{speech_prep}
    We first use a Voice Activity Detector \cite{spec_VAD} to remove non-speech parts of an input audio. Any data sample longer than $2$ seconds is split into multiple smaller $2$ second long audio samples. The resulting speech audio is then \textit{framed and windowed} into multiple smaller audio clips, called \textit{speech units}, using a hamming window of length $20$ms and stride $10$ms, as shown in Figure~\ref{fig:speech_patch}. Therefore, each speech unit of duration $20$ms sampled at $8000$Hz is represented by an audio vector of length $160$. The running window extracts a \textit{speech unit} every $10$ms from a $2$sec long input audio, thereby extracting around $200$ \textit{speech units} per $2$ second long audio sample. These \textit{speech units} are then stacked horizontally to form a two-dimensional speech audio representation called \textit{speech frame}, of dimension $160 \times 200$. The extracted speech frames are then made into \textit{speech frame triplets} and fed into DeepVOX.

    \subsubsection{Speech Frame Triplets}
    The authors in \cite{schroff2015facenet} introduced the idea of triplet based CNNs. As illustrated in Figure~\ref{fig:arch}, our DeepVOX architecture takes a \textit{speech frame triplet} $D_t$ as input. A \textit{speech frame triplet} $D_t$ is defined as a tuple of three speech frames: $D_t = (S_a,S_p,S_n)$
    Here, $S_a$, the anchor sample, and $S_p$, the positive sample, are two different speech samples from a subject `X'. $S_n$, the negative sample, is a speech sample from another subject `Y', such that $X \neq Y$.
 
    \subsubsection{DeepVOX}~\label{sec:DeepVOX_method}
    The DeepVOX architecture, as given in Figure~\ref{fig:arch}, takes as speech frame triplet as input. DeepVOX processes each speech frame in the triplet to produce a corresponding short term spectral representation, thereby generating a corresponding triplet of \textit{DeepVOX features}. The design of the DeepVOX architecture primarily comprises of 1D Dilated Convolutional Layers~\cite{chowdhury2020fusing} and SELU~\cite{klambauer2017self} (Scaled Exponential Linear Units) non-linearity. The one dimensional filters are so designed that they only learn features from within \textit{speech units} in a \textit{speech frame} and not across them. This follows the assumption that the speaker dependent characteristics within each speech unit is independent of other speech units in the speech frame. Each $160$ dimensional speech unit within a speech frame is processed by layers of 1D Dilated Convolutional Layers to generate $40$ filter responses, which constitute the corresponding short-term spectral representation. These 1D Dilated Convolutional Layers interlaced with SELU non-linearity here are designed to jointly represent a filterbank, which unlike the Mel-filterbank or the Gammatone filterbank, is specifically learned for extracting speaker dependent characteristics.
    
    \subsubsection{1D-Triplet-CNN} \label{1D-Triplet-CNN}
    The 1D-Triplet-CNN's architecture comprises of interlaced 1D-Dilated-Convolutional layers and SELU non-linearity, followed by alpha dropout and pooling layers. The use of \textit{`dilated convolutions'} over \textit{`convolutions followed by pooling layers'} is motivated by the work done in Wavenet~\cite{oord2016wavenet}, where the authors use dilated convolutions to increase the receptive field size nonlinearly with a linear increase in number of parameters. In context of 1D-Triplet-CNN, 1D dilated convolutions allow the network to learn sparse relationships between the feature values within a speech unit leading to significant performance benefits. The 1D-Triplet-CNN architecture~\cite{chowdhury2020fusing} is designed for learning speaker dependent speech embedding from triplets of \textit{DeepVOX features}. The three parallel network branches in the 1D-Triplet-CNN architecture learn and share a common set of weights (see Figure~\ref{fig:arch}). The aim of the 1D-Triplet-CNN architecture is to transform the \textit{DeepVOX feature} triplet input into a triplet of embeddings, where the intra-class samples are embedded closer to each other and inter-class samples are embedded farther apart. This embedding learning process is ensured by the cosine triplet embedding loss.

    \subsubsection{Cosine Triplet Embedding Loss} \label{lossfunction}
    The cosine triplet embedding loss~\cite{chowdhury2020fusing} is a modification upon the triplet loss intially introduced in~\cite{schroff2015facenet} by replacing the euclidean distance metric with cosine similarity. The triplet loss is designed to learn an embedding $g(f(x)) \in \Re^d$, where $f(x)$ is DeepVOX feature of speech frame $x$. In this work, $d$ is set to $128$. The embedding is so learned that the intra-class samples are embedded closer to each other than the inter-class samples. The mathematical formulation of cosine triplet embedding loss is given by :
    
    \begin{dmath}
     L(S_a,S_p,S_n) = \sum_{a,p,n}^{N}\left(\cos(g(f(S_a)),g(f(S_n))) - \cos(g(f(S_a)),g(f(S_p))) + \alpha_{m}\right)
    \end{dmath}
    
    Here, $L(\cdot,\cdot,\cdot)$ is the cosine triplet embedding loss function. $S_a$ (the anchor sample) and $S_p$ (the positive sample) are two different speech samples from a subject `X'. $S_n$ (the negative sample) is a speech sample from another subject `Y', such that $X \neq Y$. {$N$} refers to the total number of triplets in the training set. $\alpha_{m}$ is the margin of the minimum distance between positive and negative samples and is a user tunable hyper-parameter.
    
    In the training phase, the triplet loss helps the network learn the similarity between the anchor and the positive samples and dissimilarity between the anchor and the negative samples. As illustrated in Figure~\ref{fig:arch}, we train both the DeepVOX and the 1D-Triplet-CNN networks together thus simultaneously learning the embedding space using the 1D-Triplet-CNN and the feature space using the DeepVOX.
    
    In the testing phase (see Figure~\ref{fig:arch}) we arrange the trained DeepVOX and 1D-Triplet-CNN networks into a siamese network, i.e. only two identical copies of the trained networks are needed. During testing, we input a pair of speech samples into the siamese network to extract a corresponding pair of speech embeddings. The speech embedding pair is then compared using the cosine similarity metric to render a match score. Under ideal conditions, the match score for a genuine pair should be close to $1$, while the match score for an impostor pair should be close to $-1$.
    
    
    


	\subsubsection{Adaptive Triplet Mining for Online Triplet Selection}~\label{sec:triplet_mining}
	The effectiveness and generalizability of any network trained using the triplet learning paradigm, such as 1D-Triplet-CNN~\cite{chowdhury2020fusing}, depends on the difficulty of the training triplets. The authors in~\cite{chowdhury2020fusing} trained their proposed 1D-Triplet-CNN algorithm using offline-generated triplets for performing their speaker recognition experiments. However, the effectiveness and computational-feasibility of offline-triplet generation for evenly sampling a speech dataset drastically reduces with the increase in the number of training samples. Online-triplet generation is, therefore, chosen to effectively train the 1D-Triplet-CNN for our experiments. While the majority of online-triplet generation techniques use either hard or semi-hard triplet mining~\cite{schroff2015facenet}, we propose a curriculum learning-based~\cite{bengio2009curriculum} \textit{adaptive triplet mining} technique. 
	
	In adaptive triplet mining, at a given epoch $i$, the goal is to select a negative sample $S_n^i$, such that:
	
    \begin{dmath}
    \resizebox{.0141\hsize}{!}{$\cos(g(f(S_a^i)),g(f(S_p^i))) > \cos(g(f(S_a^i)),g(f(S_n^i))) + \alpha_{m}.$}
    \end{dmath}
    
     \begin{dmath}
    \tau_{S_n^i} > \tau_{S_n^{i-1}}
    \end{dmath}
    
    Here, $S_a^i$ is the anchor speech sample, $S_p^i$ is the positive speech sample , and $\alpha_{m}$ is the margin. Here, $\tau_{S_n^i}$ is a parameter that denotes the average difficulty of $S_n^i$ (a negative sample), chosen at epoch $i$. The difficulty of a negative sample is computed using its cosine similarity to the corresponding anchor speech sample in the triplet. Harder negative samples typically have higher cosine similarity to the corresponding anchor samples, making them harder to separate from the anchor samples. A value of $\tau = 0$ yields the easiest negative sample and $\tau = 1$ yields the hardest negative sample. In our experiments, the value of $\tau$ is determined by the current stage (or epoch) of the training process. We initialize the training with the value of $\tau$ at $0.4$ (empirically chosen) and increase it gradually to $1.0$ through the course of the training. This is done to ensure a minimum difficulty of the training triplets at the beginning of the training which is gradually increased as the training proceeds. This helps in avoiding the problem of bad local minima caused by introducing harder negative triplets directly at the beginning of the training~\cite{schroff2015facenet}. It is also observed that learning only on easy and semi-hard triplets lead to poor generalization capability of the model on harder evaluation pairs. Additionally, the model is pre-trained in the identification mode to ensure easier initialization of the training process.
    

    \subsection{Analysis of the DeepVOX Architecture}
    
    In Section~\ref{sec:DeepVOX_method}, we introduced the DeepVOX architecture for extracting short-term speech features. In this section, we mathematically analyze the proposed architecture and compare DeepVOX's feature learning process with some popular short-term spectral feature extraction algorithms such as MFCC, PNCC, PLP and MHEC. However, before proceeding with the mathematical analysis of DeepVOX's network architecture, we first draw a visual comparison with some popular short-term spectral feature extraction algorithms in Figure~\ref{fig:features}. The main purpose of this comparison is to identify the building blocks of different short-term spectral features and develop an understanding of their individual roles in the feature extraction process. We further use this comparative study to explain the similarities and dissimilarities between our proposed algorithm and some of the existing short-term spectral feature extraction algorithms. 
    
    Furthermore, please note that the DeepVOX method is proposed as an alternative for short-term spectral features such as MFCC and LPC and is intended to be used alongside feature embedding methods such as xVector, iVector, or 1D-Triplet-CNN for performing speaker recognition. Therefore, DeepVOX, similar to MFCC and LPC, is strictly a short-term time-domain feature extraction method, whereas xVector, iVector, and 1D-Triplet-CNN are speech feature embedding methods. Additionally, DeepVOX features, unlike the xVector embeddings, are not a mid-level representation drawn from an end-to-end speaker recognition neural network. Instead, DeepVOX is an independent neural network model carefully designed to learn a time-domain speech filterbank directly from raw audio data. Such an approach makes the DeepVOX features, unlike existing deep learning-based speech embedding networks~\cite{snyder2018x,jung2020improved}, a direct alternative for short-term spectral features such as MFCC and LPC in speaker recognition models. We specifically trained xVector and iVector models using DeepVOX features to demonstrate its compatibility with existing deep learning-based and classical speaker recognition methods. The experimental results given Section~\ref{results} show its performance benefits over MFCC, LPC, and MFCC-LPC features.
    
    \subsubsection{Building Blocks of Short-term Spectral Feature Extraction Algorithms}
    
    The comparison in Figure~\ref{fig:features} highlights some key components, given below, important for designing a short-term spectral feature extraction algorithm.
    
    \begin{figure*}[t]		
    	\centering
    	\includegraphics[scale = 0.38]{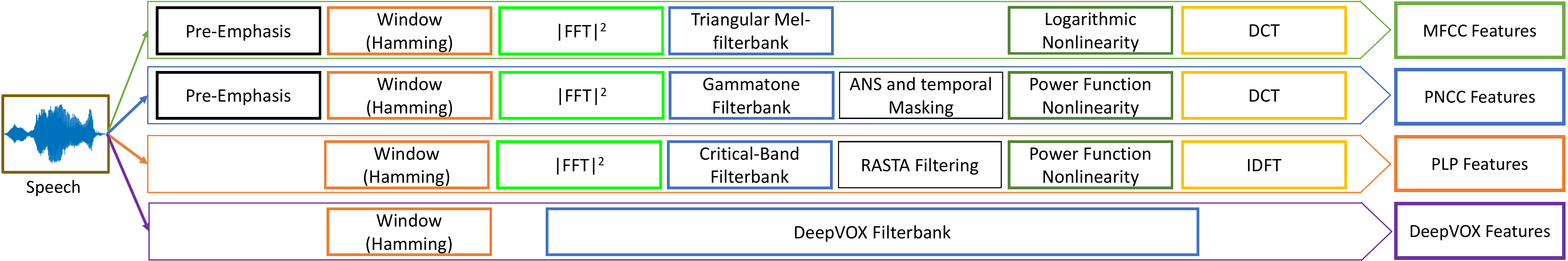}
    	\caption{A visual comparison of different short-term spectral feature extraction algorithms with our proposed DeepVOX algorithm. Boxes outlined in same colors perform similar types of operations in the corresponding feature extraction processes. }
    	\vspace{-0.4cm}
    	\label{fig:features}
    \end{figure*}
    
    \begin{itemize} [leftmargin=0cm,itemindent=.5cm,labelwidth=\itemindent,labelsep=0cm,align=left]
    	\item Pre-emphasis: In the pre-emphasis phase, the speech signal is high-pass filtered to compensate for the natural suppression of high frequency components in the human voicebox. However, this step can degrade the speech quality if the input audio has high-frequency noise and is therefore skipped in our proposed method.
    	
    	\item Framing and Windowing: Next, the speech signal is split into smaller short-term audio frames, typically $20$-$30$ms long. This is done to reliably extract speaker-dependent vocal characteristics, which are stable only within such short-term frames. In our case, we use a hamming window of length of 20ms and a stride of 10ms for framing and windowing the speech signal.
    	
    	\item Fourier Transform: FFT (Fast Fourier Transform) is performed to decompose a speech signal based on its frequency content. However, in our case, instead of decomposing the speech frames into their frequency components using FFT, DeepVOX learns speech features in the time domain itself.
    	
    	\item Filterbank Integration: The FFT response is usually processed through varied handcrafted filterbanks (such as Mel filterbank) for extracting the individual speech features. However, for DeepVOX, instead of using handcrafted filterbanks, a non-linear combination of multiple convolutional filters is used to learn filterbanks specifically suited for performing speaker recognition.
    	
    	\item Nonlinear Rectification: This step is done to compress the dynamic range of filterbank energies to improve speaker recognition performance~\cite{zhao2013analyzing}. However, for the DeepVOX there is no need for an explicit non-linear rectification step due to the inherent non-linearity in the network architecture.    	
    	
    \end{itemize}

    \subsubsection{Mathematical Analysis of the DeepVOX Architecture}
    
     Majority of the popular short-term spectral feature extraction algorithms such as MFCC and PNCC extract the speaker dependent features from a speech signal using handcrafted filterbanks. To this effect, the Fourier Transform is used to decompose a speech signal into its constituent frequencies, thereby, making filtering operation easier, both semantically and computationally. As the computationally-expensive convolution operation, between the signal and the filter, in time domain is replaced by pointwise multiplication in the frequency domain. The Fourier Transform is usually implemented using the Fast Fourier Transform (FFT) algorithm which makes the filtering of 1D audio signals even more computationally efficient, $\mathcal{O}(n\log{}n)$, as compared to general convolution operation, $\mathcal{O}(n^2)$. However, FFT only provides a close approximation of time domain filtering and is often inconsistent across different implementations~\cite{schatzman1996accuracy}, thereby enforcing a trade-off between computational complexity and accuracy. Furthermore, the recent development of efficient GPU-driven implementations of the convolution operation makes Convolutional Neural Networks (CNN) extremely well-suited for performing time domain filtering. Therefore, we use CNN in our algorithm to learn time-domain filters efficiently from raw speech audio.

	 As discussed earlier and illustrated in Figures \ref{fig:speech_patch} and \ref{fig:arch}, our proposed DeepVOX architecture takes a 2D \textit{speech frame} $S$ derived from raw speech waveform,  as input to the network. A speech frame $S$ can be represented as:
	 \begin{dmath}
	   S = \lbrack u_1, u_2, \cdots, u_i, \cdots, u_n \rbrack.
	 \end{dmath}
	  
	 Here $u_i$ is the $i^{th}$ speech unit and $n$ is the total number of speech units, in the speech frame $S$. Furthermore, the network outputs a $40$ channel filter response $f_i$ corresponding to every speech unit $u_i$. Therefore, DeepVOX's output $\textbf{O}$ can be given by:
	  
	 \begin{dmath}
	   \textbf{O} = \lbrack f_1, f_2, \cdots, f_i, \cdots, f_n \rbrack.
	 \end{dmath}
	 
	\vspace{-0.3cm}
	\begin{align}
	f_i &= \begin{bmatrix}
	x_{i,1}, x_{i,2}, \dots, x_{i,j}, \dots, x_{i,40}.
	\end{bmatrix}^\top
	\end{align}
	 
	 Here, $x_{i,j}$ is the $j^{th}$ channel filter output for $i^{th}$ speech unit $u_i$. In the DeepVOX model, channel outputs at the final layer are results of multiple convolutions of the input data with different convolution filters in the network. Therefore, the network output $f_i$ corresponding to speech unit $u_i$ can be written as:
	 
	 \begin{dmath}\label{eq:layer_output}
	 	f_{i} = (l_m(l_{m-1}( \cdots l_k(\cdots l_1(u_i)).
	 \end{dmath}
	  
     Here, $l_k()$ is the $k^{th}$ layer output of the DeepVOX model and $m$ is the total number of layers. Each layer of DeepVOX learns a multi-channel convolutional filter $C_k$. We can represent $l_k()$ as:
      
     \begin{dmath}\label{eq:time_conv}
     	l_k(u_i) = C_k \circledast u_i,
     \end{dmath}
      
     Here $C_k$ is the convolutional filter for the $k^{th}$ layer. The operation in Eq.(\ref{eq:time_conv}) is equivalent to time-domain filtering of input signal $u_i$ with filter $C_k$. Hence, we can rewrite Eq.(\ref{eq:layer_output}) as:

     \begin{dmath}\label{eq:time_conv_associative}
     f_i = (C_m \circledast (C_{m-1} \circledast ( \cdots C_k \circledast (\cdots C_1 \circledast (u_i)).
     \end{dmath}

     Since the convolution operation is associative, we can rewrite Eq.(\ref{eq:time_conv_associative}) as:
     \begin{dmath}
     f_i = \underbrace{(C_m \circledast C_{m-1} \circledast  \cdots \circledast C_k \circledast \cdots C_1 )}_{\text{learned DeepVOX filterbank}} \circledast u_i;     	     	
	 \end{dmath}	    
	    
	 The $DeepVOX_{filterbank}$, therefore, is designed to learn a $40$ channel convolution filter through a combination of multi-channel time-domain filters learned in different layers of the DeepVOX model. Here, each of the $40$ channels represents an individual time-domain speech filter in the $DeepVOX_{filterbank}$. 
	 
	 \vspace{-0.3cm}
	\section{Datasets and Experiments}\label{sec:experiments}
	In this work, we perform multiple speaker verification experiments on a variety of datasets and protocols. Primarily, we use the VOXCeleb2~\cite{chung2018voxceleb2}, Fisher English Training Speech Part 1~\cite{cieri2004fisher}, and NIST SRE (2008~\cite{SRE08}, 2010~\cite{SRE10}, and 2018~\cite{SRE18}) datasets for training and evaluating the proposed and baseline speaker verification algorithms. We also create degraded versions of the Fisher and NIST SRE 2008 speech datasets by adding diverse noise data from the NOISEX-92~\cite{noisex} dataset under varying levels of (signal-to-noise ratio) SNR (0 to 20 dB) and reverberations. This is done to evaluate the robustness of our proposed method to diverse audio degradations. Additionally, all the speech datasets were sampled at a rate of $8$kHz to match the NIST SRE dataset specifications~\cite{SRE08}.

		
	\begin{table*}[t]
		\fontsize{6}{8}\selectfont
		\caption{Verification Results on the VOXCeleb2 speech dataset. The proposed DeepVOX features outperform the baseline features for majority of the speaker recognition algorithms, across all the metrics.}
		\vspace{-0.2cm}
		\centering
		\scalebox{1.0}{
        \begin{tabular}{|c|c|c|c|c|c|c|c|c|c|c|c|c|c|}
    \hline 
    \multirow{2}{*}{\#} & \multirow{2}{*}{Method} & \multicolumn{4}{c|}{TMR@FMR=\{1\%, 10\%\}} & \multicolumn{4}{c|}{minDCF (ptar=\{0.001, 0.01\})} & \multicolumn{4}{c|}{EER(in \%)}\tabularnewline
    \cline{3-14} \cline{4-14} \cline{5-14} \cline{6-14} \cline{7-14} \cline{8-14} \cline{9-14} \cline{10-14} \cline{11-14} \cline{12-14} \cline{13-14} \cline{14-14} 
     &  & MFCC & LPC & \makecell{MFCC-\\LPC} & \makecell{Deep\\VOX} & MFCC & LPC & \makecell{MFCC-\\LPC} & \makecell{Deep\\VOX} & MFCC & LPC & \makecell{MFCC-\\LPC} & \makecell{Deep\\VOX}\tabularnewline
    \hline 
    \multirow{5}{*}{1} & 1D-Triplet-CNN-online & 70.72, 93.13 & 78.05, 94.93 & 82.09, 97.55 & \textbf{91.98,} \textbf{98.45} & 0.080, 0.67 & 0.067, 0.58 & 0.062, 0.43 & \textbf{0.030, 0.28} & 8.42 & 6.84 & 5.42 & \textbf{2.92}\tabularnewline
    \cline{2-14} \cline{3-14} \cline{4-14} \cline{5-14} \cline{6-14} \cline{7-14} \cline{8-14} \cline{9-14} \cline{10-14} \cline{11-14} \cline{12-14} \cline{13-14} \cline{14-14} 
     & 1D-Triplet-CNN & 69.30, 93.5 & 74.33, 94.57 & 84.70, 95.77 & \textbf{90.49, 98.09} & 0.078, 0.63 & 0.077, 0.54 & 0.075, 0.45 & \textbf{0.045, 0.37} & 8.62 & 7.06 & 6.05 & \textbf{3.46}\tabularnewline
    \cline{2-14} \cline{3-14} \cline{4-14} \cline{5-14} \cline{6-14} \cline{7-14} \cline{8-14} \cline{9-14} \cline{10-14} \cline{11-14} \cline{12-14} \cline{13-14} \cline{14-14} 
     & xVector-PLDA & 55.75, 85.96 & 73.61, 95.07 & 76.76, 94.75 & \textbf{90.76, 97.69} & 0.080, 0.78 & 0.074, 0.54 & 0.072, 0.52 & \textbf{0.048, 0.37} & 11.25 & 7.35 & 7.35 & \textbf{3.95}\tabularnewline
    \cline{2-14} \cline{3-14} \cline{4-14} \cline{5-14} \cline{6-14} \cline{7-14} \cline{8-14} \cline{9-14} \cline{10-14} \cline{11-14} \cline{12-14} \cline{13-14} \cline{14-14} 
     & iVector-PLDA & 86.16, 96.02 & 81.57, 97.1 & 92.54, \textbf{98.29} & \textbf{93.72}, 98.14 & \textbf{0.050}, 0.34 & 0.078, 0.53 & 0.056\textbf{, 0.32} & 0.063, 0.39 & 5.39 & 6.32 & \textbf{3.37} & 3.63\tabularnewline
    \cline{2-14} \cline{3-14} \cline{4-14} \cline{5-14} \cline{6-14} \cline{7-14} \cline{8-14} \cline{9-14} \cline{10-14} \cline{11-14} \cline{12-14} \cline{13-14} \cline{14-14} 
     & RawNet2 & \multicolumn{4}{c|}{91.75, 97.48} & \multicolumn{4}{c|}{0.056, 0.30} & \multicolumn{4}{c|}{3.91}\tabularnewline
    \hline 
    \end{tabular}

        }

		\label{tab:Experiments_VOXCeleb2}
	\end{table*}

	\begin{table*}[t]
		\fontsize{6}{8}\selectfont
		\caption{Verification Results on the degraded Fisher speech dataset. The proposed DeepVOX features outperform the baseline features for a majority of methods and data partitions, across all the metrics.}
		\vspace{-0.2cm}
		\centering
		\scalebox{1.0}{
        \begin{tabular}{|c|c|c|c|c|c|c|c|c|c|c|c|c|c|c|}
        \hline 
        \multirow{2}{*}{\#} & \multirow{2}{*}{\makecell{Training set\\ / Testing set}} & \multirow{2}{*}{Method} & \multicolumn{4}{c|}{TMR@FMR=\{1\%, 10\%\}} & \multicolumn{4}{c|}{minDCF (ptar=\{0.001, 0.01\})} & \multicolumn{4}{c|}{EER(in \%)}\tabularnewline
        \cline{4-15} \cline{5-15} \cline{6-15} \cline{7-15} \cline{8-15} \cline{9-15} \cline{10-15} \cline{11-15} \cline{12-15} \cline{13-15} \cline{14-15} \cline{15-15} 
         &  &  & MFCC & LPC & \makecell{MFCC-\\LPC} & \makecell{Deep\\VOX} & MFCC & LPC & \makecell{MFCC-\\LPC} & \makecell{Deep\\VOX} & MFCC & LPC & \makecell{MFCC-\\LPC} & \makecell{Deep\\VOX}\tabularnewline
        \hline 
        \multirow{5}{*}{2} & \multirow{5}{*}{F1/F1} & M1 & 49.13, 82.06 & 46.60, 81.87 & 59.93, 87.46 & \textbf{79.14}, \textbf{93.05} & 0.089, 0.89 & 0.094, 0.87 & 0.081, 0.81 & \textbf{0.075, 0.52} & 13.86 & 14.05 & 11.82 & \textbf{7.99}\tabularnewline
        \cline{3-15} \cline{4-15} \cline{5-15} \cline{6-15} \cline{7-15} \cline{8-15} \cline{9-15} \cline{10-15} \cline{11-15} \cline{12-15} \cline{13-15} \cline{14-15} \cline{15-15} 
         &  & M2 & 27.98, 74.62 & 31.64, 84.81 & 51.81, 84.81 & \textbf{77.27}, \textbf{92.53} & 0.095, 0.95 & 0.094, 0.93 & 0.087, 0.83 & \textbf{0.051, 0.51} & 16.50 & 17.06 & 12.65 & \textbf{8.30}\tabularnewline
        \cline{3-15} \cline{4-15} \cline{5-15} \cline{6-15} \cline{7-15} \cline{8-15} \cline{9-15} \cline{10-15} \cline{11-15} \cline{12-15} \cline{13-15} \cline{14-15} \cline{15-15} 
         &  & M3 & 20.77, 57.93 & 20.58, 63.22 & 29.10, 72.61 & \textbf{53.31, 88.63} & 0.097, 0.97 & 0.097, 0.97 & 0.096, 0.96 & \textbf{0.089, 0.87} & 22.86 & 20.43 & 17.46 & \textbf{10.92}\tabularnewline
        \cline{3-15} \cline{4-15} \cline{5-15} \cline{6-15} \cline{7-15} \cline{8-15} \cline{9-15} \cline{10-15} \cline{11-15} \cline{12-15} \cline{13-15} \cline{14-15} \cline{15-15} 
         &  & M4 & 25.42, 68.32 & 03.40, 18.01 & 29.04, 70.66 & \textbf{71.12, 90.23} & 0.098, 0.97 & 0.099, 0.99 & 0.096, 0.96 & \textbf{0.074, 0.63} & 18.47 & 43.58 & 18.13 & \textbf{9.77}\tabularnewline
        \cline{3-15} \cline{4-15} \cline{5-15} \cline{6-15} \cline{7-15} \cline{8-15} \cline{9-15} \cline{10-15} \cline{11-15} \cline{12-15} \cline{13-15} \cline{14-15} \cline{15-15} 
         &  & M5 & \multicolumn{4}{c|}{62.53, 84.50} & \multicolumn{4}{c|}{0.084, 0.65} & \multicolumn{4}{c|}{13.61}\tabularnewline
        \hline 
        \multirow{5}{*}{3} & \multirow{5}{*}{F1 / F2} & M1 & 28.36, 71.49 & 27.15, 63.86 & 39.73, 77.98 & \textbf{78.51, 93.13} & 0.094, 0.94 & 0.095, 0.95 & 0.091, 0.91 & \textbf{0.091, 0.53} & 17.75 & 20.77 & 15.72 & \textbf{7.99}\tabularnewline
        \cline{3-15} \cline{4-15} \cline{5-15} \cline{6-15} \cline{7-15} \cline{8-15} \cline{9-15} \cline{10-15} \cline{11-15} \cline{12-15} \cline{13-15} \cline{14-15} \cline{15-15} 
         &  & M2 & 14.35, 55.44 & 9.18, 46.56 & 34.74, 74.09 & \textbf{75.73, 92.33} & 0.098, 0.98 & 0.099, 0.99 & 0.094, 0.94 & \textbf{0.056, 0.49} & 23.30 & 25.98 & 17.37 & \textbf{8.42}\tabularnewline
        \cline{3-15} \cline{4-15} \cline{5-15} \cline{6-15} \cline{7-15} \cline{8-15} \cline{9-15} \cline{10-15} \cline{11-15} \cline{12-15} \cline{13-15} \cline{14-15} \cline{15-15} 
         &  & M3 & 12.65, 46.68 & 2.98, 18.84 & \textbf{12.27, 53.02} & 7.90, 36.98 & 0.099, 0.99 & \textbf{0.098, 0.98} & 0.099, 0.99 & 0.099, 0.99 & 26.59 & 44.3 & 24.02 & \textbf{31.3}\tabularnewline
        \cline{3-15} \cline{4-15} \cline{5-15} \cline{6-15} \cline{7-15} \cline{8-15} \cline{9-15} \cline{10-15} \cline{11-15} \cline{12-15} \cline{13-15} \cline{14-15} \cline{15-15} 
         &  & M4 & 5.41, 25.10 & 11.58, 42.21 & 14.78, 54.10 & \textbf{18.63, 55.50} & 0.097, 0.97 & 0.100, 0.99 & 0.099, 0.99 & \textbf{0.096, 0.96} & 37.87 & 30.93 & 23.54 & \textbf{26.10}\tabularnewline
        \cline{3-15} \cline{4-15} \cline{5-15} \cline{6-15} \cline{7-15} \cline{8-15} \cline{9-15} \cline{10-15} \cline{11-15} \cline{12-15} \cline{13-15} \cline{14-15} \cline{15-15} 
         &  & M5 & \multicolumn{4}{c|}{27.93, 59.75} & \multicolumn{4}{c|}{0.094, 0.93} & \multicolumn{4}{c|}{27.53}\tabularnewline
        \hline 
        \multirow{5}{*}{4} & \multirow{5}{*}{F2 / F2} & M1 & 47.62, 83.12 & 46.22, 82.21 & 55.78, 86.97 & \textbf{80.25, 94.08} & 0.081, 0.81 & 0.087, 0.84 & 0.085, 0.83 & \textbf{0.062, 0.57} & 13.37 & 14.24 & 11.56 & \textbf{7.25}\tabularnewline
        \cline{3-15} \cline{4-15} \cline{5-15} \cline{6-15} \cline{7-15} \cline{8-15} \cline{9-15} \cline{10-15} \cline{11-15} \cline{12-15} \cline{13-15} \cline{14-15} \cline{15-15} 
         &  & M2 & 36.40, 77.49 & 33.42, 76.02 & 50.57, 84.67 & \textbf{75.13, 92.65} & 0.099, 0.97 & 0.092, 0.92 & 0.088, 0.88 & \textbf{0.081, 0.74} & 16.16 & 16.43 & 13.03 & \textbf{8.54}\tabularnewline
        \cline{3-15} \cline{4-15} \cline{5-15} \cline{6-15} \cline{7-15} \cline{8-15} \cline{9-15} \cline{10-15} \cline{11-15} \cline{12-15} \cline{13-15} \cline{14-15} \cline{15-15} 
         &  & M3 & 20.77, 57.93 & 20.58, 63.22 & 29.10, 72.61 & \textbf{47.91, 82.00} & 0.098, 0.98 & 0.094, 0.94 & 0.097, 0.96 & \textbf{0.096, 0.86} & 22.86 & 20.43 & 17.46 & \textbf{13.9}\tabularnewline
        \cline{3-15} \cline{4-15} \cline{5-15} \cline{6-15} \cline{7-15} \cline{8-15} \cline{9-15} \cline{10-15} \cline{11-15} \cline{12-15} \cline{13-15} \cline{14-15} \cline{15-15} 
         &  & M4 & 16.19, 56.57 & 19.31, 56.84 & 29.37, 73.79 & \textbf{79.22, 92.8} & 0.097, 0.96 & 0.099, 0.99 & 0.095, 0.95 & \textbf{0.084, 0.61} & 24.08 & 23.62 & 16.65 & \textbf{7.9}\tabularnewline
        \cline{3-15} \cline{4-15} \cline{5-15} \cline{6-15} \cline{7-15} \cline{8-15} \cline{9-15} \cline{10-15} \cline{11-15} \cline{12-15} \cline{13-15} \cline{14-15} \cline{15-15} 
         &  & M5 & \multicolumn{4}{c|}{69.92, 85.85} & \multicolumn{4}{c|}{\textbf{0.066, 0.54}} & \multicolumn{4}{c|}{12.52}\tabularnewline
        \hline 
        \multirow{5}{*}{5} & \multirow{5}{*}{F2 / F1} & M1 & 20.35, 63.18 & 19.79, 53.10 & 34.71, 71.75 & \textbf{47.56, 86.53} & 0.095, 0.95 & 0.097, 0.97 & 0.098, 0.96 & \textbf{0.098, 0.94} & 21.26 & 25.57 & 19.95 & \textbf{11.91}\tabularnewline
        \cline{3-15} \cline{4-15} \cline{5-15} \cline{6-15} \cline{7-15} \cline{8-15} \cline{9-15} \cline{10-15} \cline{11-15} \cline{12-15} \cline{13-15} \cline{14-15} \cline{15-15} 
         &  & M2 & 10.57, 39.80 & 6.80, 36.18 & 18.16, 62.31 & \textbf{45.93, 86.17} & 0.100, 0.99 & 0.099, 0.99 & 0.099, 0.99 & \textbf{0.099, 0.90} & 30.97 & 31.76 & 22.85 & \textbf{12.18}\tabularnewline
        \cline{3-15} \cline{4-15} \cline{5-15} \cline{6-15} \cline{7-15} \cline{8-15} \cline{9-15} \cline{10-15} \cline{11-15} \cline{12-15} \cline{13-15} \cline{14-15} \cline{15-15} 
         &  & M3 & 7.61, 29.29 & 7.04, 28.83 & \textbf{9.51, 44.39} & 6.98, 31.19 & 0.099, 0.99 & 0.099, 0.99 & 0.099, 0.99 & \textbf{0.097, 0.97} & 37.39 & 31.57 & \textbf{27.23} & 36.59\tabularnewline
        \cline{3-15} \cline{4-15} \cline{5-15} \cline{6-15} \cline{7-15} \cline{8-15} \cline{9-15} \cline{10-15} \cline{11-15} \cline{12-15} \cline{13-15} \cline{14-15} \cline{15-15} 
         &  & M4 & 11.03, 36.78 & 3.25, 22.58 & \textbf{11.71, 41.62} & 3.89, 37.74 & \textbf{0.098, 0.98} & 0.099, 0.99 & 0.099, 0.99 & 0.100, 0.99 & 31.46 & 41.35 & 29.00 & \textbf{25.6}\tabularnewline
        \cline{3-15} \cline{4-15} \cline{5-15} \cline{6-15} \cline{7-15} \cline{8-15} \cline{9-15} \cline{10-15} \cline{11-15} \cline{12-15} \cline{13-15} \cline{14-15} \cline{15-15} 
         &  & M5 & \multicolumn{4}{c|}{23.75, 66.18} & \multicolumn{4}{c|}{0.0100, 1.00,} & \multicolumn{4}{c|}{22.32}\tabularnewline
        \hline 
        \end{tabular}

        }
        
    \vspace{0.1cm}
    \centering
    \begin{tabular}{|c|c|c|c|c|c|}
    \hline 
    Method & M1 & M2 & M3 & M4 & M5\tabularnewline
    \hline 
    \hline 
    Algorithm & 1D-Triplet-CNN-online & 1D-Triplet-CNN & xVector-PLDA & iVector-PLDA & RawNet2\tabularnewline
    \hline 
    \end{tabular} 
    \quad
    \vspace{0.1cm}
    \centering
    \begin{tabular}{|c|c|c|}
    \hline 
    Data Subset & F1 & F2\tabularnewline
    \hline 
    \hline 
    Noise Characteristics & Babble, R1,V1 & F16, R1, V1\tabularnewline
    \hline 
    \end{tabular} 
	
	\label{tab:Experiments_Fisher}
	\vspace{-0.2cm}
	\end{table*}

    \begin{table*}[t]
    \fontsize{6}{8}\selectfont
    \caption{Verification Results on the original and degraded, NIST SRE 2008, 2010, and 2018 datasets. The proposed DeepVOX features outperform the baseline features for a majority of methods and data partitions, across all the metrics.}
    \vspace{-0.2cm}
    \centering
    
    \scalebox{0.96}{
    \begin{tabular}{|c|c|c|c|c|c|c|c|c|c|c|c|c|c|c|}
    \hline 
    \multirow{2}{*}{\#} & \multirow{2}{*}{\makecell{Train set\\ / Test set}} & \multirow{2}{*}{Method} & \multicolumn{4}{c|}{TMR@FMR=\{1\%, 10\%\}} & \multicolumn{4}{c|}{minDCF (ptar=\{0.001, 0.01\})} & \multicolumn{4}{c|}{Equal Error Rate (EER, in \%)}\tabularnewline
    \cline{4-15} \cline{5-15} \cline{6-15} \cline{7-15} \cline{8-15} \cline{9-15} \cline{10-15} \cline{11-15} \cline{12-15} \cline{13-15} \cline{14-15} \cline{15-15} 
     &  &  & MFCC & LPC & \makecell{MFCC-\\LPC} & \makecell{Deep\\VOX} & MFCC & LPC & \makecell{MFCC-\\LPC} & \makecell{Deep\\VOX} & MFCC & LPC & \makecell{MFCC-\\LPC} & \makecell{Deep\\VOX}\tabularnewline
    \hline 
    \multirow{5}{*}{6} & \multirow{5}{*}{P1 / P1} & M1 & 55.21, 93.06 & 41.49, 87.25 & 52.50, 93.22 & \textbf{81.05, 97.63} &   0.097, 0.76 &   0.084, 0.84 &   0.095, 0.89 &   \textbf{0.081, 0.60} & 8.74 & 11.18 & 8.18 & \textbf{4.45}\tabularnewline
    \cline{3-15} \cline{4-15} \cline{5-15} \cline{6-15} \cline{7-15} \cline{8-15} \cline{9-15} \cline{10-15} \cline{11-15} \cline{12-15} \cline{13-15} \cline{14-15} \cline{15-15} 
     &  & M2 & 53.17, 89.12 & 49.17, 86.65 & 60.21, 93.36 & \textbf{81.37, 97.30} &   0.082, 0.82 &   0.085, 0.83 &   0.079, 0.76 &   \textbf{0.066, 0.59} & 10.55 & 11.62 & 8.34  & \textbf{4.77}\tabularnewline
    \cline{3-15} \cline{4-15} \cline{5-15} \cline{6-15} \cline{7-15} \cline{8-15} \cline{9-15} \cline{10-15} \cline{11-15} \cline{12-15} \cline{13-15} \cline{14-15} \cline{15-15} 
     &  & M3 & 25.20, 78.60 & 22.96, 76.47 & \textbf{24.00, 85.21} & 23.97, 78.72 &   0.099, 0.99 &   0.098, 0.98 &   \textbf{0.098, 0.98} &   0.099, 0.99 & 14.15 & 15.15 & \textbf{11.95} & 14.68\tabularnewline
    \cline{3-15} \cline{4-15} \cline{5-15} \cline{6-15} \cline{7-15} \cline{8-15} \cline{9-15} \cline{10-15} \cline{11-15} \cline{12-15} \cline{13-15} \cline{14-15} \cline{15-15} 
     &  & M4 & 48.70, 85.13 & 30.64, 78.20 & 42.16, 88.35 & \textbf{37.63, 96.12} &  \textbf{ 0.087, 0.87} &   0.097, 0.97 &   0.093, 0.93 &   0.094, 0.93 & 12.37 & 15.85 & 10.81 & \textbf{6.85} \tabularnewline
    \cline{3-15} \cline{4-15} \cline{5-15} \cline{6-15} \cline{7-15} \cline{8-15} \cline{9-15} \cline{10-15} \cline{11-15} \cline{12-15} \cline{13-15} \cline{14-15} \cline{15-15} 
     &  & M5 & \multicolumn{4}{c|}{81.62, 93.57} & \multicolumn{4}{c|}{0.047, 0.47} & \multicolumn{4}{c|}{7.53}\tabularnewline
    \hline 
    \multirow{5}{*}{7} & \multirow{5}{*}{P1 / P2} & M1 & 8.40, 24.93 & 7.58, 23.56 & \textbf{8.40, 24.47} & 4.84, 21.00 & 0.096, 0.96  & 0.098, 0.98   & \textbf{0.096, 0.96 }  & 0.098, 0.98  & \textbf{43.29} & 43.65 & 43.74 & 47.31\tabularnewline
    \cline{3-15} \cline{4-15} \cline{5-15} \cline{6-15} \cline{7-15} \cline{8-15} \cline{9-15} \cline{10-15} \cline{11-15} \cline{12-15} \cline{13-15} \cline{14-15} \cline{15-15} 
     &  & M2 & 2.28, 21.64 & 2.65, 18.54 & 4.13, \textbf{25.20} & \textbf{6.57}, 23.19 & 0.099, 0.99   & 0.099, 0.99   & 0.099, 0.99   & \textbf{0.098, 0.98 }  & 45.02 & 44.11 & \textbf{39.40} & 46.57\tabularnewline
    \cline{3-15} \cline{4-15} \cline{5-15} \cline{6-15} \cline{7-15} \cline{8-15} \cline{9-15} \cline{10-15} \cline{11-15} \cline{12-15} \cline{13-15} \cline{14-15} \cline{15-15} 
     &  & M3 & 3.01, \textbf{19.27} & 1.74, 15.62 & 2.10, 17.17 & \textbf{4.01, }19.17 & 0.099, 0.99   & 0.099, 0.99   & 0.099, 0.99   & \textbf{0.097, 0.97 }  & \textbf{43.84} & 46.39 & 45.57 & 46.66\tabularnewline
    \cline{3-15} \cline{4-15} \cline{5-15} \cline{6-15} \cline{7-15} \cline{8-15} \cline{9-15} \cline{10-15} \cline{11-15} \cline{12-15} \cline{13-15} \cline{14-15} \cline{15-15} 
     &  & M4 & 3.29, 16.35 & 3.74, 17.26 & 1.19, 10.14 & \textbf{3.37, 19.54} & \textbf{0.098, 0.98}   & 0.099, 0.99   & 0.099, 0.99   & 0.099, 0.99 & 44.75 & \textbf{44.29} & 47.40 & 46.30\tabularnewline
    \cline{3-15} \cline{4-15} \cline{5-15} \cline{6-15} \cline{7-15} \cline{8-15} \cline{9-15} \cline{10-15} \cline{11-15} \cline{12-15} \cline{13-15} \cline{14-15} \cline{15-15} 
     &  & M5 & \multicolumn{4}{c|}{0, 15.35} & \multicolumn{4}{c|}{0.100, 1.00} & \multicolumn{4}{c|}{44.46}\tabularnewline
    \hline 
    \multirow{5}{*}{8} & \multirow{5}{*}{P1 / P3} & M1 & 9.92, 32.07 & 6.73, 24.73 & \textbf{10.46, 32.09} & 8.06, 29.53 & 0.099, 0.99 & 0.099, 0.99 & \textbf{0.098, 0.98} & 0.099, 0.99 & 38.95 & 42.39 & \textbf{38.43} & 39.04\tabularnewline
    \cline{3-15} \cline{4-15} \cline{5-15} \cline{6-15} \cline{7-15} \cline{8-15} \cline{9-15} \cline{10-15} \cline{11-15} \cline{12-15} \cline{13-15} \cline{14-15} \cline{15-15} 
     &  & M2 & 8.45, 29.69 & 5.74, 22.99 & \textbf{9.75, 30.17} & 6.73, 26.27 & 0.099, 0.99 & 0.099, 0.99 & 0.099, 0.99 & \textbf{0.099, 0.99} & \textbf{38.98} & 42.67 & 39.78 & 40.30\tabularnewline
    \cline{3-15} \cline{4-15} \cline{5-15} \cline{6-15} \cline{7-15} \cline{8-15} \cline{9-15} \cline{10-15} \cline{11-15} \cline{12-15} \cline{13-15} \cline{14-15} \cline{15-15} 
     &  & M3 & 1.89, 15.44 & 1.47, 12.02 & 1.34, 13.95  & \textbf{4.41, 19.14} & 0.099, 0.99 & 0.099, 0.99 & 0.100, 1.00 & \textbf{0.099, 0.99} & 45.32 & 48.30 & 46.63 & \textbf{45.24}\tabularnewline
    \cline{3-15} \cline{4-15} \cline{5-15} \cline{6-15} \cline{7-15} \cline{8-15} \cline{9-15} \cline{10-15} \cline{11-15} \cline{12-15} \cline{13-15} \cline{14-15} \cline{15-15} 
     &  & M4 & 5.35, 24.57 & 1.02, 12.04 & 4.18, 20.64 & \textbf{5.72, 24.57} & 0.099, 0.99 & 0.099, 0.99 & 0.100, 1.00 & \textbf{0.099, 0.99} & \textbf{40.16} & 47.98 & 42.32 & 41.20\tabularnewline
    \cline{3-15} \cline{4-15} \cline{5-15} \cline{6-15} \cline{7-15} \cline{8-15} \cline{9-15} \cline{10-15} \cline{11-15} \cline{12-15} \cline{13-15} \cline{14-15} \cline{15-15} 
     &  & M5 & \multicolumn{4}{c|}{2.50, 21.54} & \multicolumn{4}{c|}{0.100, 1.00} & \multicolumn{4}{c|}{41.36}\tabularnewline
    \hline 
    \multirow{5}{*}{9} & \multirow{5}{*}{P4 / P4} & M1 & 35.28, 83.49 & 38.01, 81.19 & 35.25, 86.86 & 70.16, 94.46 & 0.088, 0.88 & 0.090, 0.90 & 0.096, 0.96 & \textbf{0.058, 0.58} & 12.47  & 13.44 & 11.40 & 7.44\tabularnewline
    \cline{3-15} \cline{4-15} \cline{5-15} \cline{6-15} \cline{7-15} \cline{8-15} \cline{9-15} \cline{10-15} \cline{11-15} \cline{12-15} \cline{13-15} \cline{14-15} \cline{15-15} 
     &  & M2 & 39.28, 84.26  & 35.48, 80.49 & 53.92, 90.00 & \textbf{69.22, 95.36} & 0.090, 0.90 & 0.097, 0.94 & 0.075, 0.75 & \textbf{0.073, 0.68} & 12.94 & 14.24 & 10.00 & \textbf{7.10}\tabularnewline
    \cline{3-15} \cline{4-15} \cline{5-15} \cline{6-15} \cline{7-15} \cline{8-15} \cline{9-15} \cline{10-15} \cline{11-15} \cline{12-15} \cline{13-15} \cline{14-15} \cline{15-15} 
     &  & M3 & 22.44, 75.09  & 20.81, 65.42 & 23.64, 72.66 & \textbf{24.17, 63.72} & 0.099, 0.99 & \textbf{0.095, 0.95} & 0.099, 0.99 & 0.099, 0.99 & \textbf{15.24} & 19.24 & 16.17 & 21.19\tabularnewline
    \cline{3-15} \cline{4-15} \cline{5-15} \cline{6-15} \cline{7-15} \cline{8-15} \cline{9-15} \cline{10-15} \cline{11-15} \cline{12-15} \cline{13-15} \cline{14-15} \cline{15-15} 
     &  & M4 & \textbf{39.57}, 82.87  & 31.58, 72.46 & 11.70, 41.25 & 31.30, \textbf{83.67} & 0.099, 0.99 & \textbf{0.093, 0.93} & 0.099, 0.99 & 0.099, 0.99 & 13.53 & 17.34 & 28.34 & \textbf{12.31}\tabularnewline
    \cline{3-15} \cline{4-15} \cline{5-15} \cline{6-15} \cline{7-15} \cline{8-15} \cline{9-15} \cline{10-15} \cline{11-15} \cline{12-15} \cline{13-15} \cline{14-15} \cline{15-15} 
     &  & M5 & \multicolumn{4}{c|}{67.85, 89.68} & \multicolumn{4}{c|}{0.091, 0.66} & \multicolumn{4}{c|}{10.24}\tabularnewline
    \hline 
    \multirow{5}{*}{10} & \multirow{5}{*}{P5 / P5} & M1 & 26.70, 68.28 & 22.21, 61.86 & 20.01, 59.52 & \textbf{62.40, 95.19} & 0.097, 0.97 & 0.098, 0.98 & 0.093, 0.93 & \textbf{0.080, 0.80} & 19.63 & 21.24 & 22.64 & \textbf{7.25}\tabularnewline
    \cline{3-15} \cline{4-15} \cline{5-15} \cline{6-15} \cline{7-15} \cline{8-15} \cline{9-15} \cline{10-15} \cline{11-15} \cline{12-15} \cline{13-15} \cline{14-15} \cline{15-15} 
     &  & M2 & 35.34, 75.31 & 29.39, 73.41 & 43.02, 84.97 & \textbf{71.36, 94.68} & 0.097, 0.97 & 0.095, 0.95 & 0.092, 0.89 & \textbf{0.067, 0.64} & 16.29 & 17.19  & 12.67 & \textbf{6.99}\tabularnewline
    \cline{3-15} \cline{4-15} \cline{5-15} \cline{6-15} \cline{7-15} \cline{8-15} \cline{9-15} \cline{10-15} \cline{11-15} \cline{12-15} \cline{13-15} \cline{14-15} \cline{15-15} 
     &  & M3 & 17.15, 58.77 & 17.58, 54.97 & 22.03, 66.63 & \textbf{36.20, 77.43} & 0.096, 0.96 & 0.097, 0.97 & 0.098, 0.98 & \textbf{0.084, 0.84} & 20.88 & 22.28  & 19.27 & \textbf{15.57}\tabularnewline
    \cline{3-15} \cline{4-15} \cline{5-15} \cline{6-15} \cline{7-15} \cline{8-15} \cline{9-15} \cline{10-15} \cline{11-15} \cline{12-15} \cline{13-15} \cline{14-15} \cline{15-15} 
     &  & M4 & 22.73, 60.57 & 6.10, 28.74  & 4.45, 23.00  & \textbf{27.30, 86.43} & \textbf{0.095, 0.95} & 0.098, 0.98 & 0.099, 0.99 & 0.099, 0.99 & 21.13 & 36.96  & 37.89 & \textbf{11.15}\tabularnewline
    \cline{3-15} \cline{4-15} \cline{5-15} \cline{6-15} \cline{7-15} \cline{8-15} \cline{9-15} \cline{10-15} \cline{11-15} \cline{12-15} \cline{13-15} \cline{14-15} \cline{15-15} 
     &  & M5 & \multicolumn{4}{c|}{63.15, 90.81} & \multicolumn{4}{c|}{0.071, 0.71} & \multicolumn{4}{c|}{9.50}\tabularnewline
    \hline 
    \multirow{5}{*}{11} & \multirow{5}{*}{P4 / P5} & M1 & 8.00, 34.59 & 9.65, 36.92 & 8.83, 38.86 & \textbf{15.46, 58.06} & 0.099, 0.99 & \textbf{0.098, 0.98} & 0.099, 0.99 & 0.099, 0.99 & 31.97 & 33.55 & 29.49 & \textbf{22.46}\tabularnewline
    \cline{3-15} \cline{4-15} \cline{5-15} \cline{6-15} \cline{7-15} \cline{8-15} \cline{9-15} \cline{10-15} \cline{11-15} \cline{12-15} \cline{13-15} \cline{14-15} \cline{15-15} 
     &  & M2 & 14.42, 49.12 & 14.78, 47.04 & \textbf{18.41, 55.36} & 11.37, 47.75 & 0.099, 0.99 & 0.099, 0.99 & \textbf{0.097, 0.97} & 0.099, 0.99 & 26.01 & 28.13 & \textbf{23.29} & 26.08\tabularnewline
    \cline{3-15} \cline{4-15} \cline{5-15} \cline{6-15} \cline{7-15} \cline{8-15} \cline{9-15} \cline{10-15} \cline{11-15} \cline{12-15} \cline{13-15} \cline{14-15} \cline{15-15} 
     &  & M3 & 7.71, 31.97  & 8.22, 35.06  & 14.53, \textbf{53.00} & \textbf{15.97}, 40.98 & 0.097, 0.97 & 0.099, 0.99 & \textbf{0.096, 0.96} & 0.099, 0.99 & 34.95 & 31.43 & \textbf{22.46} & 31.83\tabularnewline
    \cline{3-15} \cline{4-15} \cline{5-15} \cline{6-15} \cline{7-15} \cline{8-15} \cline{9-15} \cline{10-15} \cline{11-15} \cline{12-15} \cline{13-15} \cline{14-15} \cline{15-15} 
     &  & M4 & 6.03, 27.92  & 3.70, 20.85  & 2.22, 15.97  & \textbf{6.09}, \textbf{28.34} & 0.099, 0.99 & 0.099, 0.99 & 0.099, 0.99 & \textbf{0.099, 0.99} & 35.24 & 41.51 & 43.24 & \textbf{34.76}\tabularnewline
    \cline{3-15} \cline{4-15} \cline{5-15} \cline{6-15} \cline{7-15} \cline{8-15} \cline{9-15} \cline{10-15} \cline{11-15} \cline{12-15} \cline{13-15} \cline{14-15} \cline{15-15} 
     &  & M5 & \multicolumn{4}{c|}{13.85, 47.32} & \multicolumn{4}{c|}{0.099, 0.99} & \multicolumn{4}{c|}{25.97}\tabularnewline
    \hline 
    \multirow{5}{*}{12} & \multirow{5}{*}{P5 / P4} & M1 & 19.14, 58.55 & 7.10, 40.01 & 19.14, 58.55 & \textbf{35.05, 78.74} & 0.0947, 0.94 & 0.0995, 0.99 & 0.0986, 0.98 & \textbf{0.0945, 0.94} & 22.67 & 28.74 & 22.67 & \textbf{15.22}\tabularnewline
    \cline{3-15} \cline{4-15} \cline{5-15} \cline{6-15} \cline{7-15} \cline{8-15} \cline{9-15} \cline{10-15} \cline{11-15} \cline{12-15} \cline{13-15} \cline{14-15} \cline{15-15} 
     &  & M2 & 11.34, 37.08 & 4.57 , 27.84  & 19.34, 56.59  & \textbf{21.09, 68.32} & 0.0972, 0.97 & 0.0998, 0.99 & \textbf{0.0972, 0.97} & 0.0976, 0.97 & 32.28 & 37.55  & 23.61 & \textbf{18.29}\tabularnewline
    \cline{3-15} \cline{4-15} \cline{5-15} \cline{6-15} \cline{7-15} \cline{8-15} \cline{9-15} \cline{10-15} \cline{11-15} \cline{12-15} \cline{13-15} \cline{14-15} \cline{15-15} 
     &  & M3 & 12.17, 45.38 & 12.77, 52.82  & \textbf{14.54, 47.35 } & 12.98, 40.42 & 0.0999, 0.99 & 0.0986, 0.98 & 0.0988, 0.98 & \textbf{0.0981, 0.98} & \textbf{27.54} & 22.87  & 27.64 & 31.01\tabularnewline
    \cline{3-15} \cline{4-15} \cline{5-15} \cline{6-15} \cline{7-15} \cline{8-15} \cline{9-15} \cline{10-15} \cline{11-15} \cline{12-15} \cline{13-15} \cline{14-15} \cline{15-15} 
     &  & M4 & \textbf{9.50}, 36.15  & 3.60, 21.51   & 3.33, 20.21   & 7.54, \textbf{37.95} & \textbf{0.0990, 0.99} & 0.0995, 0.99 & 0.0999, 0.99 & 0.0997, 0.99 & 34.11 & 40.88  & 41.71 & \textbf{32.0}\tabularnewline
    \cline{3-15} \cline{4-15} \cline{5-15} \cline{6-15} \cline{7-15} \cline{8-15} \cline{9-15} \cline{10-15} \cline{11-15} \cline{12-15} \cline{13-15} \cline{14-15} \cline{15-15} 
     &  & M5 & \multicolumn{4}{c|}{9.04, 41.75} & \multicolumn{4}{c|}{0.100, 0.99} & \multicolumn{4}{c|}{27.16}\tabularnewline
    \hline 
    \end{tabular}
    }
    \vspace{0.2cm}
    \centering
    \begin{tabular}{|c|c|c|c|c|c|}
    \hline 
    Method & M1 & M2 & M3 & M4 & M5\tabularnewline
    \hline 
    \hline 
    Algorithm & \makecell{1D-Triplet-CNN-\\online} & 1D-Triplet-CNN & \makecell{xVector-\\PLDA} & \makecell{iVector-\\PLDA} & RawNet2\tabularnewline
    \hline 
    \end{tabular}
    \quad
    \vspace{0.2cm}
    \centering
    \begin{tabular}{|c|c|c|c|c|c|}
    \hline 
    Data Subset & P1 & P2 & P3 & P4 & P5\tabularnewline
    \hline 
    \hline 
    Noise Type & NIST SRE 08 & NIST SRE 10 & NIST SRE 18 & P1 + Babble & P1 + F16\tabularnewline
    \hline 
    \end{tabular}
            
    \label{tab:Experiments_SRE}
    \vspace{-0.6cm}
    \end{table*}

    \begin{table*}[t]
    \fontsize{6}{8}\selectfont
    \caption{Verification Results on multi-lingual speakers from the NIST SRE 2008 dataset. The proposed DeepVOX features outperform the baseline features for a majority of methods and data partitions, across all the metrics.}
    \vspace{-0.2cm}
    \centering
    
    \scalebox{0.96}{
    \begin{tabular}{|c|c|c|c|c|c|c|c|c|c|c|c|c|c|c|}
    \hline 
    \multirow{2}{*}{\#} & \multirow{2}{*}{\makecell{Train set\\ / Test set}} & \multirow{2}{*}{Method} & \multicolumn{4}{c|}{TMR@FMR=\{1\%, 10\%\}} & \multicolumn{4}{c|}{minDCF (ptar=\{0.001, 0.01\})} & \multicolumn{4}{c|}{Equal Error Rate (EER, in \%)}\tabularnewline
    \cline{4-15} \cline{5-15} \cline{6-15} \cline{7-15} \cline{8-15} \cline{9-15} \cline{10-15} \cline{11-15} \cline{12-15} \cline{13-15} \cline{14-15} \cline{15-15} 
     &  &  & MFCC & LPC & \makecell{MFCC-\\LPC} & \makecell{Deep\\VOX} & MFCC & LPC & \makecell{MFCC-\\LPC} & \makecell{Deep\\VOX} & MFCC & LPC & \makecell{MFCC-\\LPC} & \makecell{Deep\\VOX}\tabularnewline
    \hline 
    \multirow{5}{*}{13} & \multirow{5}{*}{L1 / L1} & M1 & 47.88, 85.30 & 45.26, 85.26 & 55.94, 90.34 & \textbf{80.30, 99.16} & 0.095, 0.89 & 0.088, 0.85 & 0.092, 0.85 & \textbf{0.062, 0.56} & 11.90 & 12.58 & 9.80 & \textbf{3.98}\tabularnewline
    \cline{3-15} \cline{4-15} \cline{5-15} \cline{6-15} \cline{7-15} \cline{8-15} \cline{9-15} \cline{10-15} \cline{11-15} \cline{12-15} \cline{13-15} \cline{14-15} \cline{15-15} 
     &  & M2 & 33.44, 79.70 & 36.34, 77.88 & 47.54, 86.70 & \textbf{77.60, 99.30} & 0.094, 0.91 & 0.089, 0.89 & 0.093, 0.90 & \textbf{0.075, 0.63} & 13.92  & 14.78 & 11.30 & \textbf{4.32}\tabularnewline
    \cline{3-15} \cline{4-15} \cline{5-15} \cline{6-15} \cline{7-15} \cline{8-15} \cline{9-15} \cline{10-15} \cline{11-15} \cline{12-15} \cline{13-15} \cline{14-15} \cline{15-15} 
     &  & M3 & 47.88, 85.30 & 45.26, 85.26 & 55.94, 90.34 & \textbf{72.84, 97.94} & 0.090, 0.90 & 0.091, 0.87 & 0.090, 0.81 & \textbf{0.089, 0.66} & 11.90  & 12.58 & 9.80  & \textbf{5.64}\tabularnewline
    \cline{3-15} \cline{4-15} \cline{5-15} \cline{6-15} \cline{7-15} \cline{8-15} \cline{9-15} \cline{10-15} \cline{11-15} \cline{12-15} \cline{13-15} \cline{14-15} \cline{15-15} 
     &  & M4 & 46.86, 83.58 & 41.46, 83.24 & 60.06, 93.76 & \textbf{76.54, 98.42} & 0.094, 0.88 & 0.098, 0.87 & 0.078, 0.75 & \textbf{0.089, 0.65} & 12.74  & 12.96 & 8.14  & \textbf{5.00}\tabularnewline
    \cline{3-15} \cline{4-15} \cline{5-15} \cline{6-15} \cline{7-15} \cline{8-15} \cline{9-15} \cline{10-15} \cline{11-15} \cline{12-15} \cline{13-15} \cline{14-15} \cline{15-15} 
     &  & M5 & \multicolumn{4}{c|}{71.54, 95.64} & \multicolumn{4}{c|}{0.084, 0.75} & \multicolumn{4}{c|}{6.86}\tabularnewline
    \hline 
    \multirow{5}{*}{14} & \multirow{5}{*}{L1 / L2} & M1 & 39.52, 82.03 & 43.40, 79.60 & 47.95, 86.53 & \textbf{77.26, 97.87} & 0.096, 0.88 & 0.089, 0.86 & 0.083, 0.78 & \textbf{0.063, 0.60} & 13.56 & 14.7 & 11.61 & \textbf{5.04}\tabularnewline
    \cline{3-15} \cline{4-15} \cline{5-15} \cline{6-15} \cline{7-15} \cline{8-15} \cline{9-15} \cline{10-15} \cline{11-15} \cline{12-15} \cline{13-15} \cline{14-15} \cline{15-15} 
     &  & M2 & 32.39, 74.86 & 35.80, 75.04 & 41.67, 83.09 & \textbf{66.91, 97.70} & 0.097, 0.97 & 0.095, 0.91 & 0.089, 0.84 & \textbf{0.075, 0.64} & 16.21 & 16.77 & 13.1 & \textbf{5.17}\tabularnewline
    \cline{3-15} \cline{4-15} \cline{5-15} \cline{6-15} \cline{7-15} \cline{8-15} \cline{9-15} \cline{10-15} \cline{11-15} \cline{12-15} \cline{13-15} \cline{14-15} \cline{15-15} 
     &  & M3 & 39.52, 82.03 & 43.40, 79.60 & 47.90, 86.50 & \textbf{72.49, 97.57} & 0.095, 0.92 & 0.094, 0.83 & 0.090, 0.90 & \textbf{0.079, 0.66} & 13.56 & 14.7 & 11.61 & \textbf{5.96}\tabularnewline
    \cline{3-15} \cline{4-15} \cline{5-15} \cline{6-15} \cline{7-15} \cline{8-15} \cline{9-15} \cline{10-15} \cline{11-15} \cline{12-15} \cline{13-15} \cline{14-15} \cline{15-15} 
     &  & M4 & 40.48, 80.17 & 39.58, 78.17 & 56.23, 88.30 & \textbf{77.64, 98.39} & 0.098, 0.96 & 0.085, 0.85 & 0.090, 0.78 & \textbf{0.061, 0.55} & 14.1 & 15.02 & 10.74 & \textbf{4.78}\tabularnewline
    \cline{3-15} \cline{4-15} \cline{5-15} \cline{6-15} \cline{7-15} \cline{8-15} \cline{9-15} \cline{10-15} \cline{11-15} \cline{12-15} \cline{13-15} \cline{14-15} \cline{15-15} 
     &  & M5 & \multicolumn{4}{c|}{67.30, 93.18} & \multicolumn{4}{c|}{0.091, 0.69} & \multicolumn{4}{c|}{8.03}\tabularnewline
    \hline 
    \multirow{5}{*}{15} & \multirow{5}{*}{L1 / L3} & M1 & 29.06, 70.46 & 28.10, 64.68 & 33.14, 74.82 & \textbf{62.24, 88.82} & 0.095, 0.94 & 0.098, 0.97 & 0.092, 0.90 & \textbf{0.081, 0.74} & 17.64 & 21.26 & 16.52 & \textbf{10.72}\tabularnewline
    \cline{3-15} \cline{4-15} \cline{5-15} \cline{6-15} \cline{7-15} \cline{8-15} \cline{9-15} \cline{10-15} \cline{11-15} \cline{12-15} \cline{13-15} \cline{14-15} \cline{15-15} 
     &  & M2 & 25.78, 64.28 & 18.38, 57.04 & 30.82, 67.60 & \textbf{55.96, 89.02} & 0.097, 0.97 & 0.098, 0.98 & 0.094, 0.92 & \textbf{0.098, 0.88} & 20.30 & 23.04 & 18.80 & \textbf{10.60}\tabularnewline
    \cline{3-15} \cline{4-15} \cline{5-15} \cline{6-15} \cline{7-15} \cline{8-15} \cline{9-15} \cline{10-15} \cline{11-15} \cline{12-15} \cline{13-15} \cline{14-15} \cline{15-15} 
     &  & M3 & 29.06, 70.46 & 28.10, 64.68 & 47.95, 86.53 & \textbf{54.42, 87.88} & 0.093, 0.93 & 0.097, 0.97 & 0.094, 0.94 & \textbf{0.091, 0.84} & 17.64 & 21.26 & 11.61 & \textbf{11.20}\tabularnewline
    \cline{3-15} \cline{4-15} \cline{5-15} \cline{6-15} \cline{7-15} \cline{8-15} \cline{9-15} \cline{10-15} \cline{11-15} \cline{12-15} \cline{13-15} \cline{14-15} \cline{15-15} 
     &  & M4 & 26.30, 66.30 & 20.72, 61.40 & 38.70, 74.80 & \textbf{56.90, 88.06} & 0.094, 0.94 & 0.096, 0.96 & 0.092, 0.89 & \textbf{0.098, 0.86} & 19.52 & 22.00 & 16.86 & \textbf{11.16}\tabularnewline
    \cline{3-15} \cline{4-15} \cline{5-15} \cline{6-15} \cline{7-15} \cline{8-15} \cline{9-15} \cline{10-15} \cline{11-15} \cline{12-15} \cline{13-15} \cline{14-15} \cline{15-15} 
     &  & M5 & \multicolumn{4}{c|}{50.40, 81.44} & \multicolumn{4}{c|}{0.090, 0.85} & \multicolumn{4}{c|}{14.58}\tabularnewline
    \hline 
    \end{tabular}
    
    }
    
    \vspace{0.1cm}
    \centering
    \begin{tabular}{|c|c|c|c|c|c|}
    \hline 
    Method & M1 & M2 & M3 & M4 & M5\tabularnewline
    \hline 
    \hline 
    Algorithm & 1D-Triplet-CNN-online & 1D-Triplet-CNN & xVector-PLDA & iVector-PLDA & RawNet2\tabularnewline
    \hline 
    \end{tabular}
    \quad
    \vspace{0.1cm}
    \centering
    \begin{tabular}{|c|c|c|c|}
    \hline 
    Data Subset & L1 & L2 & L3\tabularnewline
    \hline 
    \hline 
    Language Characteristics & English Only & Multi-Lingual & Cross-Lingual\tabularnewline
    \hline 
    \end{tabular}
            
    \label{tab:Experiments_SRE_lang}
    \vspace{-0.4cm}
    \end{table*}
	
	\begin{table*}[htp]
    \fontsize{6}{8}\selectfont
    \caption{Verification Results under varying audio length on the NIST SRE 2008 dataset. The proposed DeepVOX features outperform the baseline features for a majority of methods and data partitions, across all the metrics.}
    \vspace{-0.2cm}
    \centering
    
    \scalebox{1.0}{
    \begin{tabular}{|c|c|c|c|c|c|c|c|c|c|c|c|c|c|}
    \hline 
    \multirow{2}{*}{\makecell{Length\\(secs)}} & \multirow{2}{*}{Method} & \multicolumn{4}{c|}{TMR@FMR=\{1\%, 10\%\}} & \multicolumn{4}{c|}{minDCF (ptar=\{0.001,0.01\})} & \multicolumn{4}{c|}{Equal Error Rate (EER, in \%)}\tabularnewline
    \cline{3-14} \cline{4-14} \cline{5-14} \cline{6-14} \cline{7-14} \cline{8-14} \cline{9-14} \cline{10-14} \cline{11-14} \cline{12-14} \cline{13-14} \cline{14-14} 
     &  & MFCC & LPC & \makecell{MFCC-\\LPC} & \makecell{Deep\\VOX} & MFCC & LPC & \makecell{MFCC-\\LPC} & \makecell{Deep\\VOX} & MFCC & LPC & \makecell{MFCC-\\LPC} & \makecell{Deep\\VOX}\tabularnewline
    \hline 
    \multirow{5}{*}{3.5} & M1 & 55.20, 93.05 & 42.28, 86.84 & 49.43, 92.32 & \textbf{80.59, 97.63} & 0.094, 0.78 & 0.087, 0.85 & 0.090, 0.83 & \textbf{0.079, 0.62} & 8.74 & 11.61 &  8.57 & \textbf{4.52}\tabularnewline
    \cline{2-14} \cline{3-14} \cline{4-14} \cline{5-14} \cline{6-14} \cline{7-14} \cline{8-14} \cline{9-14} \cline{10-14} \cline{11-14} \cline{12-14} \cline{13-14} \cline{14-14} 
     & M2 & 59.61, 90.72 & 52.67, 88.58 & 65.99, 94.53 & \textbf{79.87, 97.74} & 0.088, 0.72 & 0.083, 0.79 & 0.080,\textbf{ 0.69} & \textbf{0.076}, 0.71 & 9.65 & 10.71 &  7.64 & \textbf{4.59}\tabularnewline
    \cline{2-14} \cline{3-14} \cline{4-14} \cline{5-14} \cline{6-14} \cline{7-14} \cline{8-14} \cline{9-14} \cline{10-14} \cline{11-14} \cline{12-14} \cline{13-14} \cline{14-14} 
     & M3 & 27.10, 78.81 & 19.26, 74.70 & 24.57, \textbf{81.21} & \textbf{29.81}, 77.39 & 0.099, 0.99 & 0.099, 0.99 & \textbf{0.097, 0.97} & 0.099, 0.99 & 14.39 & 15.45 & \textbf{12.92} & 15.24\tabularnewline
    \cline{2-14} \cline{3-14} \cline{4-14} \cline{5-14} \cline{6-14} \cline{7-14} \cline{8-14} \cline{9-14} \cline{10-14} \cline{11-14} \cline{12-14} \cline{13-14} \cline{14-14} 
     & M4 & 44.89, 78.60 & 25.50, 75.70 & 37.48, 86.28 & \textbf{51.34, 95.87} & 0.092, 0.92 & 0.098, 0.98 & 0.096, 0.96 & \textbf{0.078, 0.78} & 14.82 & 16.49 & 11.92 & \textbf{6.9}\tabularnewline
    \cline{2-14} \cline{3-14} \cline{4-14} \cline{5-14} \cline{6-14} \cline{7-14} \cline{8-14} \cline{9-14} \cline{10-14} \cline{11-14} \cline{12-14} \cline{13-14} \cline{14-14} 
     & M5 & \multicolumn{4}{c|}{82.23, 93.86} & \multicolumn{4}{c|}{0.056, 0.47} & \multicolumn{4}{c|}{7.39}\tabularnewline
    \hline 
    \multirow{5}{*}{3.0} & M1 & 55.90, 91.02 & 41.48, 85.14 & 52.80, 92.15 & \textbf{80.05, 97.48} & 0.093, 0.80 & 0.089, 0.88 & 0.094, 0.83 & \textbf{0.077, 0.62} & 9.47 & 12.04 &  8.87 & \textbf{4.73}\tabularnewline
    \cline{2-14} \cline{3-14} \cline{4-14} \cline{5-14} \cline{6-14} \cline{7-14} \cline{8-14} \cline{9-14} \cline{10-14} \cline{11-14} \cline{12-14} \cline{13-14} \cline{14-14} 
     & M2 & 57.58, 90.22 & 50.63, 88.58 & 65.49, 94.13 & \textbf{76.89, 97.74}  & \textbf{0.075}, 0.74 & 0.085, 0.77 & 0.078, 0.70 & 0.083\textbf{, 0.64} & 9.85 & 10.75 &  7.71 & \textbf{4.63}\tabularnewline
    \cline{2-14} \cline{3-14} \cline{4-14} \cline{5-14} \cline{6-14} \cline{7-14} \cline{8-14} \cline{9-14} \cline{10-14} \cline{11-14} \cline{12-14} \cline{13-14} \cline{14-14} 
     & M3 & 24.63, 76.50 & 18.46, 71.16 & 23.66, \textbf{79.11} & \textbf{28.99}, 75.60 & \textbf{0.098, 0.97} & 0.099, 0.99 & 0.098, 0.98 & 0.099, 0.99 & 15.15 & 17.12 & \textbf{14.12} & 15.89\tabularnewline
    \cline{2-14} \cline{3-14} \cline{4-14} \cline{5-14} \cline{6-14} \cline{7-14} \cline{8-14} \cline{9-14} \cline{10-14} \cline{11-14} \cline{12-14} \cline{13-14} \cline{14-14} 
     & M4 & 41.62, 77.27 & 25.03, 71.50 & 35.11, 84.71 & \textbf{51.66, 95.19} & 0.093, 0.92 & 0.098, 0.98 & 0.096, 0.96 & \textbf{0.080, 0.80} & 16.19 & 17.86 & 12.65 & \textbf{7.03}\tabularnewline
    \cline{2-14} \cline{3-14} \cline{4-14} \cline{5-14} \cline{6-14} \cline{7-14} \cline{8-14} \cline{9-14} \cline{10-14} \cline{11-14} \cline{12-14} \cline{13-14} \cline{14-14} 
     & M5 & \multicolumn{4}{c|}{81.16, 94.15} & \multicolumn{4}{c|}{0.046, 0.46} & \multicolumn{4}{c|}{7.28}\tabularnewline
    \hline 
    \multirow{5}{*}{2.5} & M1 & 54.17, 89.19 & 41.98, 85.41 & 54.33, 91.78 & \textbf{77.11, 97.31} & 0.090, 0.82 & 0.087, 0.87 & 0.091, 0.78 & \textbf{0.059, 0.59} & 10.04 & 12.24 & 9.17 & \textbf{5.10}\tabularnewline
    \cline{2-14} \cline{3-14} \cline{4-14} \cline{5-14} \cline{6-14} \cline{7-14} \cline{8-14} \cline{9-14} \cline{10-14} \cline{11-14} \cline{12-14} \cline{13-14} \cline{14-14} 
     & M2 & 54.44, 89.95 & 47.50, 88.15 & 66.86, 94.23 & \textbf{74.56, 97.34} & 0.080, 0.80 & 0.081, 0.81 & 0.086, 0.73 & \textbf{0.071, 0.61} & 10.01 & 11.11 & 7.74 & \textbf{5.10}\tabularnewline
    \cline{2-14} \cline{3-14} \cline{4-14} \cline{5-14} \cline{6-14} \cline{7-14} \cline{8-14} \cline{9-14} \cline{10-14} \cline{11-14} \cline{12-14} \cline{13-14} \cline{14-14} 
     & M3 & \textbf{39.92,} 70.83 & 20.23, 67.49 & 31.98, \textbf{82.04} & 28.88, 72.37 & \textbf{0.097, 0.97} & 0.099, 0.99 & 0.099, 0.99 & 0.099, 0.99 & 17.76 & 19.93 & \textbf{13.79} & 17.22\tabularnewline
    \cline{2-14} \cline{3-14} \cline{4-14} \cline{5-14} \cline{6-14} \cline{7-14} \cline{8-14} \cline{9-14} \cline{10-14} \cline{11-14} \cline{12-14} \cline{13-14} \cline{14-14} 
     & M4 & 20.46, 69.96 & 16.79, 66.59 & 24.13, 75.33 & \textbf{49.73, 94.90} & 0.094, 0.87 & 0.098, 0.98 & 0.095, 0.95 & \textbf{0.079, 0.78} & 17.09 & 18.79 & 15.32 & \textbf{7.60}\tabularnewline
    \cline{2-14} \cline{3-14} \cline{4-14} \cline{5-14} \cline{6-14} \cline{7-14} \cline{8-14} \cline{9-14} \cline{10-14} \cline{11-14} \cline{12-14} \cline{13-14} \cline{14-14} 
     & M5 & \multicolumn{4}{c|}{77.03, 93.21} & \multicolumn{4}{c|}{0.063, 0.51} & \multicolumn{4}{c|}{8.14}\tabularnewline
    \hline 
    \multirow{5}{*}{2.0} & M1 & 51.73, 86.41 & 42.05, 83.84 & 51.26, 89.68 & \textbf{74.74, 96.91} & 0.090, 0.80 & 0.092, 0.87 & 0.087, 0.84 & \textbf{0.075, 0.68} & 11.34 & 13.08 & 10.14 & \textbf{5.45}\tabularnewline
    \cline{2-14} \cline{3-14} \cline{4-14} \cline{5-14} \cline{6-14} \cline{7-14} \cline{8-14} \cline{9-14} \cline{10-14} \cline{11-14} \cline{12-14} \cline{13-14} \cline{14-14} 
     & M2 & 55.77, 87.98 & 48.20, 85.78 & 60.01, 93.16 & \textbf{71.91, 97.24} & 0.085, 0.77 & 0.085\textbf{, 0.72} & 0.075, 0.75 & \textbf{0.075}, 0.75 & 10.81 & 12.18 & 8.28  & \textbf{5.53}\tabularnewline
    \cline{2-14} \cline{3-14} \cline{4-14} \cline{5-14} \cline{6-14} \cline{7-14} \cline{8-14} \cline{9-14} \cline{10-14} \cline{11-14} \cline{12-14} \cline{13-14} \cline{14-14} 
     & M3 & 17.82, 61.58 & 13.68, 57.38 & 20.69, 66.62 & \textbf{23.28, 68.17} & \textbf{0.098, 0.98} & 0.098, 0.98 & 0.099, 0.99 & 0.099, 0.99 & 20.46 & 21.83 & \textbf{18.32} & 19.66\tabularnewline
    \cline{2-14} \cline{3-14} \cline{4-14} \cline{5-14} \cline{6-14} \cline{7-14} \cline{8-14} \cline{9-14} \cline{10-14} \cline{11-14} \cline{12-14} \cline{13-14} \cline{14-14} 
     & M4 & 30.77, 66.99 & 17.69, 59.78 & 24.73, 78.14 & \textbf{44.31, 93.72} & 0.097, 0.95 & 0.097, 0.97 & 0.097, 0.97 & \textbf{0.090, 0.89} & 20.43 & 22.50 & 15.29 & \textbf{8.14}\tabularnewline
    \cline{2-14} \cline{3-14} \cline{4-14} \cline{5-14} \cline{6-14} \cline{7-14} \cline{8-14} \cline{9-14} \cline{10-14} \cline{11-14} \cline{12-14} \cline{13-14} \cline{14-14} 
     &  & \multicolumn{4}{c|}{69.86, 89.84} & \multicolumn{4}{c|}{0.068, 0.66} & \multicolumn{4}{c|}{10.08}\tabularnewline
    \hline 
    \multirow{5}{*}{1.5} & M1 & 44.89, 82.17 & 36.21, 77.77 & 45.52, 86.21 & \textbf{71.33, 96.30 } & 0.095, 0.91 & 0.088, 0.88 & 0.086, 0.85 & \textbf{0.085, 0.63} & 13.71 & 15.08 & 11.71 & \textbf{6.03}\tabularnewline
    \cline{2-14} \cline{3-14} \cline{4-14} \cline{5-14} \cline{6-14} \cline{7-14} \cline{8-14} \cline{9-14} \cline{10-14} \cline{11-14} \cline{12-14} \cline{13-14} \cline{14-14} 
     & M2 & 45.56, 86.42 & 49.70, 84.95 & 56.11, 91.66 & \textbf{63.08, 96.27} & 0.093, 0.88 & 0.092, 0.85 & 0.085, 0.79 & \textbf{0.082, 0.72} & 11.75 & 12.25 & 9.01  & \textbf{6.17}\tabularnewline
    \cline{2-14} \cline{3-14} \cline{4-14} \cline{5-14} \cline{6-14} \cline{7-14} \cline{8-14} \cline{9-14} \cline{10-14} \cline{11-14} \cline{12-14} \cline{13-14} \cline{14-14} 
     & M3 & 14.59, 52.00 & 11.62, 47.80 & 15.99, 57.01 & \textbf{17.68, 57.98} & 0.098, 0.98 & 0.099, 0.99 & \textbf{0.097, 0.97} & 0.099, 0.99 & 24.73 & 26.30 & \textbf{22.56} & 23.07\tabularnewline
    \cline{2-14} \cline{3-14} \cline{4-14} \cline{5-14} \cline{6-14} \cline{7-14} \cline{8-14} \cline{9-14} \cline{10-14} \cline{11-14} \cline{12-14} \cline{13-14} \cline{14-14} 
     & M4 & 19.13, 58.41 & 13.35, 49.00 & 20.33, 68.89 & \textbf{33.04, 89.91} & 0.097, 0.97 & 0.098, 0.98 & 0.098, 0.98 & \textbf{0.092, 0.92} & 24.37 & 27.24 & 18.42 & 10.08\tabularnewline
    \cline{2-14} \cline{3-14} \cline{4-14} \cline{5-14} \cline{6-14} \cline{7-14} \cline{8-14} \cline{9-14} \cline{10-14} \cline{11-14} \cline{12-14} \cline{13-14} \cline{14-14} 
     & M5 & \multicolumn{4}{c|}{64.15, 86.65} & \multicolumn{4}{c|}{0.083, 0.62} & \multicolumn{4}{c|}{12.05}\tabularnewline
    \hline 
    \multirow{5}{*}{1.0} & M1 & 33.74, 70.42 & 29.00, 69.85 & 40.02, 79.93 & \textbf{62.68, 94.40} & 0.086, 0.86 & 0.089, 0.89 & 0.087, 0.87 & \textbf{0.078, 0.78} & 18.82 & 18.72 & 14.51 & \textbf{7.43}\tabularnewline
    \cline{2-14} \cline{3-14} \cline{4-14} \cline{5-14} \cline{6-14} \cline{7-14} \cline{8-14} \cline{9-14} \cline{10-14} \cline{11-14} \cline{12-14} \cline{13-14} \cline{14-14} 
     & M2 & 39.32, 80.37 & 35.65, 79.04 & 50.93, 87.75 & \textbf{53.35, 94.26} & 0.093, 0.91 & 0.097, 0.95 & 0.089, 0.87 & \textbf{0.099, 0.85} & 13.72 & 14.89 & 11.05 & \textbf{7.61}\tabularnewline
    \cline{2-14} \cline{3-14} \cline{4-14} \cline{5-14} \cline{6-14} \cline{7-14} \cline{8-14} \cline{9-14} \cline{10-14} \cline{11-14} \cline{12-14} \cline{13-14} \cline{14-14} 
     & M3 & 8.71, 37.51  & 7.76, 34.75  & 9.74, 41.20  & \textbf{11.87, 47.11} & \textbf{0.097, 0.97} & 0.099, 0.99 & 0.099, 0.99 & 0.099, 0.99 & 31.91 & 32.66 & 29.31 & \textbf{27.77}\tabularnewline
    \cline{2-14} \cline{3-14} \cline{4-14} \cline{5-14} \cline{6-14} \cline{7-14} \cline{8-14} \cline{9-14} \cline{10-14} \cline{11-14} \cline{12-14} \cline{13-14} \cline{14-14} 
     & M4 & 12.92, 40.82 & 8.31, 33.51  & 15.65, 54.41 & \textbf{28.45, 82.31} & 0.096, 0.96 & 0.099, 0.99 & 0.097, 0.97 & \textbf{0.096, 0.96} & 30.54 & 33.71 & 24.33 & \textbf{12.98}\tabularnewline
    \cline{2-14} \cline{3-14} \cline{4-14} \cline{5-14} \cline{6-14} \cline{7-14} \cline{8-14} \cline{9-14} \cline{10-14} \cline{11-14} \cline{12-14} \cline{13-14} \cline{14-14} 
     & M5 & \multicolumn{4}{c|}{44.27, 73.51} & \multicolumn{4}{c|}{0.093, 0.82} & \multicolumn{4}{c|}{18.47}\tabularnewline
    \hline 
    \multirow{5}{*}{0.5} & M1 & 18.42, 47.56 & 18.49, 52.26 & 22.73, 59.47 & \textbf{48.22, 87.01} & 0.095, 0.95 & 0.094, 0.94 & \textbf{0.091, 0.91} & 0.094, 0.93 & 28.13 & 26.06 & 23.29 & \textbf{11.41}\tabularnewline
    \cline{2-14} \cline{3-14} \cline{4-14} \cline{5-14} \cline{6-14} \cline{7-14} \cline{8-14} \cline{9-14} \cline{10-14} \cline{11-14} \cline{12-14} \cline{13-14} \cline{14-14} 
     & M2 & 21.33, 65.02 & 23.50, 63.05 & 34.71, 76.37 & \textbf{47.36, 85.83} & 0.098, 0.98 & 0.099, 0.99 & 0.095, 0.95 & \textbf{0.098, 0.94} & 20.56 & 20.66 & 15.99 & \textbf{12.27}\tabularnewline
    \cline{2-14} \cline{3-14} \cline{4-14} \cline{5-14} \cline{6-14} \cline{7-14} \cline{8-14} \cline{9-14} \cline{10-14} \cline{11-14} \cline{12-14} \cline{13-14} \cline{14-14} 
     & M3 & 4.48, 19.38  & 3.50, 20.04  & 3.73, 20.04  & \textbf{6.56, 30.35} & 0.099, 0.99 & 0.099, 0.99 & 0.099, 0.99 & \textbf{0.099, 0.99} & 43.15 & 42.62 & 40.80 & \textbf{35.48}\tabularnewline
    \cline{2-14} \cline{3-14} \cline{4-14} \cline{5-14} \cline{6-14} \cline{7-14} \cline{8-14} \cline{9-14} \cline{10-14} \cline{11-14} \cline{12-14} \cline{13-14} \cline{14-14} 
     & M4 & 4.14, 22.73  & 3.70, 19.73  & 7.04, 31.41  & \textbf{17.54, 55.47} & 0.099, 0.99 & 0.099, 0.99 & 0.099, 0.99 & \textbf{0.097, 0.97} & 41.72 & 44.29 & 35.88 & \textbf{22.64}\tabularnewline
    \cline{2-14} \cline{3-14} \cline{4-14} \cline{5-14} \cline{6-14} \cline{7-14} \cline{8-14} \cline{9-14} \cline{10-14} \cline{11-14} \cline{12-14} \cline{13-14} \cline{14-14} 
     & M5 & \multicolumn{4}{c|}{23.35, 45.35} & \multicolumn{4}{c|}{0.099, 0.99} & \multicolumn{4}{c|}{31.79}\tabularnewline
    \hline 
    \end{tabular}

    }
                
    \vspace{0.05cm}
    \centering
    \begin{tabular}{|c|c|c|c|c|c|}
    \hline 
    Method & M1 & M2 & M3 & M4 & M5\tabularnewline
    \hline 
    \hline 
    Algorithm & 1D-Triplet-CNN-online & 1D-Triplet-CNN & xVector-PLDA & iVector-PLDA & RawNet2\tabularnewline
    \hline 
    \end{tabular}
            
    \label{tab:Experiments_al}
    \vspace{-0.4cm}
    \end{table*}

	\subsection{Datasets}
	\label{datasets}
	
	\subsubsection{VOXCeleb2 Dataset}
	The VoxCeleb2~\cite{chung2018voxceleb2} dataset contains short interview video clips of $6,112$ celebrities recorded in unconstrained scenarios. The entire VOXCeleb2 dataset contains $145,569$ video samples from $5,994$ celebrities in the training set and $4,911$ videos from the remaining $118$ speakers in the evaluation set. However, for keeping the triplet-based training process computationally tractable, we only use speech data from one randomly selected video for each subject. In our experiments, each video in the dataset is processed to extract the speech audio, sampled at $8000$Hz, from its audio track. Speech samples longer than $5$ seconds are split into multiple non-overlapping $5$ second long speech samples.
	
	
	\subsubsection{Fisher English Training Speech Part 1 Dataset}
	The Fisher dataset contains pair-wise conversational speech data, collected over telephone channels, from a set of around $12000$ speakers. Since the amount of speech data per speaker varies in the dataset, in order to ensure data balance across different speakers, we choose to work with a subset of $6991$ speakers, each having at least $250$ seconds of speech audio, across $50$ samples, after performing voice activity detection. Further, a random subset of $4500$ speakers is chosen to train the models and the remaining speakers form the testing set. As mentioned earlier, we have also added the `F-16' and `Babble' noise from the NOISEX-92~\cite{noisex} noise dataset to the Fisher speech dataset. The resultant `degraded-Fisher' speech dataset was maintained at a SNR level of 10dB. We also added reverberations to the speech data generated in a simulated cubical room of side length $4$m.
	
	\subsubsection{NIST SRE 2008, 2010, and 2018 Datasets}
\begin{figure*}[htp]
	\centering
	\subfloat [Experiment 1] {\includegraphics[scale=0.13, clip]{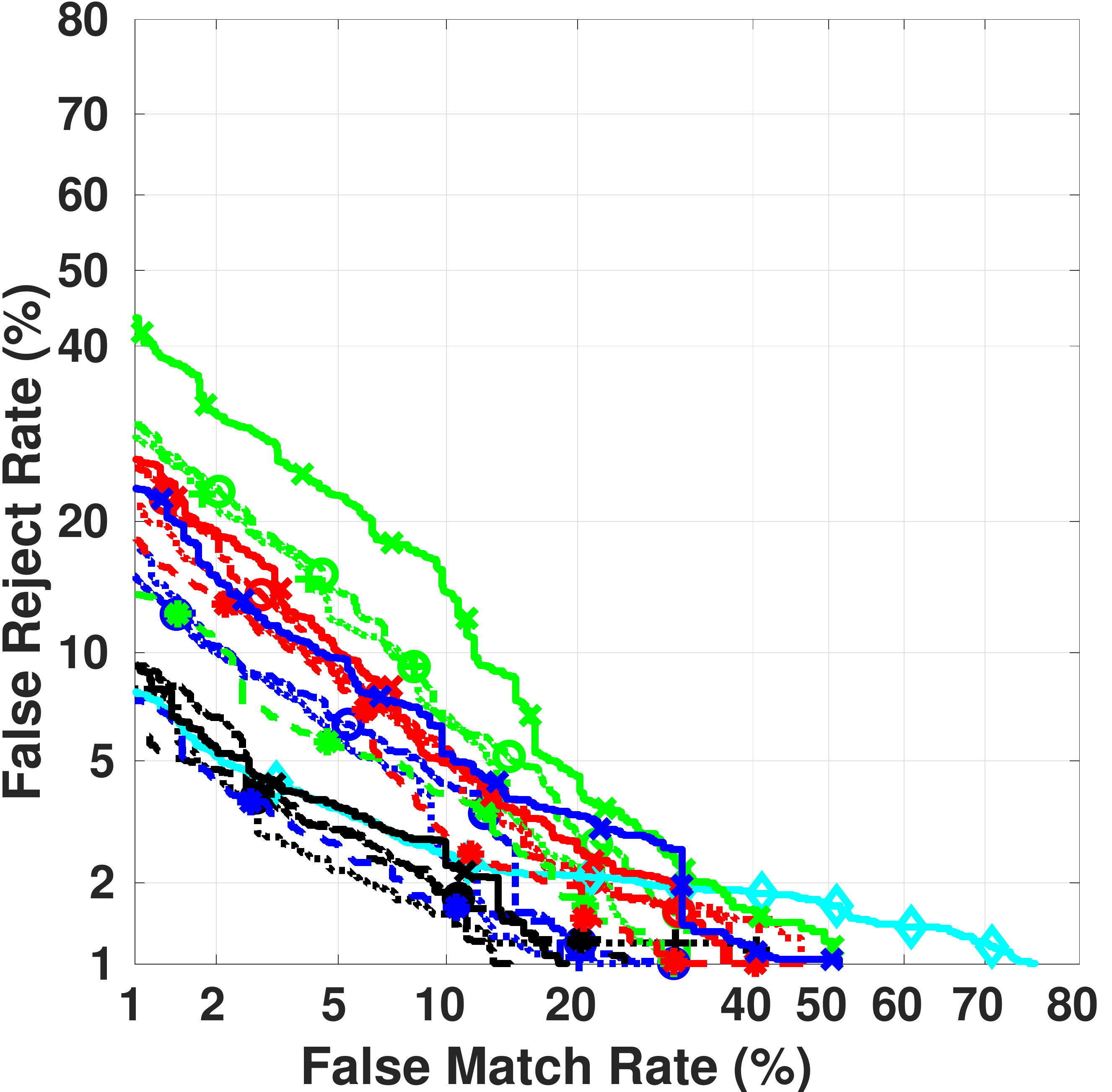}}
	\subfloat [Experiment 2] {\includegraphics[scale=0.13, clip]{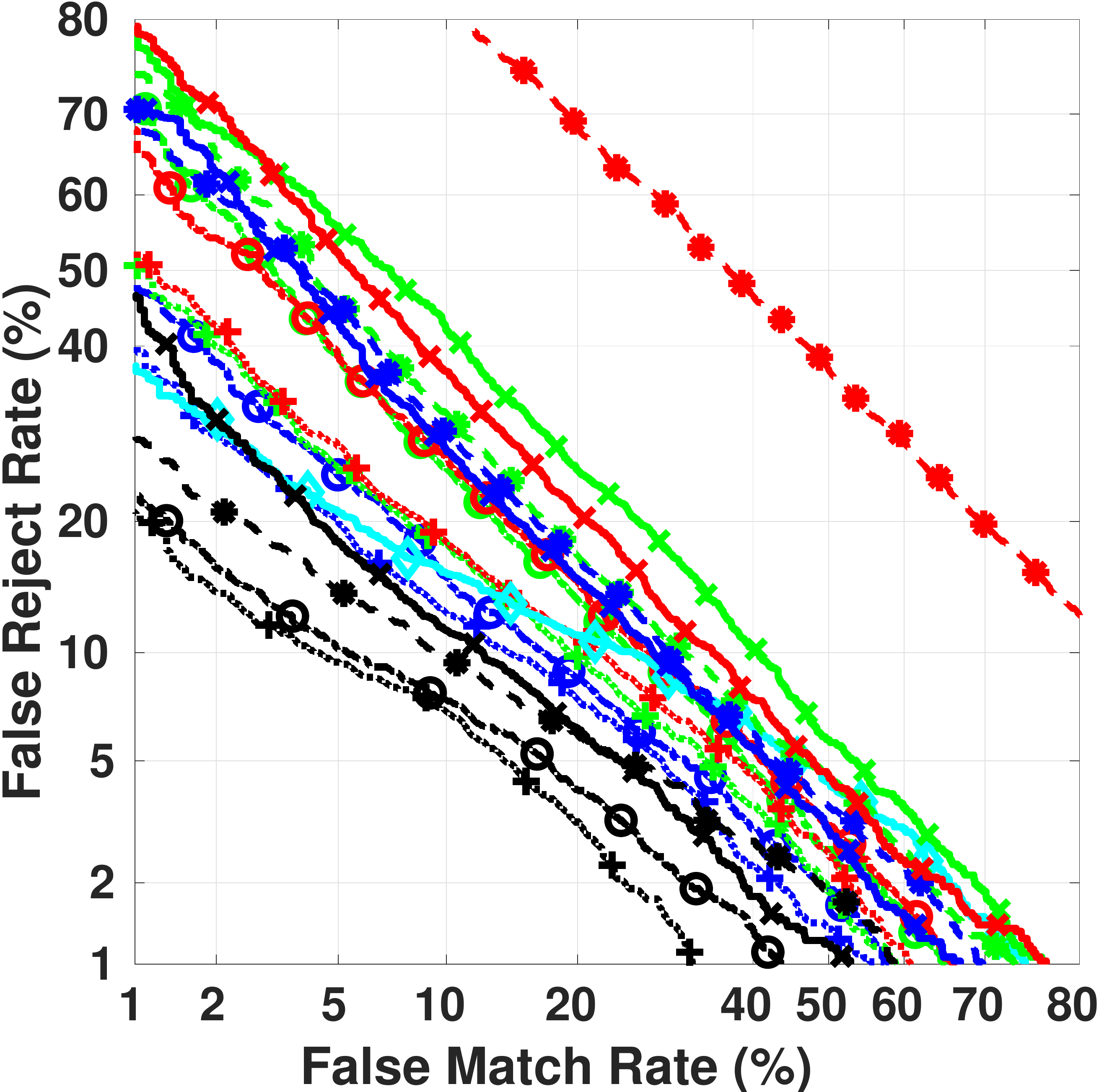}}
	\subfloat [Experiment 3] {\includegraphics[scale=0.13, clip]{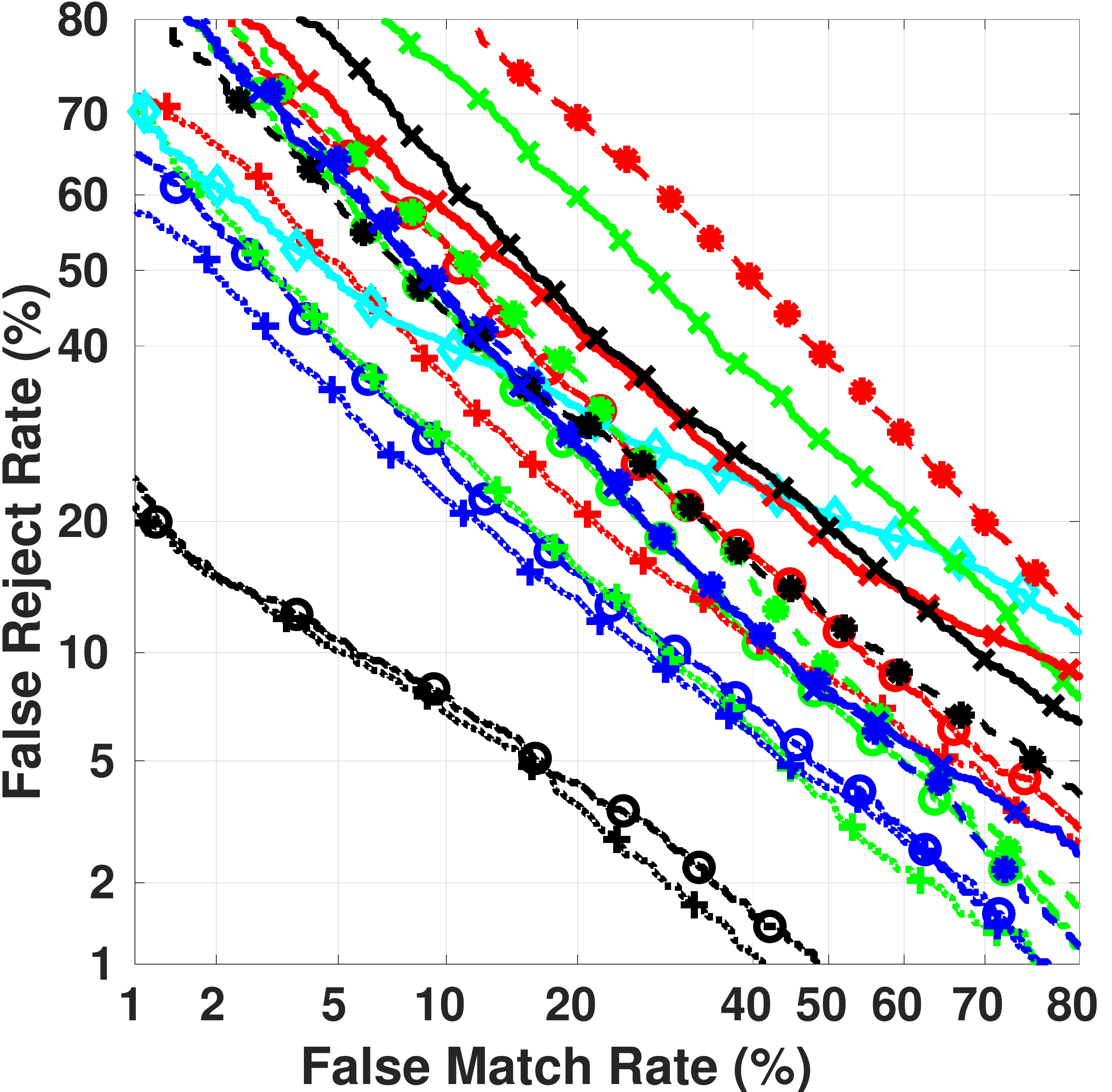}}
	\subfloat [Experiment 4] {\includegraphics[scale=0.13, clip]{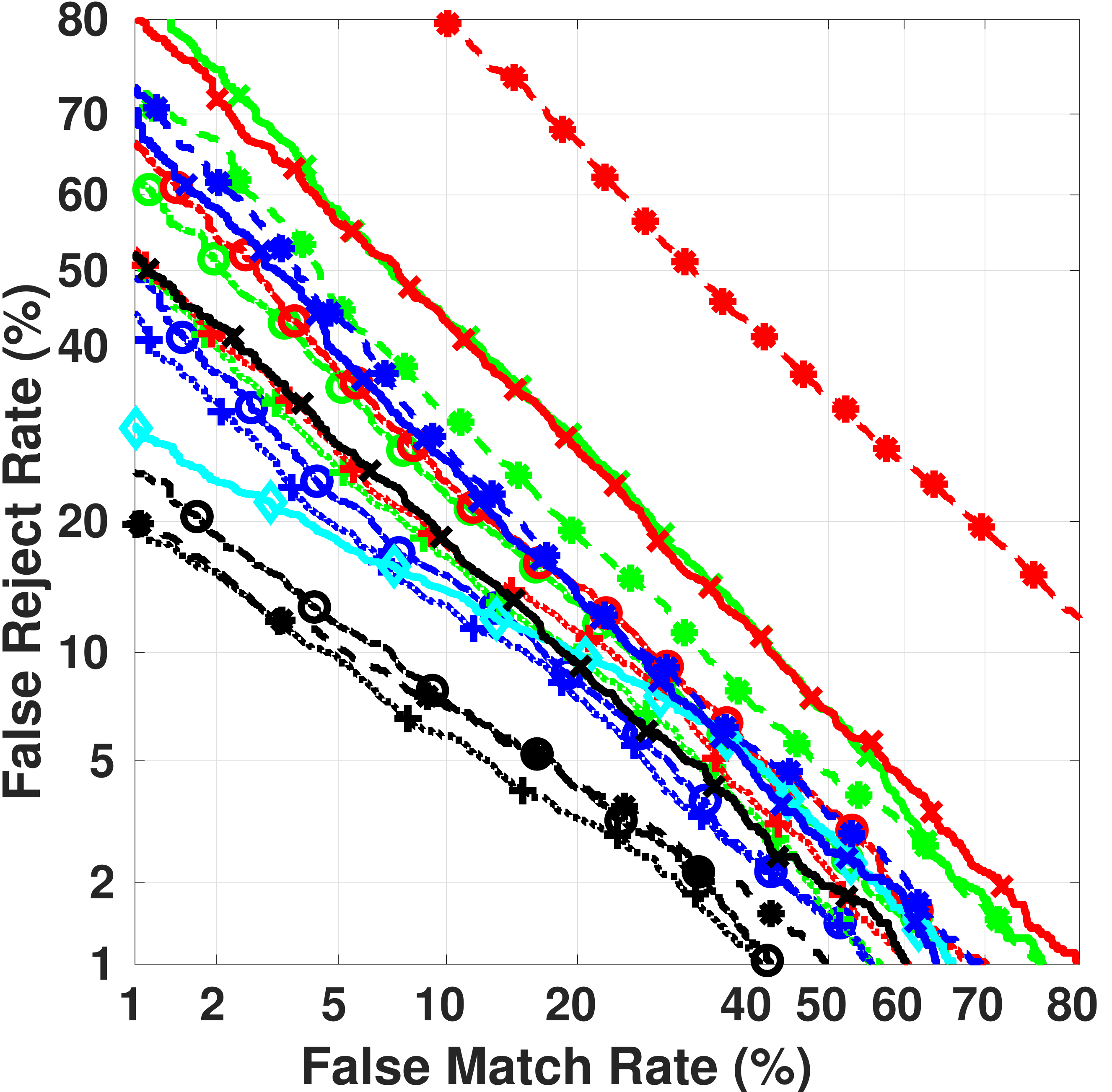}}
	\subfloat [Experiment 5] {\includegraphics[scale=0.13, clip]{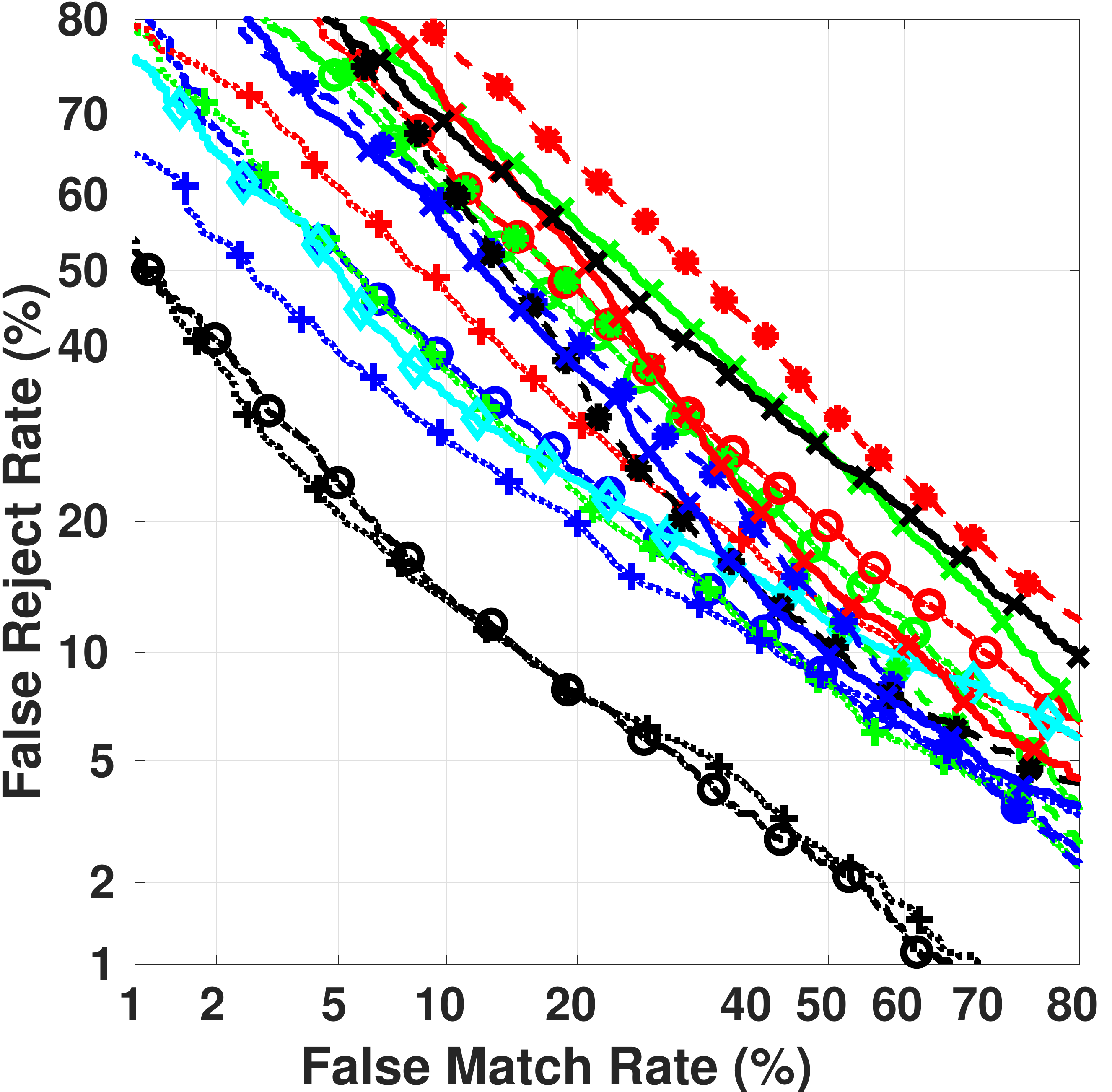}}
	\\
	\vspace{-0.3cm}
	\subfloat [Experiment 6] {\includegraphics[scale=0.13, clip]{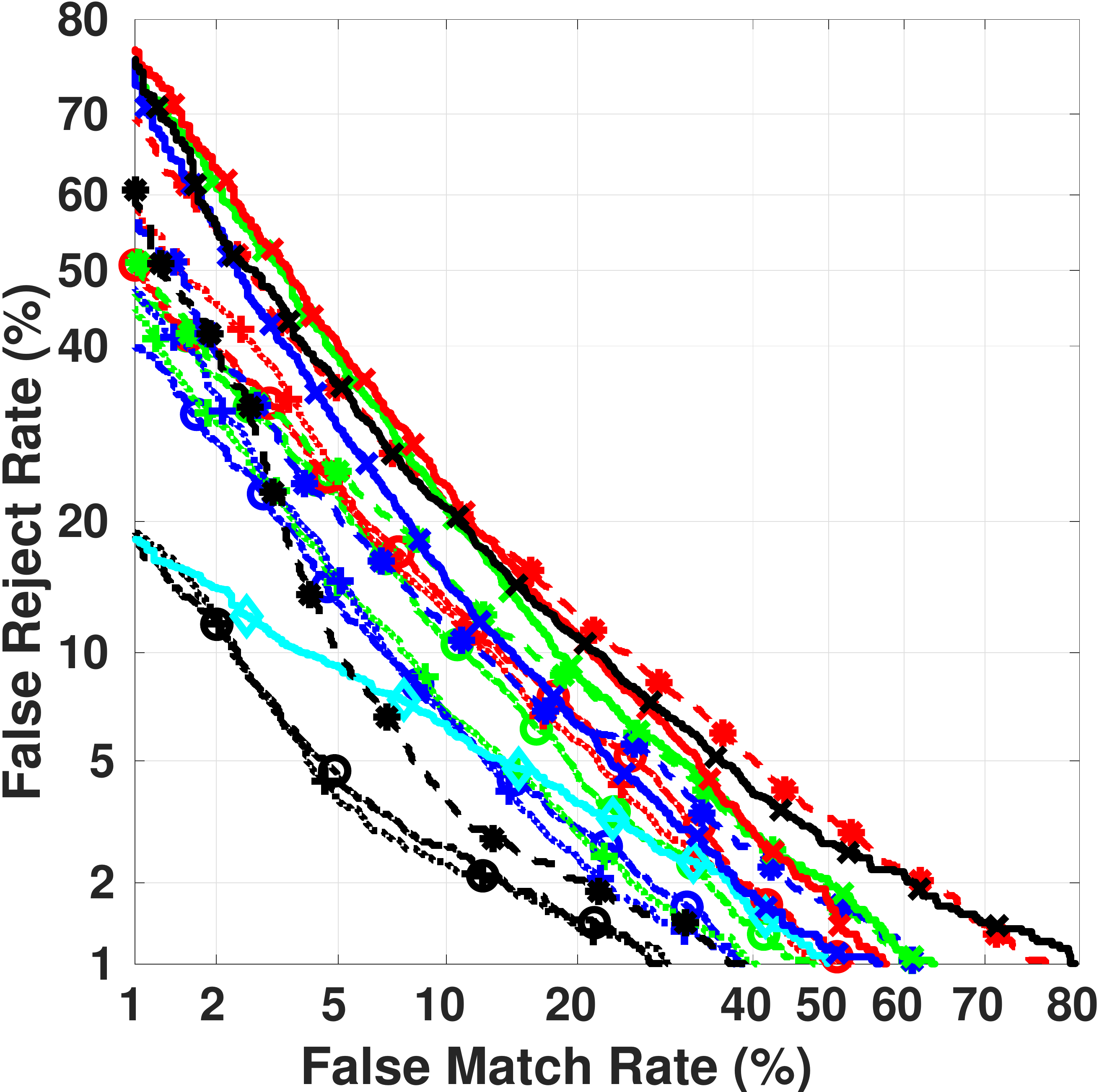}}
	\subfloat [Experiment 7] {\includegraphics[scale=0.13, clip]{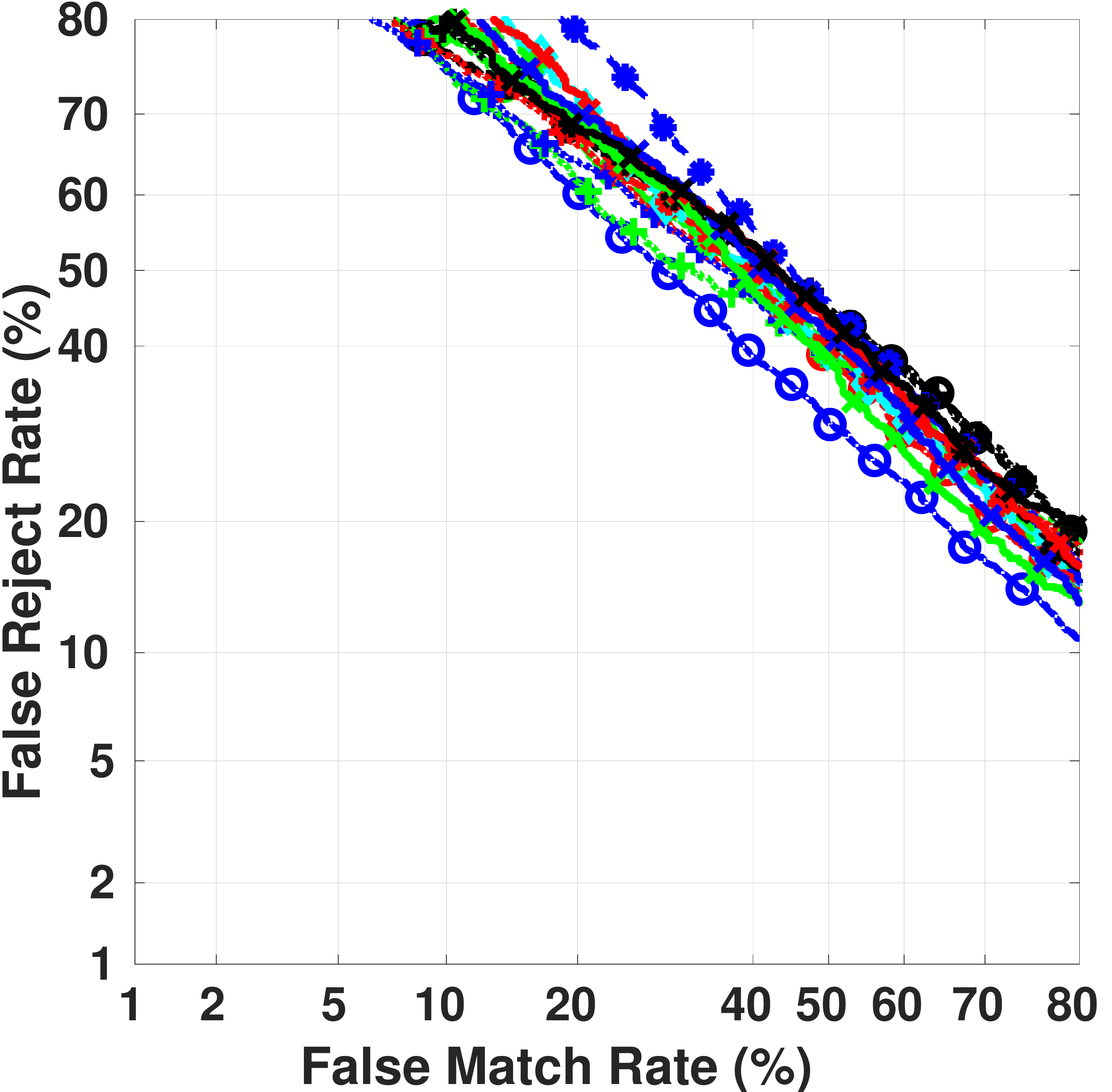}}
	\subfloat [Experiment 8] {\includegraphics[scale=0.13, clip]{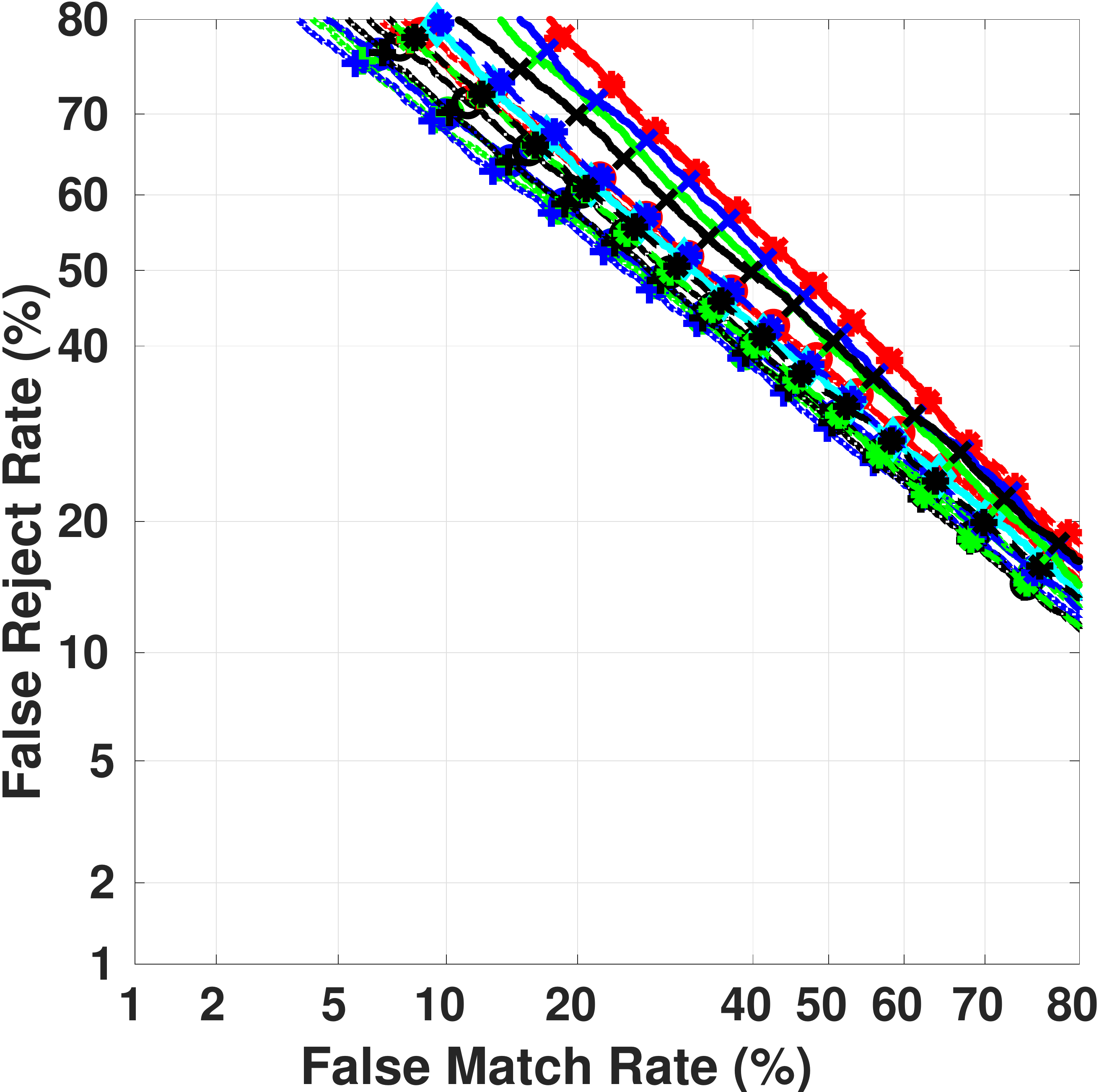}}
	\subfloat [Experiment 9] {\includegraphics[scale=0.13, clip]{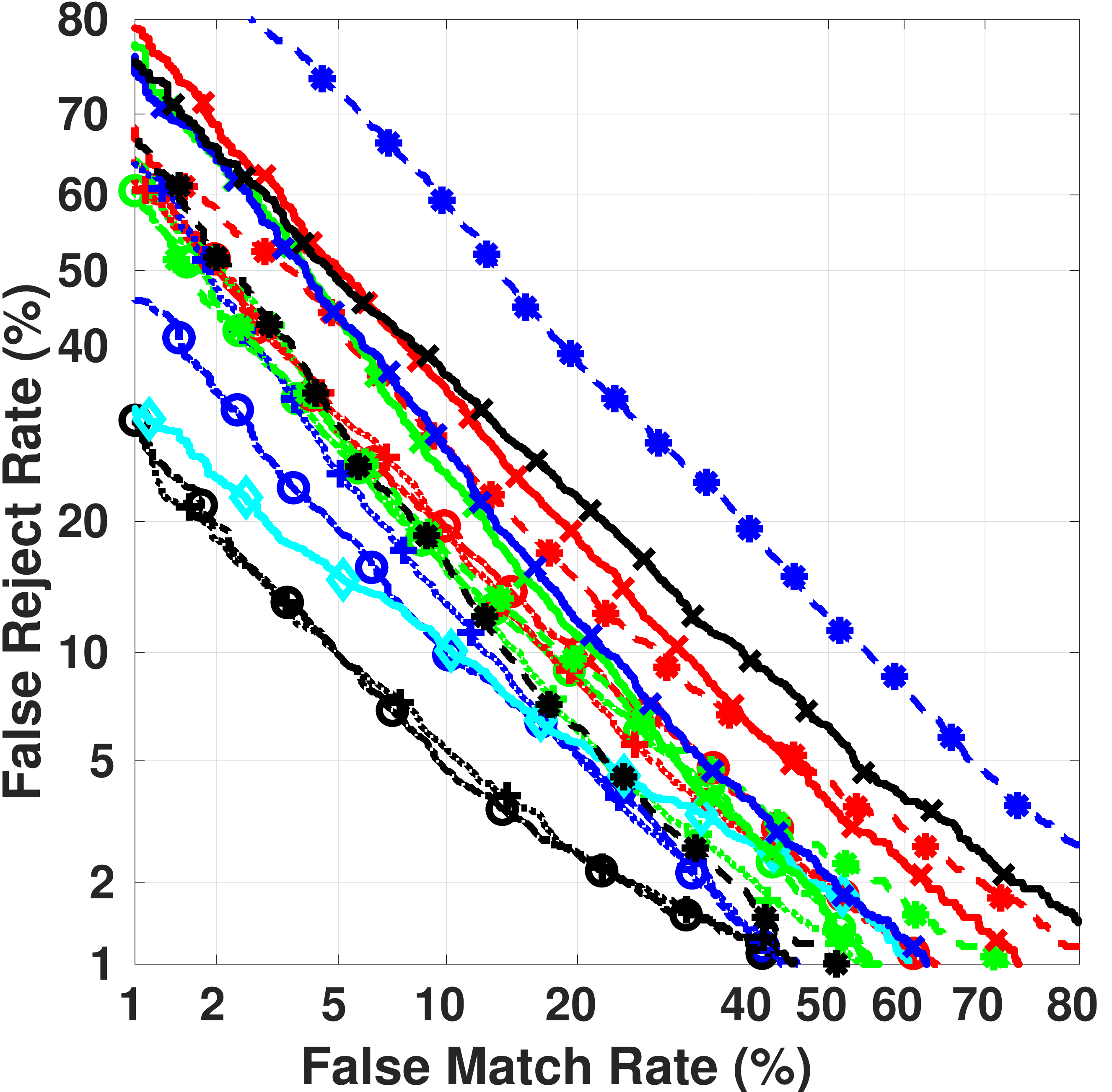}}
	\subfloat [Experiment 10] {\includegraphics[scale=0.13, clip]{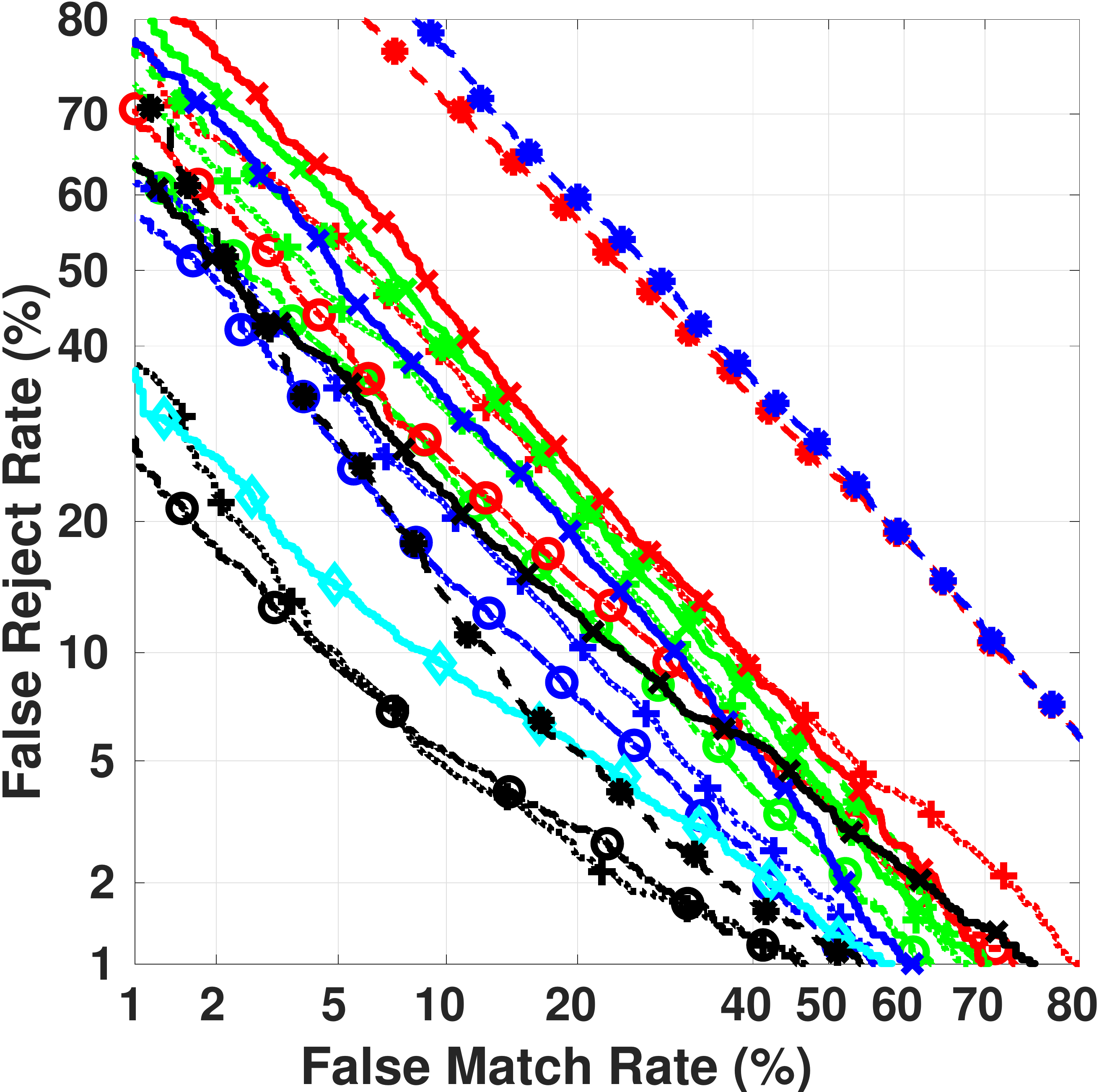}}
	\\
	\vspace{-0.3cm}
	\subfloat [Experiment 11] {\includegraphics[scale=0.13, clip]{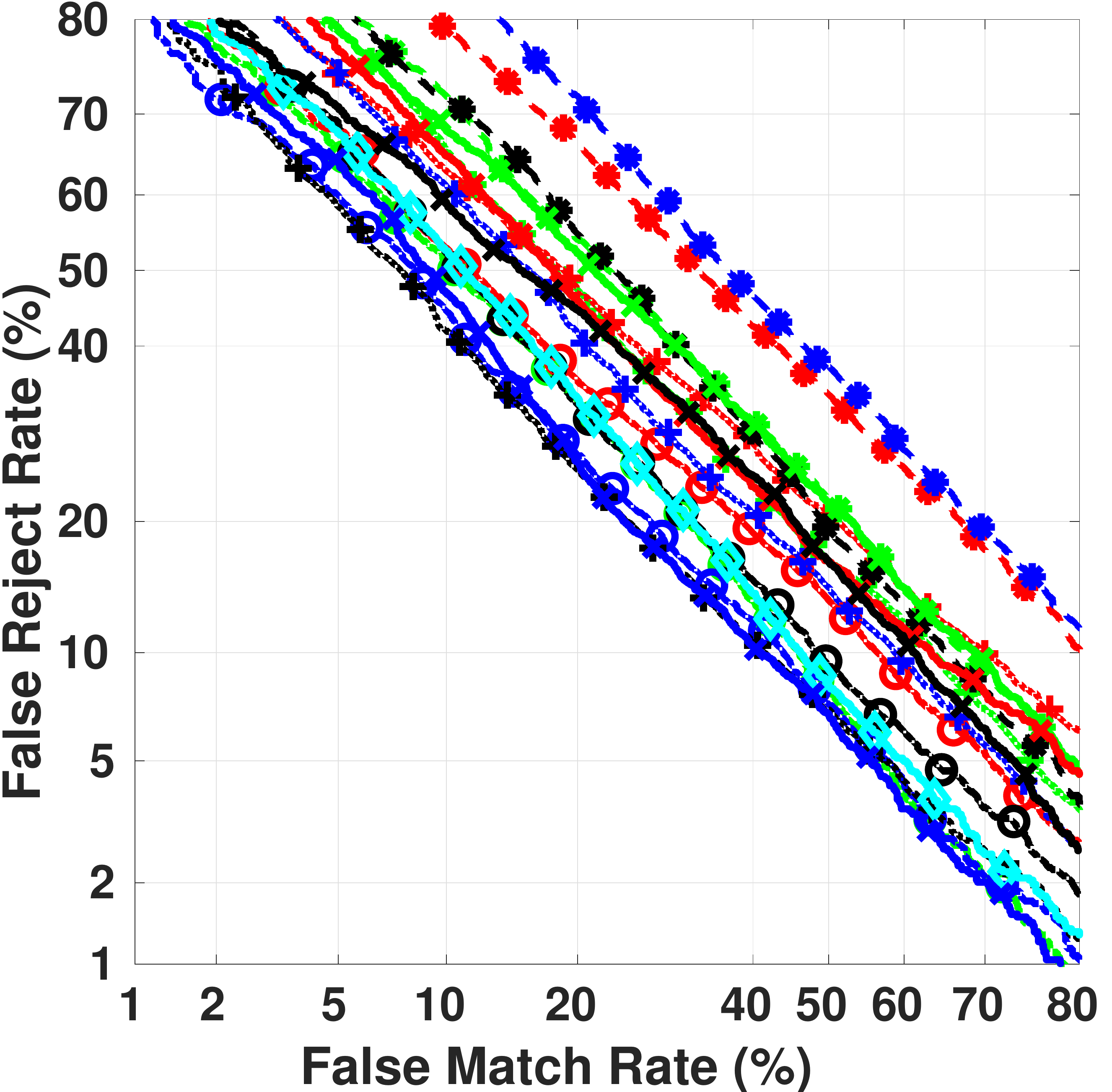}}
	\subfloat [Experiment 12] {\includegraphics[scale=0.13, clip]{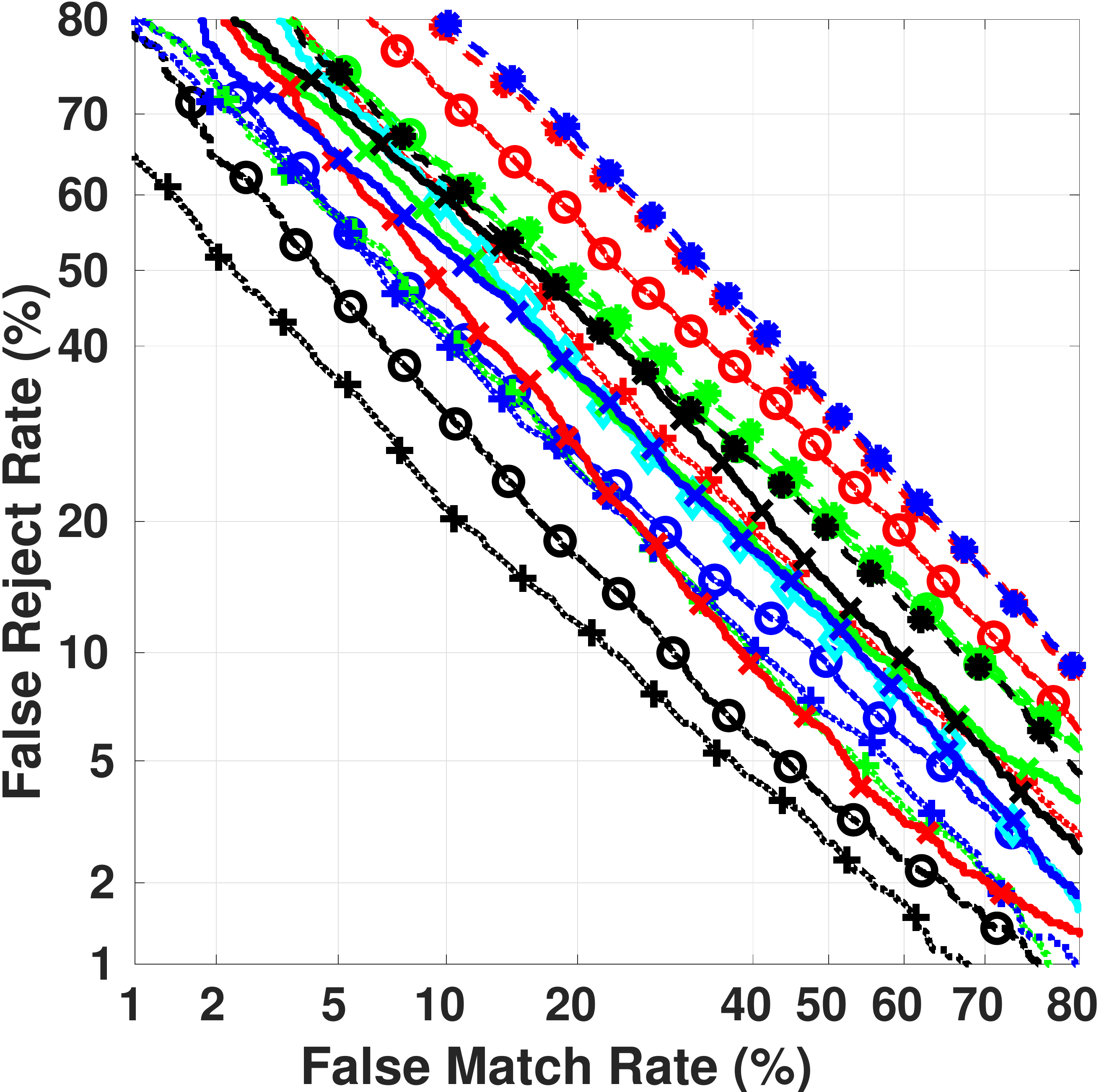}}
	\subfloat [Experiment 13] {\includegraphics[scale=0.13, clip]{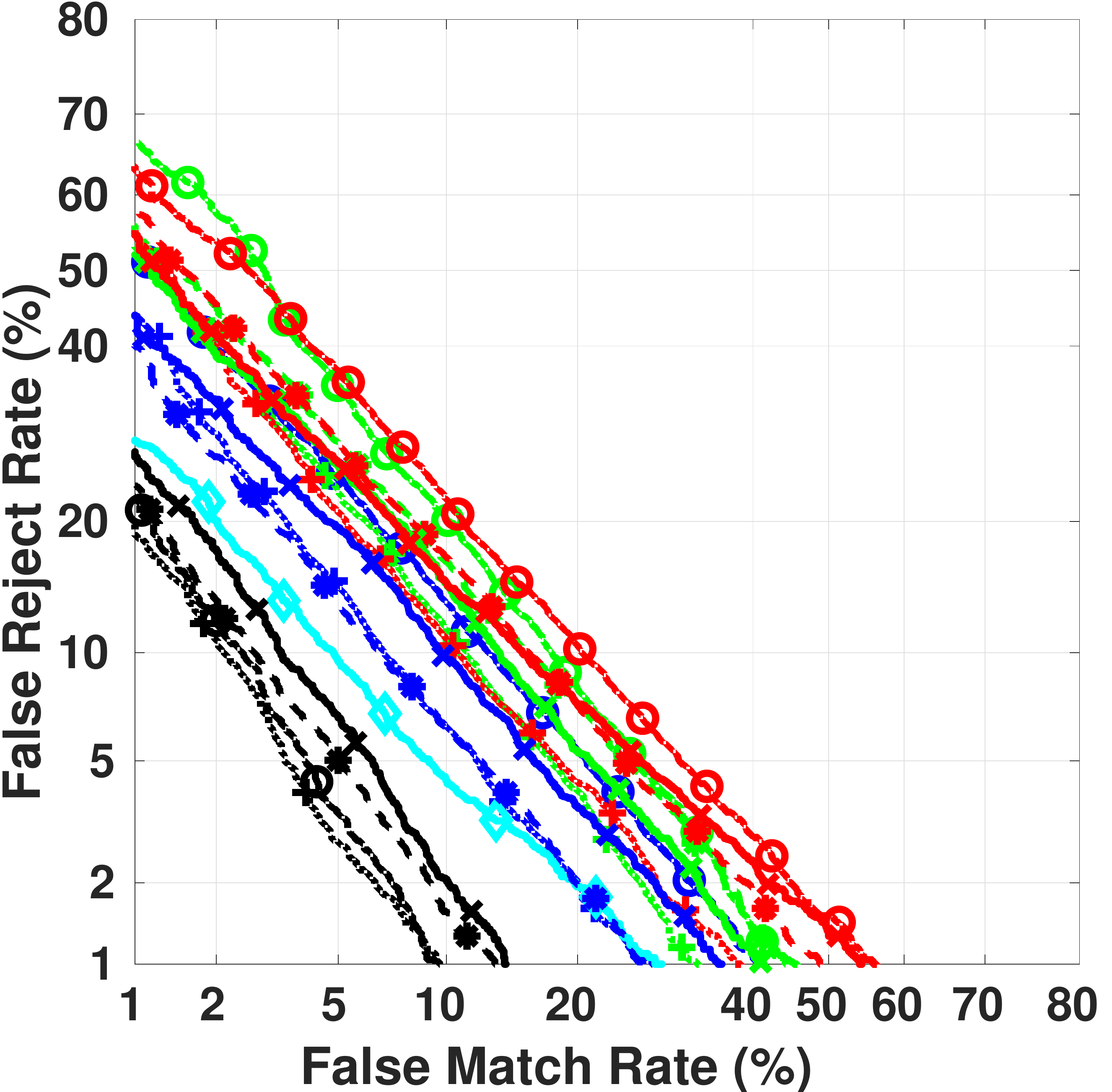}}
	\subfloat [Experiment 14] {\includegraphics[scale=0.13, clip]{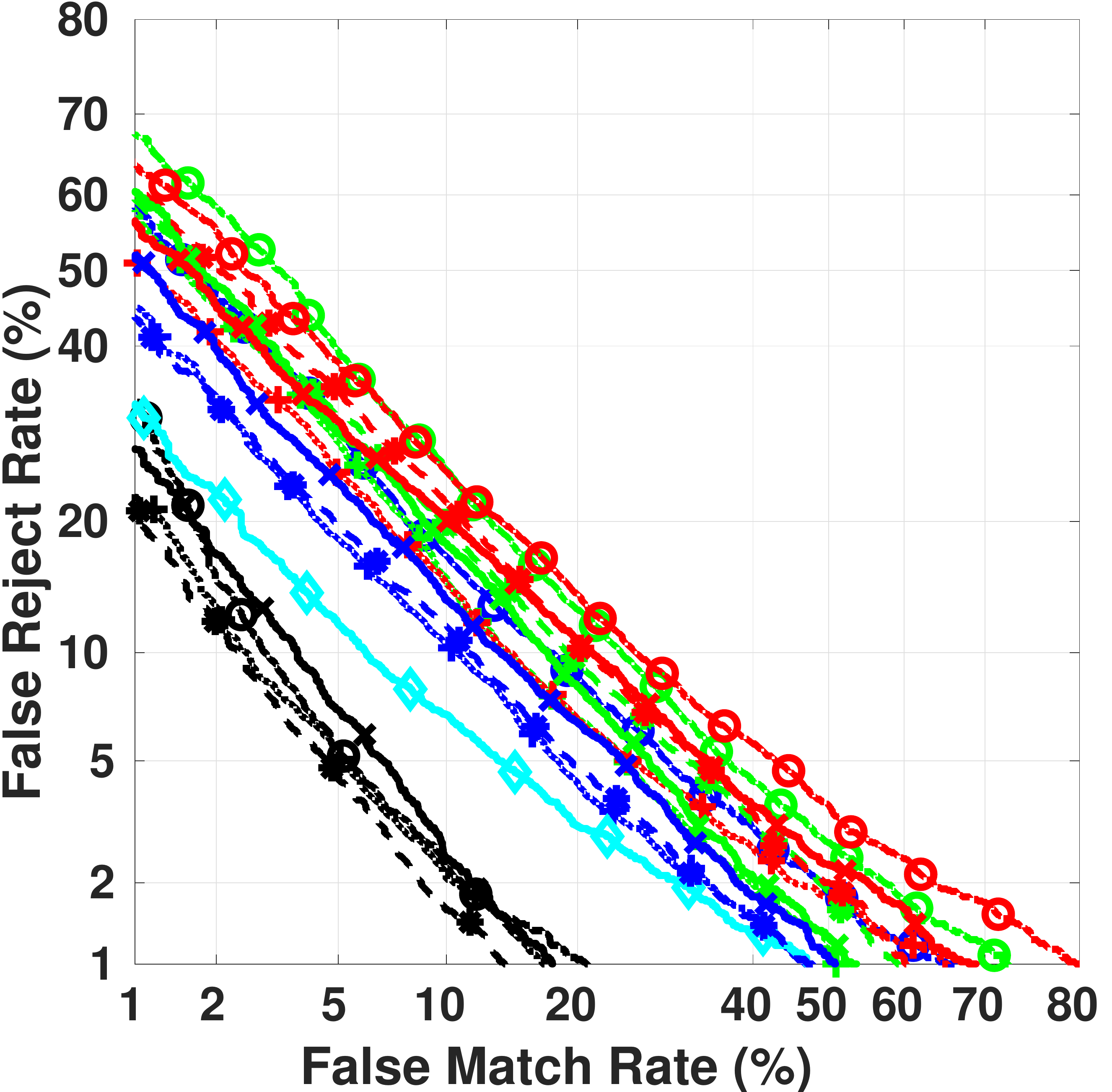}}
	\subfloat [Experiment 15] {\includegraphics[scale=0.13, clip]{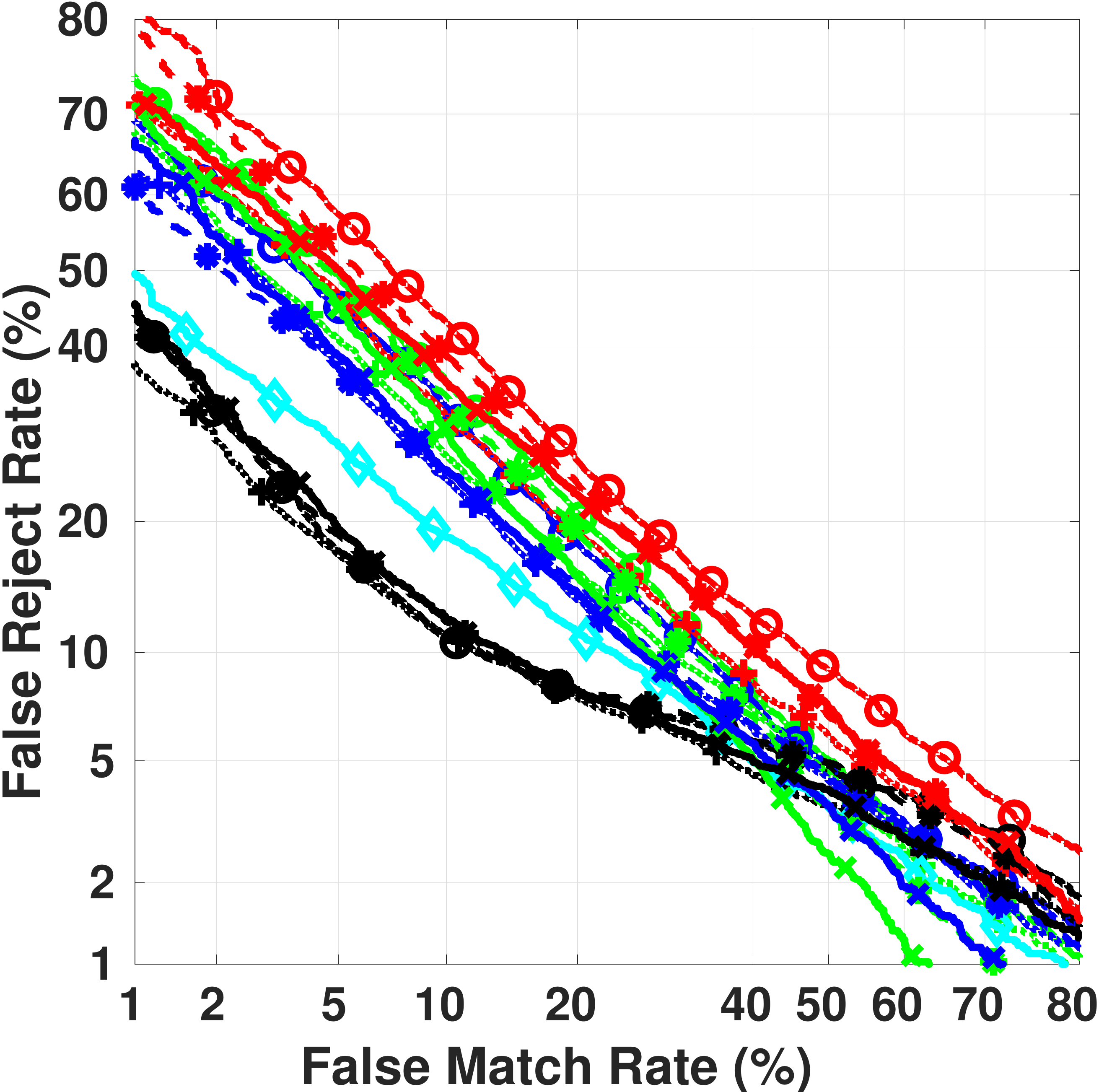}}
	\vspace{-0.001cm}
	\subfloat {\includegraphics[scale=0.35, clip]{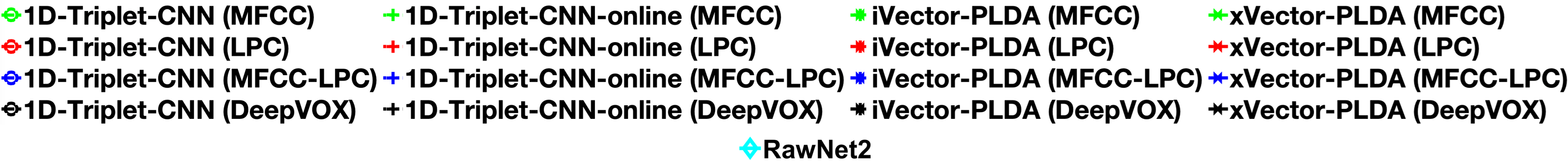}}
	
	\vspace{-0.4cm}
	\caption{\label{fig:ROC} DET curves for the speaker verification experiments on the VOXCeleb2 (Exp. 1), degraded Fisher (Exp. 2 to 5, the clean and degraded NIST SRE 2008, 2010, and 2018 datasets (Exp. 6 to 12), and the multilingual subset of NIST SRE 2008 dataset (Exp. 13 to 15) using RawNet2, iVector-PLDA, xVector-PLDA, 1D-Triplet-CNN, and 1D-Triplet-CNN-online algorithms on MFCC, LPC, MFCC-LPC, and DeepVOX feature sets.\vspace{-0.5cm}}
	\end{figure*}

	We also use the NIST SRE 2008~\cite{SRE08} dataset in our experiments, given in Table~\ref{tab:Experiments_SRE} and Figure~\ref{fig:ROC}, to evaluate the performance of our proposed algorithm in the presence of multi-lingual data. For our experiments, we choose a subset of speech data from the `phonecall' and `interview' speech types collected under audio conditions labeled as `10-sec', `long' and `short2'. The chosen data subset contains speech from $1336$ speakers out of which a randomly chosen subset of $200$ speakers is reserved for evaluation purposes. As mentioned earlier, we also add F-16 and Babble noise at a resultant SNR of 0dB to the NIST SRE 2008 dataset to vastly increase the difficulty of the task. We also perform cross-dataset speaker verification performance evaluation using speech data from all the speakers in the evaluation sets of the NIST SRE 2010~\cite{SRE10} and NIST SRE 2018~\cite{SRE18} datasets.
	
	
	\begin{figure}[htp]
    \vspace{-0.5cm}
	\centering
	\subfloat [TMR@FMR=1\%] {\includegraphics[scale=0.13, clip]{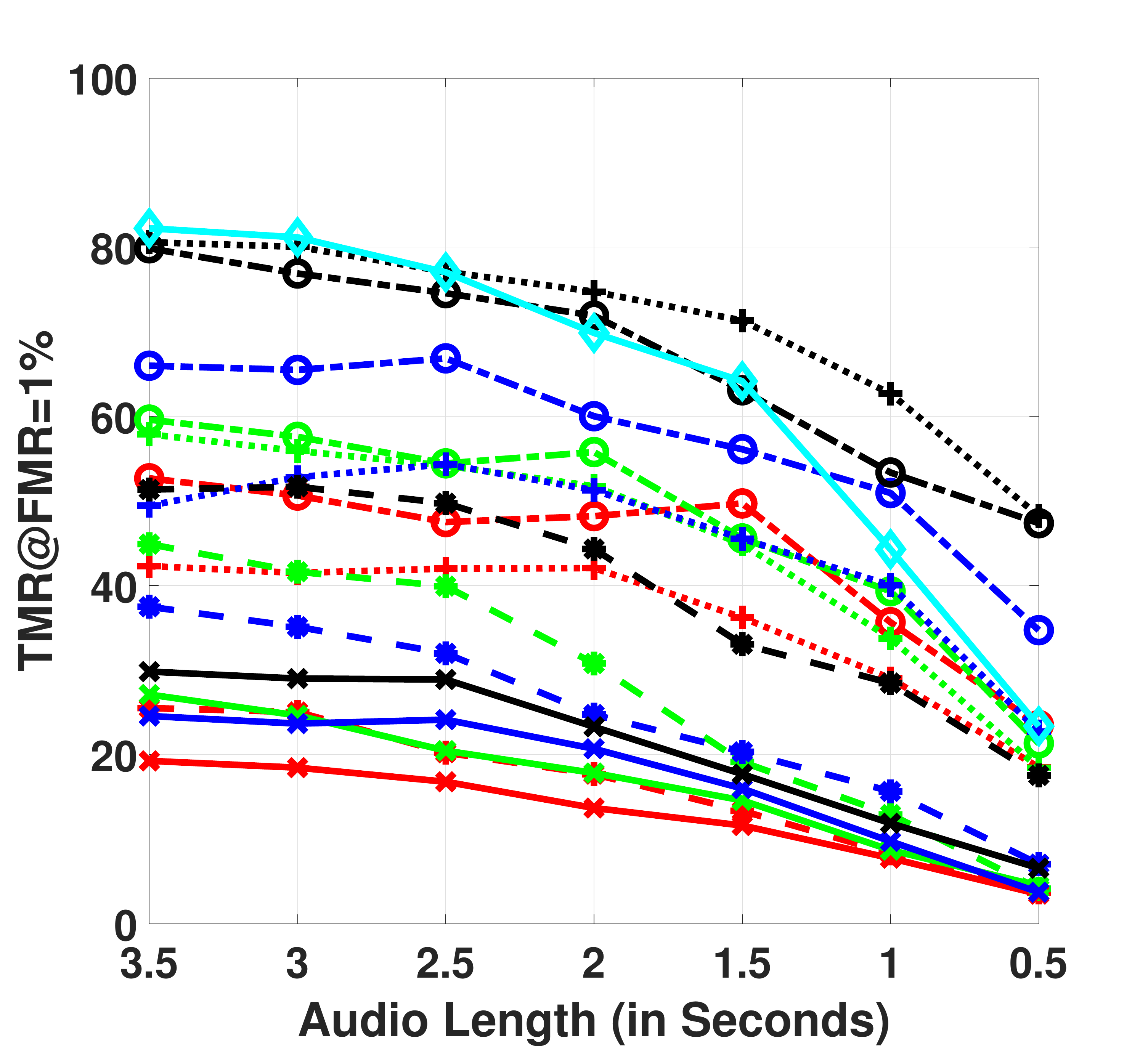}}
    \subfloat [EER] {\includegraphics[scale=0.13, clip]{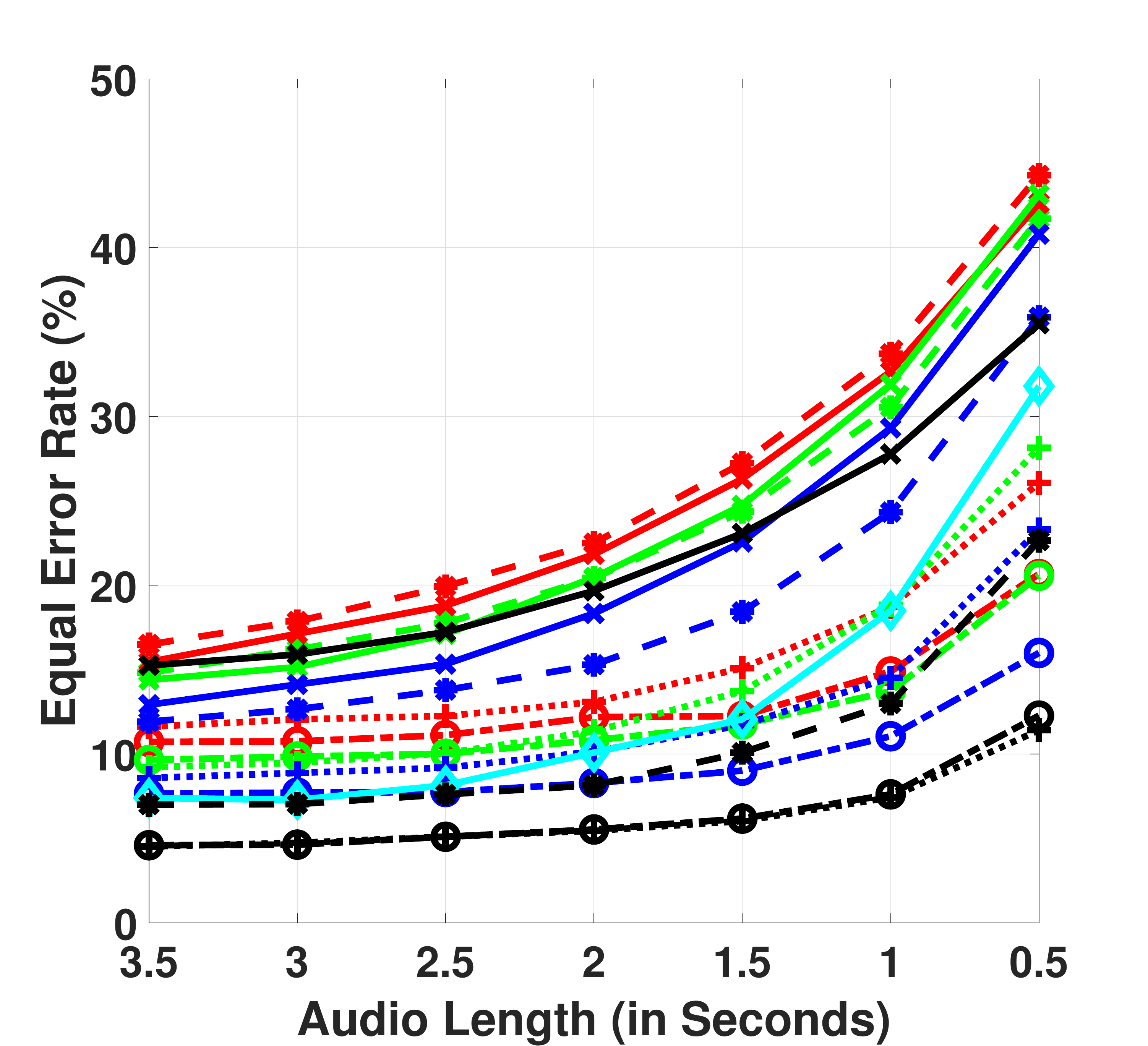}}
    \\
    \subfloat {\includegraphics[scale=0.19, clip]{Figures/TPAMI2019/legend.png}}
	\\
	\caption{\label{fig:ROC_AL} (a) TMR@FMR=1\% and (b) EER under varying audio length on the clean NIST SRE 2008 dataset. 1D-Triplet-CNN(DeepVOX) performs the best across varying lengths of test audio.}
	\vspace{-0.5cm}
    \end{figure}
    
    \vspace{-0.3cm}
	\subsection{Experimental Protocols} \label{sec:exp}
	In all the experiments, we ensure disjoint set of speakers in the training and testing sets. For evaluating robustness of our models we perform same-noise, cross-noise and  cross-dataset experiments as shown in Tables \ref{tab:Experiments_VOXCeleb2}, \ref{tab:Experiments_Fisher}, and \ref{tab:Experiments_SRE}. The noise characteristics of the training and testing sets used in the different experiments are given alongside in Tables~\ref{tab:Experiments_VOXCeleb2}, \ref{tab:Experiments_Fisher}, and \ref{tab:Experiments_SRE}. For example, in Experiment $3$ given in Table~\ref{tab:Experiments_Fisher}, the model was trained on speech data from the training set of Fisher Speech Dataset degraded with Babble noise, and the evaluation was done on speech data from testing set of Fisher Speech Dataset degraded with F16 noise. Note that, no mention of a noise type, such as in Experiment $1$ given in Table \ref{tab:Experiments_VOXCeleb2}, indicates usage of un-altered speech data from the original dataset. Additionally, we have also conducted speaker verification experiments on a subset of multi-lingual speakers from the NIST SRE 2008 dataset, as shown in Table~\ref{tab:Experiments_SRE_lang}, for evaluating the effect of speech language on speaker verification performance. Finally, as illustrated in Figure~\ref{fig:PSD} and discussed in Section~\ref{sec:ablation}, we have performed Guided Backpropagation~\cite{springenberg2014striving} based ablation study of the features extracted by trained DeepVOX models, to understand the type of audio features considered important for performing speaker recognition by the DeepVOX model.
	
	\subsubsection{Baseline Speaker Verification Experiments}
	For establishing baseline speaker verification performance on the VOXCeleb2, Fisher, NIST SRE 2008, 2010, and 2018 speech datasets, we choose iVector-PLDA~\cite{garcia2011analysis} and xVector-PLDA~\cite{snyder2018x} algorithms trained on the baseline features (MFCC, LPC, MFCC-LPC) and DeepVOX features separately. This is done to evaluate and compare the effectiveness of DeepVOX features, with respect to baseline features, in both classical and deep learning-based speaker recognition algorithms. However, unlike the baseline features, DeepVOX feature extraction process requires a DeepVOX model to be trained. For each of the experiments in Tables~\ref{tab:Experiments_VOXCeleb2}, ~\ref{tab:Experiments_Fisher}, and \ref{tab:Experiments_SRE} we use speech data only from corresponding training set to train the DeepVOX model, ensuring disjoint data and subjects in the training and testing sets for the DeepVOX feature extraction process. We also use the RawNet2~\cite{jung2020improved} algorithm for establishing baseline raw audio-based speaker recognition performance.
	\noindent
	\begin{itemize}[leftmargin=0cm,itemindent=.5cm,labelwidth=\itemindent,labelsep=0cm,align=left]
		
		\item \textit{iVector-PLDA~\cite{garcia2011analysis}-based Speaker Verification Experiments}:
		We use MSR Identity Toolkit's~\cite{sadjadi2013msr} iVector-PLDA implementation as our first baseline speaker verification method. A Gaussian-PLDA (gPLDA)-based matcher~\cite{sadjadi2013msr} is used to compare the extracted i-Vector embeddings of a pair of speech samples.


		\item \textit{xVector-PLDA~\cite{snyder2018x}-based Speaker Verification Experiments}:
        We use the PyTorch-based implementation~\cite{chowdhury2020fusing} of the xVector algorithm as our second baseline speaker verification method. A gPLDA-based matcher~\cite{sadjadi2013msr} is used to compare the extracted xVector embeddings of a pair of speech samples.
		
		\item \textit{RawNet2{~\cite{jung2020improved}}-based Speaker Verification Experiments}:
		We use the RawNet2 algorithm to establish a baseline raw audio-based speaker recognition performance. We use the authors'~\cite{jung2020improved} original implementation of the RawNet2 method for performing the RawNet2-based experiments.
        
	\end{itemize}
	
	\subsubsection{Speaker Verification Experiments on 1D-Triplet-CNN Algorithm Using MFCC-LPC Feature Fusion}
	We also perform speaker recognition experiments using the 1D-Triplet-CNN~\cite{chowdhury2020fusing} algorithm. These experiments provide benchmark results (given in Tables~\ref{tab:Experiments_VOXCeleb2},\ref{tab:Experiments_Fisher}, and \ref{tab:Experiments_SRE}) to directly compare the performance of the DeepVOX feature to MFCC, LPC, and MFCC-LPC features in a deep learning framework.
	For training the 1D-Triplet-CNN, speech audio triplets are formed using the speakers from the training set. The speech audio triplets are then processed to extract $40$ dimensional MFCC and LPC features separately. The extracted MFCC and LPC features are then stacked together to form a two-channel input feature patch for the 1D-Triplet-CNN. For evaluation, speech audio pairs are fed to the trained model to generate pairs of speech embeddings. The speech embeddings are then matched using the cosine similarity metric.
	
	\subsubsection{Speaker Verification Experiments on 1D-Triplet-CNN Algorithm Using DeepVOX Features (Proposed Algorithm)}
	In these set of experiments, we evaluate the performance of our proposed approach on multiple training and testing splits (given in the Tables~\ref{tab:Experiments_VOXCeleb2},\ref{tab:Experiments_Fisher}, and \ref{tab:Experiments_SRE}) drawn from different datasets and noise types and compare it with the baseline algorithms. Similar to the MFCC-LPC-based 1D-Triplet-CNN~\cite{chowdhury2020fusing} algorithm, our algorithm also trains on speech audio triplets. However, instead of extracting hand-crafted features like MFCC or LPC, our algorithm trains the DeepVOX and 1D-Triplet-CNN modules together to learn both the DeepVOX-based feature representation and 1D-Triplet-CNN-based speech feature embedding simultaneously. For evaluation, speech audio pairs are fed to the trained DeepVOX model to extract pairs of DeepVOX features which are then fed into the trained 1D-Triplet-CNN model to extract pairs of speech embeddings and compare them using the cosine similarity metric.
	
	\subsubsection{1D-Triplet-CNN-based Speaker Recognition Experiments Using Adaptive Triplet Mining}
	The proposed adaptive triplet mining technique is evaluated by repeating all the 1D-Triplet-CNN based speaker verification experiments on MFCC, LPC, MFCC-LPC, and DeepVOX features, referred to as \textit{1D-Triplet-CNN-online} in Tables~\ref{tab:Experiments_VOXCeleb2},~\ref{tab:Experiments_Fisher}, and \ref{tab:Experiments_SRE}. In our experiments, the 1D-Triplet-CNN models are pretrained in identification mode for 50 epochs followed by 800 epochs of training in verification mode using adaptive triplet mining. As also mentioned in Section~\ref{sec:triplet_mining}, the difficulty ($\tau$) of the mined negative samples is gradually increased from $0.4$ to $1.0$ linearly over $800$ epochs. Also, it is important to note that the triplet mining is done in mini-batches of $6$ randomly chosen samples drawn from each of the $25$ randomly chosen training subjects.

	\subsubsection{Effect of Language on Speaker Verification}
	The effect of language on speaker recognition performance, also known as the language-familiarity effect (LFE), of both humans and machines, has been studied in the literature~\cite{lu2009effect}. According to LFE, human listeners perform speaker recognition better when they understand the language being spoken. Similar trends have been noticed in the performance of automatic speaker recognition systems~\cite{lu2009effect}. In this work, we perform additional speaker recognition experiments (Exp. \#  12 to 14 in Table~\ref{tab:Experiments_SRE_lang}) on a subset of the NIST SRE 2008 dataset for evaluating the robustness of the DeepVOX features compared to MFCC, LPC, and MFCC-LPC features in the presence of multi-lingual speech data. In all the experiments (Exp. \# 12 to 14), the models are trained on English speech data spoken by a subset of $1076$ English-speaking subjects in NIST SRE 2008's training set and evaluated on a subset of $59$ multi-lingual subjects, containing speech data from 15 different languages, in NIST SRE 2008's test set. The evaluation trials in experiments 12 to 14 varied as follows:
	
    \begin{description}[leftmargin=0cm,itemindent=.5cm,labelwidth=\itemindent,labelsep=0cm,align=left]
        \item[Same language, english only trials : ] In Exp. \# $12$, the trained models are evaluated on same-language (English Only) trials. This experiment establishes the baseline same-language (English to English) speaker verification performance of all the algorithms.
        \item[Same language, non-english trials: ] In Exp. \# $13$, the trained models are evaluated on same-language (Multi-lingual) trials. This experiment aims to investigate the performance of speaker recognition models trained on English-only speech data for matching Non-English same-language (e.g: Hindi to Hindi) speech trials.
        \item[Cross-lingual trials: ] In Exp. \# $14$, the trained models are evaluated on different-language (Cross-lingual) trials. This experiment aims to investigate the performance of speaker recognition models trained on English-only speech data for matching Non-English different-language (e.g., Chinese to Russian) speech trials.
    \end{description}
	\vspace{-0.2cm}
	\subsubsection{Effect of Audio Length on Speaker Verification}~\label{exp_al}
	The reliability of the speaker-dependent features extracted from an audio sample depends on the amount of usable speech data present within, which is directly dependent on the length of the audio sample. Therefore, performing speaker recognition in audio samples of a small duration is a challenging task. Since in real-life scenarios, probe audios are of relatively small audio durations (1 sec - 3 secs), the feature extraction algorithm needs to be able to reliably extract speaker-dependent features from speech audio of limited duration. In this experiment (see Table~\ref{tab:Experiments_al} and Figure~\ref{fig:ROC_AL}), we compare the speaker verification performance of our proposed algorithm with the baseline algorithms on speech data of varying duration from the NIST SRE 2008 dataset. The duration of probe audio is varied between $3.5$ secs and $0.5$ secs in steps of $0.5$ secs.

    
    \vspace{-0.5cm}
	\section{Results and Analysis} \label{results}
	
	The results for all the experiments described in Section~\ref{sec:exp} are given in Tables~\ref{tab:Experiments_VOXCeleb2}, \ref{tab:Experiments_Fisher}, \ref{tab:Experiments_SRE}, \ref{tab:Experiments_SRE_lang}, \ref{tab:Experiments_al}. The content of all the tables is summarised in the DET curves given in Figures~\ref{fig:ROC}, \ref{fig:ROC_AL} to present the results in an easier-to-consume format. For all the speaker verification experiments, we report the	True Match Rate at False Match Rate of $1\%$ and $10\%$ (TMR@FMR=$\{1\%,10\%\}$), minimum Detection Cost Function (minDCF) at $C_{miss}$ (cost of a missed detection) value of $1$ and Equal Error Rate (EER, in $\%$) as our performance metrics for comparison of the baseline methods and the proposed method. The minDCF is reported at two different \textit{a priori} probability of the specified target speaker, viz., $P_{tar}$ of $0.01$ and $0.001$ (minDCF($P_{tar}=\{0.01,0.001\}$). The Detection Error Tradeoff (DET) curves are given in Figure~\ref{fig:ROC}.
	
	
	\begin{itemize}[leftmargin=0cm,itemindent=.5cm,labelwidth=\itemindent,labelsep=0cm,align=left]
	
		\item Overall, in all the speaker verification experiments given in Tables~\ref{tab:Experiments_VOXCeleb2}, \ref{tab:Experiments_Fisher}, \ref{tab:Experiments_SRE}, \ref{tab:Experiments_SRE_lang}, and \ref{tab:Experiments_al}, the 1D-Triplet-CNN algorithm using DeepVOX features trained with adaptive triplet mining, also referred to as 1D-Triplet-CNN-online(DeepVOX), performs the best. The proposed adaptive triplet mining
		method improves the verification performance (TMR@FMR=$1$\%) of the 1D-Triplet-CNN algorithm using DeepVOX features by $3.01$\%, and MFCC-LPC features by $8.71$\%. Similar performance improvements are also noticed for the MFCC and LPC features across all the performance metrics. This establishes the benefits of using the adaptive triplet mining technique over offline-triplet mining for efficiently training the 1D-Triplet-CNN based speaker recognition models.
		
		\item Across all the speaker verification experiments given in Tables~\ref{tab:Experiments_VOXCeleb2}, \ref{tab:Experiments_Fisher}, \ref{tab:Experiments_SRE}, \ref{tab:Experiments_SRE_lang}, and \ref{tab:Experiments_al}, the second-best performance, after DeepVOX features, is obtained by the feature level combination of MFCC and LPC features, referred to as MFCC-LPC features. Therefore, we choose MFCC-LPC features as our strongest baseline feature. In the upcoming discussions, all performance improvements offered by the DeepVOX features, for any particular algorithm, is reported in comparison to the MFCC-LPC features. Furthermore, we will also draw comparison with the RawNet2 model to establish DeepVOX's performance benefits over a current state-of-the art raw speech audio-based speaker recognition method.
		
	    \item In experiment \#1 on the VOXCeleb2 dataset, given in Table~\ref{tab:Experiments_VOXCeleb2} and Figure~\ref{fig:ROC}, the 1D-Triplet-CNN-online(DeepVOX) method performs the best across all the performance metrics. The DeepVOX features improve the verification performance (TMR@FMR=$\{1\%,10\%\}$), specifically for the 1D-Triplet-CNN-online algorithm, over the best performing baseline feature (MFCC-LPC) by $\{9.89\%,0.9\%\}$. It also reduces the EER by $2.5\%$ and minDCF($P_{tar}=\{0.001,0.01\}$) by $\{0.03, 0.15\}$. Similarly, for the 1D-Triplet-CNN, xVector-PLDA, and iVector-PLDA algorithm, the DeepVOX features improve verification performance over the best performing baseline feature (MFCC-LPC). The 1D-Triplet-CNN(DeepVOX) method also outperforms the RawNet2 across all the performance metrics. The TMR@FMR=$\{1\%,10\%\}$ is increased by $\{0.23\%,0.97\%\}$, EER is reduced by $0.99\%$, and minDCF($P_{tar}=\{0.001,0.01\}$) is reduced by $\{0.026,0.02\}$.

	    
    	
    	\item In all the four speaker verification experiments (Experiments $2$ to $5$) on the degraded Fisher dataset given in Table~\ref{tab:Experiments_Fisher} and Figure~\ref{fig:ROC}, the 1D-Triplet-CNN-online (DeepVOX) method performs the best across all the performance metrics. It is important to note that the performance of all the algorithms is significantly lower in case of cross-noise experiments (Experiments 3 and 5) when compared to the same-noise experiments (Experiments 2 and 4). However, the usage of the proposed DeepVOX features in all the algorithms improves their robustness to the mis-match in the training and testing noise characteristics. Also, the speaker recognition performance in the presence of babble noise, compared to the F-16 noise, is observed to be significantly lower. This indicates speech babble as one of the more disruptive speech degradations for speaker recognition tasks. All the algorithms when trained on DeepVOX features, as compared to MFCC, LPC or MFCC-LPC features, gain significant performance improvements.
    	
    	\item On an average across the four speaker verification experiments (Experiments $2$ to $5$) on the degraded Fisher dataset, the incorporation of DeepVOX features in the 1D-Triplet-CNN-online algorithm improves the verification performance (TMR@FMR=$\{1\%,10\%\}$) over the MFCC-LPC feature by $\{23.83\%,10.65\%\}$, reduces the EER by $5.98\%$, and improves minDCF($P_{tar}=\{0.001,0.01\}$) by $\{0.007,0.24\}$. Similarly, for the 1D-Triplet-CNN, xVector-PLDA, and iVector-PLDA algorithm, the DeepVOX features improve speaker verification performance over the best performing baseline feature (MFCC-LPC). The 1D-Triplet-CNN(DeepVOX) method also outperforms the RawNet2 across all the performance metrics. The TMR@FMR=$\{1\%,10\%\}$ is increased by $\{25.32\%,17.62\%\}$, EER is reduced by $10.21\%$, and minDCF($P_{tar}=\{0.001,0.01\}$) is reduced by $\{0.004,0.14\}$. Furthermore, the proposed method's performance benefits compared to the RawNet2 is even greater in the cross-noise experiments (Experiments 3 and 5), demonstrating its superior resilience to mis-matched degraded audio conditions.

    	\item On an average across the seven speaker verification experiments (Experiments $6$ to $12$), all the algorithms gain performance benefits when the MFCC, LPC and MFCC-LPC features are replaced with DeepVOX features for training the models. Replacing the best performing baseline feature (MFCC-LPC) by DeepVOX features in the 1D-Triplet-CNN-online algorithm improves the verification performance (TMR@FMR=$\{1\%,10\%\}$) by $\{14.72\%,8.7\%\}$, reduces the EER by $3.67\%$ and minDCF($P_{tar}=\{0.001,0.01\}$) by $\{0.009,0.11\}$. Similarly, for the 1D-Triplet-CNN, xVector-PLDA, and iVector-PLDA algorithm, the DeepVOX features improve speaker verification performance over the best performing baseline feature (MFCC-LPC). The 1D-Triplet-CNN(DeepVOX) method also outperforms the RawNet2 across majority of the performance metrics. The TMR@FMR=$\{1\%,10\%\}$ is increased by $\{5.57\%,10.6\%\}$, EER is reduced by $3.29\%$. However, no significant change in minDCF($P_{tar}=\{0.001,0.01\}$) was observed. The 1D-Triplet-CNN(DeepVOX) method also vastly outperforms the RawNet2 method in cross-noise experiments (Experiments $11$ and $12$) on the degraded NIST SRE 2008 dataset.

		\item In the three speaker verification experiments (Experiments $13$ to $15$, given in Table~\ref{tab:Experiments_SRE_lang}) on multi-lingual speakers from the NIST SRE 2008 dataset, DeepVOX features perform the best across all the algorithms and metrics. The incorporation of DeepVOX features, compared to the MFCC-LPC features, in the 1D-Triplet-CNN-online algorithm, improves the verification performance (TMR@FMR=$\{1\%,10\%\}$) by $\{23.95\%,8.97\%\}$, reduces the EER by $4.99\%$ and minDCF($P_{tar}=\{0.001,0.01\}$) by $\{0.02,0.21\}$. Similar performance benefits of DeepVOX features were noted for the 1D-Triplet-CNN, xVector-PLDA, and iVector-PLDA algorithms, as well. The 1D-Triplet-CNN(DeepVOX) method also outperforms the RawNet2 across all the performance metrics. The TMR@FMR=$\{1\%,10\%\}$ is increased by $\{10.18\%,5.19\%\}$, EER is reduced by $3.24\%$, and minDCF($P_{tar}=\{0.001,0.01\}$) is reduced by $\{0.019,0.12\}$.

		
		\item It is interesting to note the effect of language on verification performance in Experiments $13$ to $15$. Best speaker verification performance is achieved in Experiment $13$, where the models are trained on English speech data and evaluated on same-language English-only speech audio pairs. However, introduction of same-language multi-lingual speech audio pairs to the evaluation set (in Experiment $14$) reduces the verification performance (TMR@FMR=$1\%$) of 1D-Triplet-CNN-online by $3.70$\% for the DeepVOX features, $14.28$\% for the MFCC-LPC features, $4.11$\% for the MFCC features, and $17.46$\% for the LPC features. Furthermore, re-evaluating the same models on cross-language multi-lingual speech audio pairs in Experiment $15$ results in the largest reduction in verification performance, verifying the impact of language-familiarity effect~\cite{lu2009effect} in all algorithms and features used in our experiments. It is important to note that the detrimental effects of the language-familiarity effect (in Experiment $14$) are observed to be the weakest at $22.49$\% (performance reduction (TMR@FMR=$1\%$)) for the DeepVOX features compared to $40.76$\% for the MFCC-LPC features, $39.31$\% for the MFCC features, and $37.91$\% for the LPC features, using the best-performing 1D-Triplet-CNN-online algorithm.
		
		\item In the experimental results given in Table~\ref{tab:Experiments_al} and illustrated in Figure~\ref{fig:ROC_AL}, we notice a gradual decrease in verification performance (across all algorithms and features) with the decrease in length of audio samples in the testing data. However, the loss in performance is observed to be much lower with the usage of DeepVOX features compared to MFCC, LPC, or MFCC-LPC features across all the algorithms. The 1D-Triplet-CNN-online algorithm using DeepVOX features sufferes a performance (TMR@FMR=$10\%$) reduction of $~10$\%, compared to a reduction of $~32$\% using MFCC-LPC features, $~45$\% using MFCC features, $~34$\% using LPC features, when the audio length is reduced from $3.5$ seconds to $0.5$ seconds. Similar trends were observed for the 1D-Triplet-CNN, xVector-PLDA, and the iVector-PLDA algorithms across the DeepVOX, MFCC-LPC, MFCC, and LPC features. For the RawNet2 algorithm, a performance loss of $~48$\% is observed when the length of raw input audio is reduced from $3.5$ seconds to $0.5$ seconds. It is important to note that when compared to the 1D-Triplet-CNN based algorithms, relatively larger performance losses are observed for the iVector-PLDA, xVector-PLDA, and RawNet2 algorithms, across all the features. However, using the DeepVOX features improves the robustness of even the iVector-PLDA and xVector-PLDA algorithms when performing speaker verification on speech samples of limited duration, thereby asserting the effectiveness of the DeepVOX features in the task.

		
		
	\end{itemize}
	\vspace{-0.2cm}


    \begin{figure}[t]		
	\centering
	\includegraphics[scale=0.26, clip]{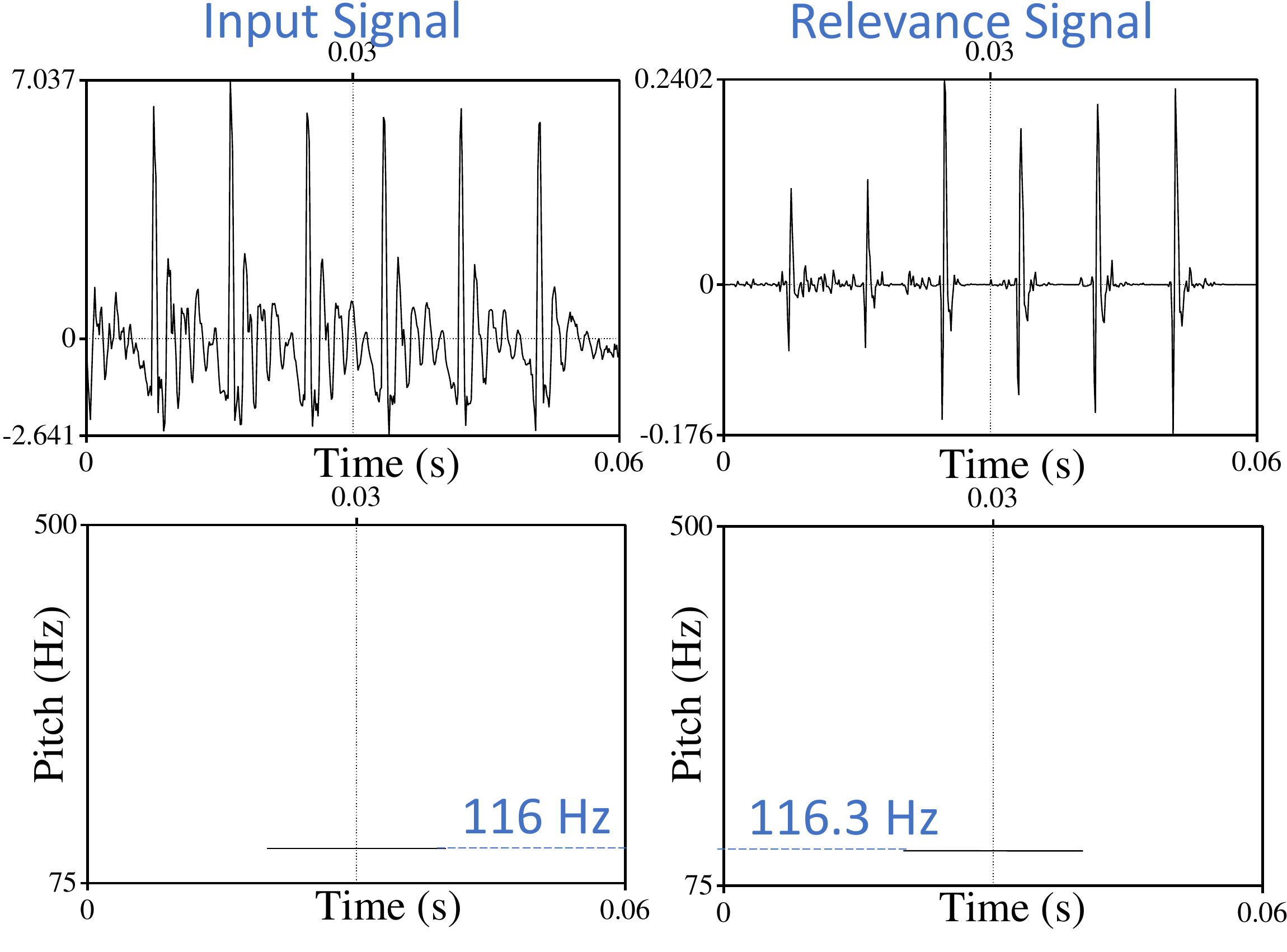}
	\vspace{-0.3cm}
	\caption{A visual comparison of the waveform (top row) and F0 contour (bottom row) for the /ah/ phoneme and its corresponding relevance signal obtained for the DeepVOX model, using the Praat~\cite{boersma2002praat} toolkit. Similar results were observed for the /eh/,/iy/,/ow/, and /uw/ phonemes.}
	\vspace{-0.5cm}
	\label{fig:F0}
    \end{figure}

    \begin{figure}[htp]
    \vspace{-0.4cm}
	\centering
	\subfloat [Clean Speech] {\includegraphics[scale=0.087, clip]{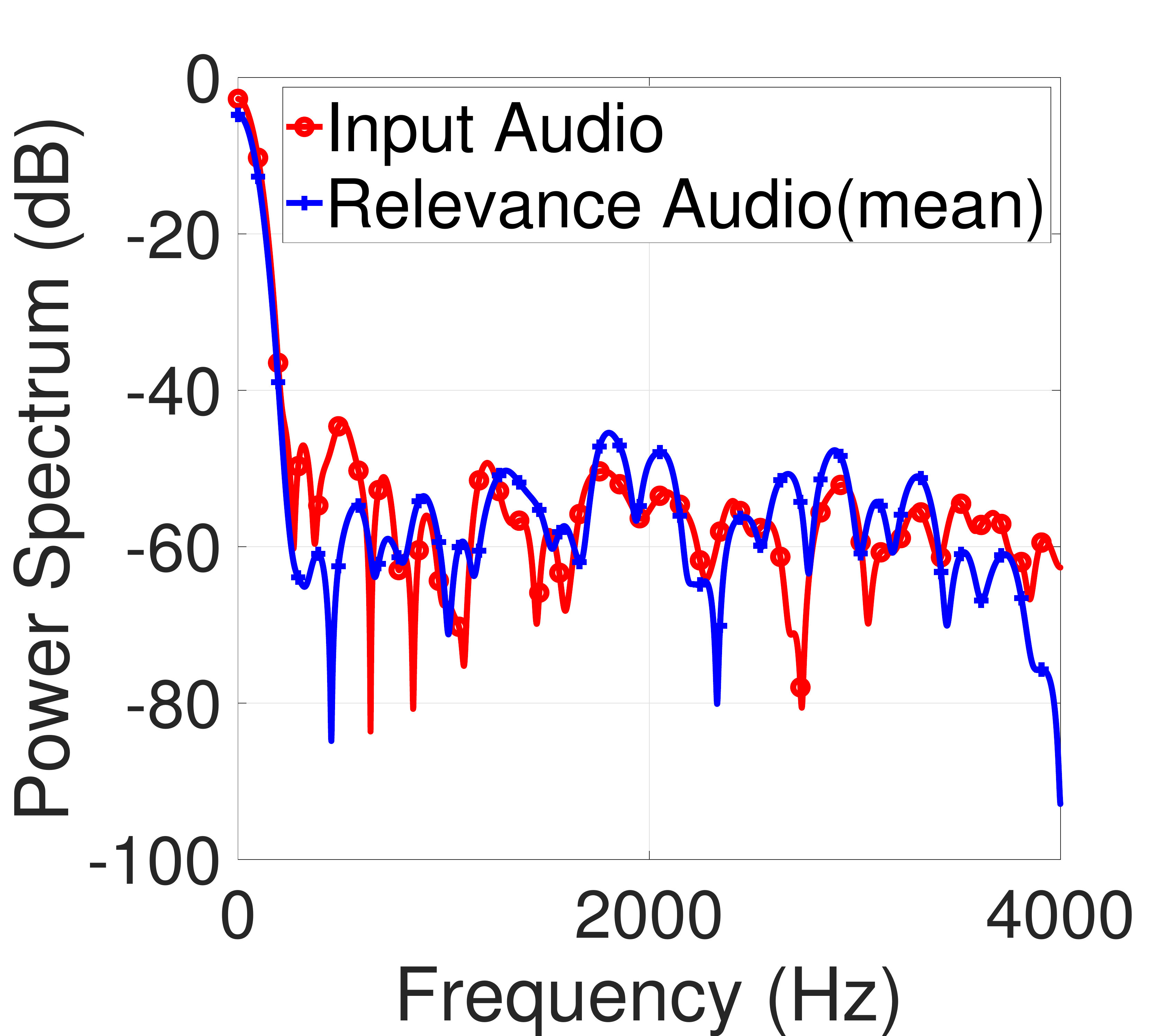}}
	\subfloat [Degraded Speech] {\includegraphics[scale=0.087, clip]{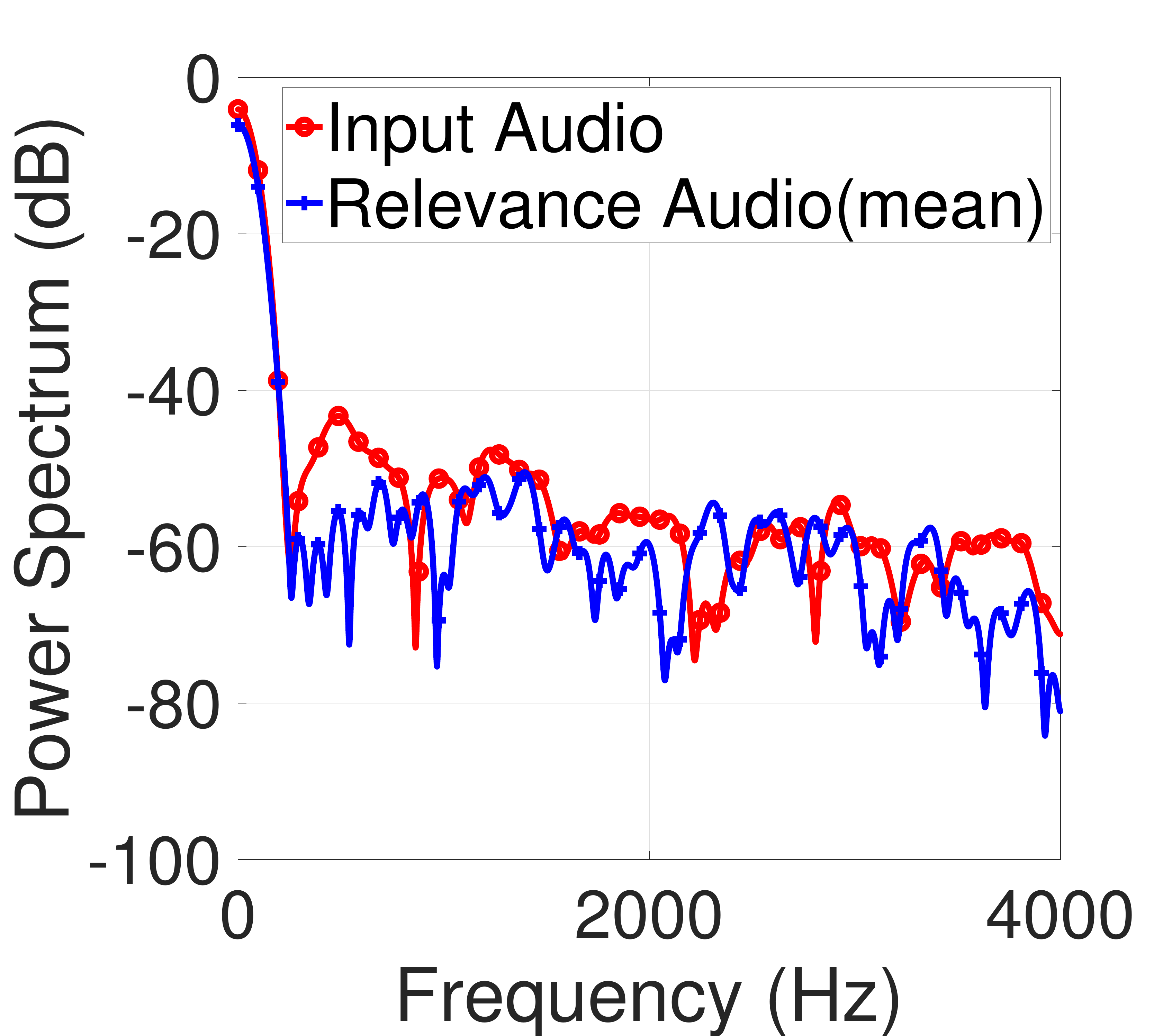}}
	\subfloat [Synthetic Car Noise] {\includegraphics[scale=0.087, clip]{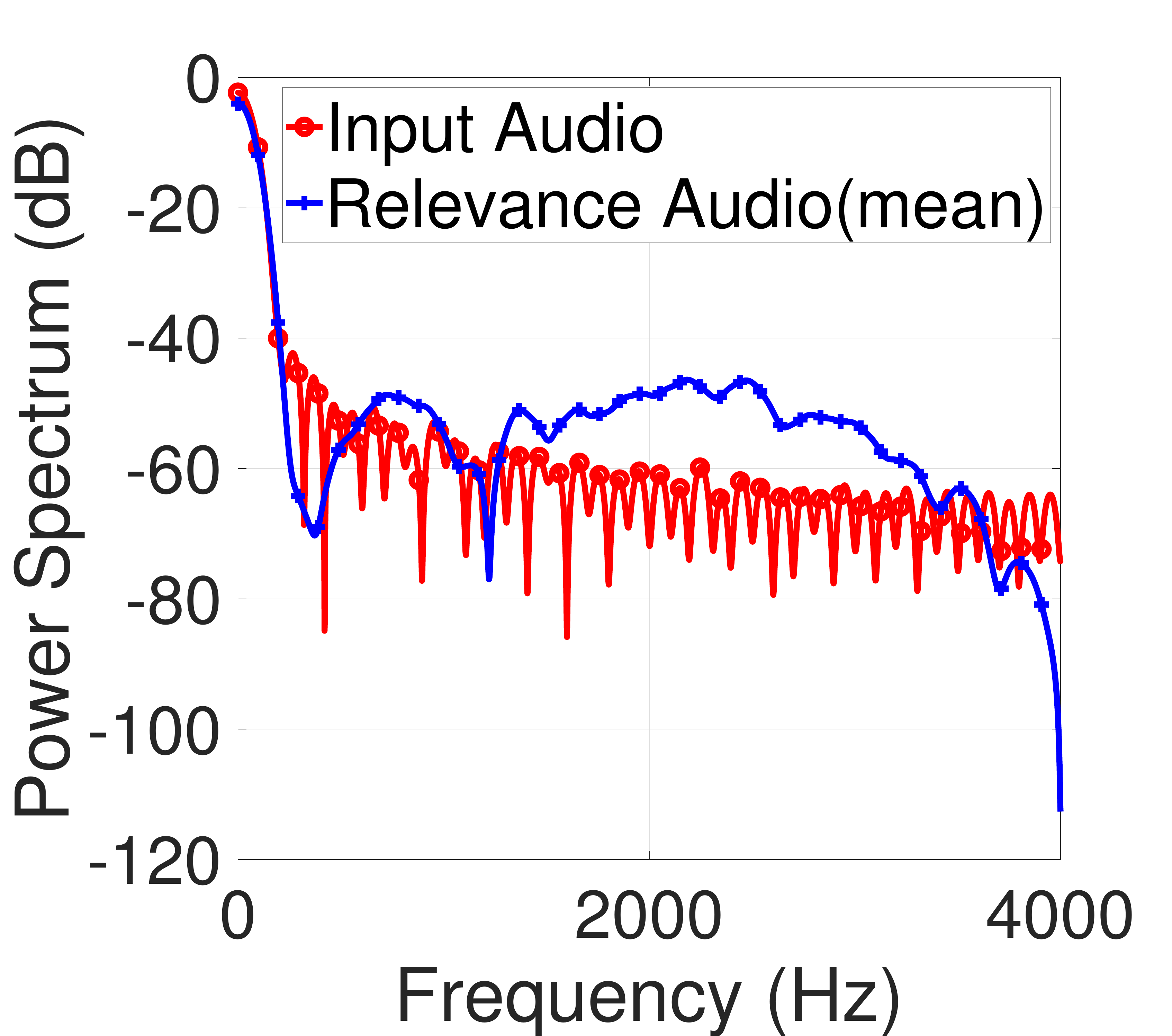}}
	\caption{\label{fig:PSD} Power Spectral Density(PSD) plots for the analysing the representation capability of the learned DeepVOX filterbank on a speech audio sample from TIMIT dataset in presence of synthetic noise audio taken from NOISEX-92 dataset.}
	\vspace{-0.6cm}
	\end{figure}
	
	\section{Ablation Study of DeepVOX}~\label{sec:ablation}
	In this section, we  use `Guided Backpropagation'~\cite{springenberg2014striving} to analyze the type of speech information being extracted by the DeepVOX feature. Such an analysis reveals the components of a speech audio that are deemed important, by the DeepVOX model, in the context of speaker recognition. In this analysis, we use the DeepVOX model trained for Experiment \#1 on the VOXCeleb2 dataset, due to diverse speakers and recording conditions in the training data. For evaluation, we choose audio samples from the TIMIT~\cite{timit} dataset due to the availability of ground-truth information for analysis of frequency sub-bands essential for speaker recognition~\cite{kinnunen2003spectral,gallardo2014spectral}. For analysing the DeepVOX method, we feed an input audio sample to the trained DeepVOX model and extract the 40-dimensional DeepVOX features. Guided backpropagation is then used individually on each of the $40$ features to estimate the corresponding relevance signals. The relevance signal in this case refers to the portion of input audio signal (in the frequency domain) that the DeepVOX model fixates on to extract a corresponding DeepVOX feature. The $40$ relevance signals corresponding to the $40$ DeepVOX features are aggregated to estimate the mean relevance signal. The mean relevance signal is then analysed, as given below, to characterize the properties of the speech signal extracted by the DeepVOX features important for performing speaker recognition:
	
	\begin{description}[leftmargin=0cm,itemindent=.5cm,labelwidth=\itemindent,labelsep=0cm,align=left]
    \item[Fundamental Frequency (F0) Extraction by the DeepVOX:] In this experiment, we extract speech utterances corresponding to the five phonemes /ah/, /eh/, /iy/, /ow/, /uw/ from a randomly chosen speaker in the TIMIT dataset. The speech audio of these phonemes is then fed to the trained DeepVOX model to extract corresponding DeepVOX features and subsequently extract the corresponding relevance signals. The input speech signal and the corresponding mean relevance signal are then compared using the Praat~\cite{boersma2002praat} toolkit (see Figure~\ref{fig:F0}). While the waveform representation of the original input signal and the corresponding mean relevance signal differ visually, pitch contour analysis of the signals reveals that the relevance signal successfully captures the F0 information from the input speech signal. This indicates that the DeepVOX architecture successfully extracts and uses fundamental frequency (F0) (a vocal source feature), for representing the human voice. This could be seen as a direct effect of the presence of phase information in the raw input speech audio, as phase information in speech audio captures rich vocal source information~\cite{vocal_tract_and_source}.
    
    \item[Operational Frequency-range of DeepVOX: ]Similar to~\cite{muckenhirn2019understanding}, we plot the input audio signal (in red color) and corresponding relevance signal (in blue color) on the Power Spectral Density (PSD) plots (given in Figure~\ref{fig:PSD}). The PSD plots are inspected for frequency-band overlap in the input audio signal and the corresponding mean relevance signal. The overlap indicates the frequency components of the input audio signal that DeepVOX captures for performing speaker recognition. As observed in Figure~\ref{fig:PSD}[a], the trained DeepVOX model reliably models a clean speech input signal in the frequency range of $0$ to $4000$Hz, with better modeling performance observed in the range of $2000$Hz to $4000$Hz, which is known to contain highly discriminative speaker-dependent information~\cite{kinnunen2003spectral,gallardo2014spectral}. This demonstrates DeepVOX's ability to use spectral information in the frequency range of $0$ to $4000$Hz for performing speaker recognition.

    \item[Effect of Audio Degradation on the DeepVOX: ] Furthermore, as shown in Figure~\ref{fig:PSD} [(b)], the trained DeepVOX model also reliably models a speech signal degraded with synthetic car noise from the NOISEX-92 dataset~\cite{noisex}. However, it fails to model the synthetic car noise in absence of speech, as shown in Figure~\ref{fig:PSD} [(c)]. This demonstrates DeepVOX's ability to selectively model the speech components and reject the background noise in an audio sample for performing speaker recognition.

    \begin{figure}[t]		
	\centering
	\includegraphics[scale = 0.45]{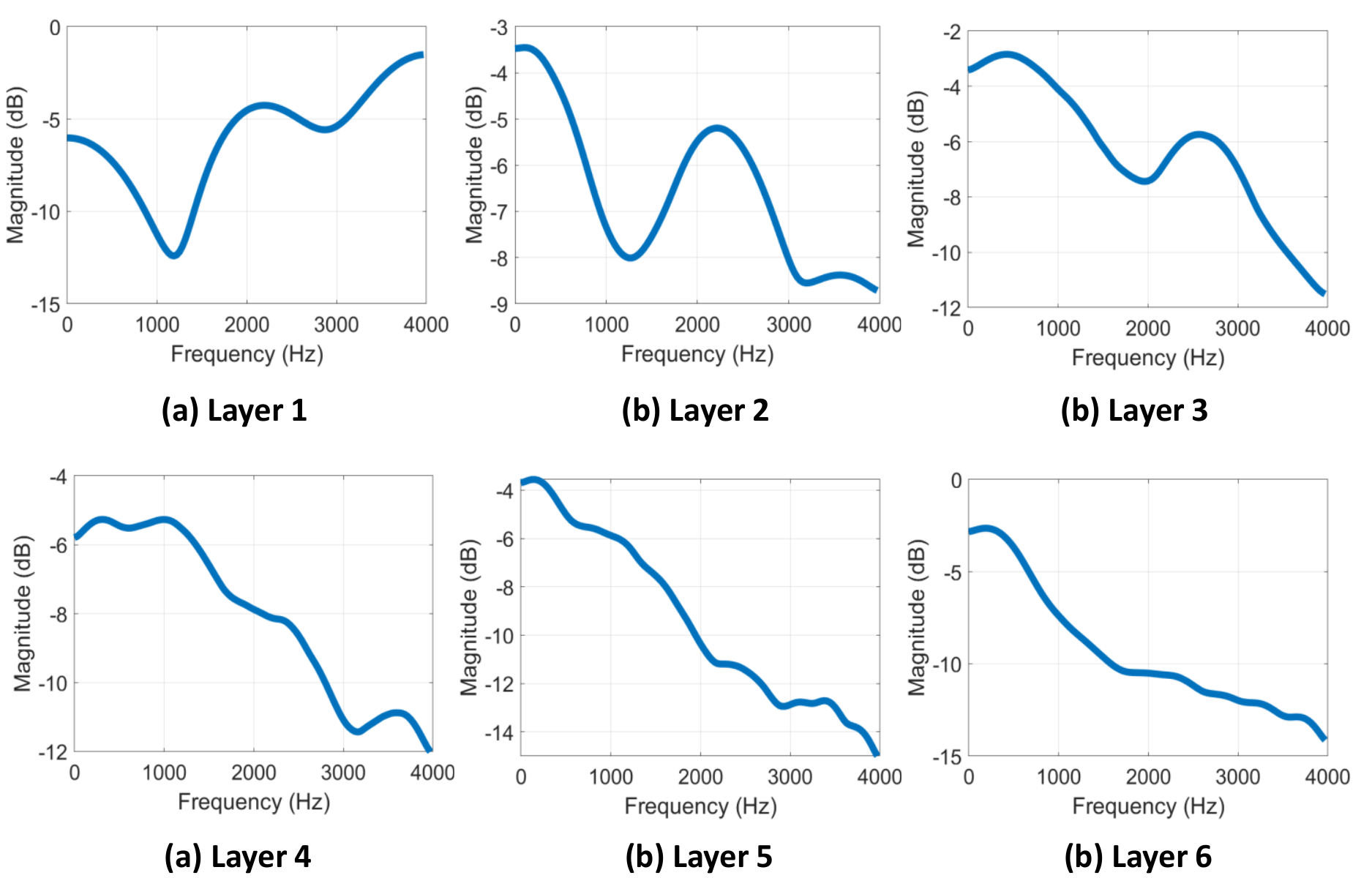}
	\vspace{-0.35cm}
	\caption{Cumulative layer-wise magnitude frequency response of the DeepVOX model trained on the VoxCeleb2 dataset}
	\label{fig:mag_res}
	\vspace{-0.3cm}
    \end{figure}
    
    \item[Layer-wise magnitude frequency response of the DeepVOX:] Finally, we also plotted (see Figure{~\ref{fig:mag_res}}) the layer-wise cumulative magnitude frequency response of the convolution filters in the DeepVOX model trained on the VoxCeleb2 dataset. Here we observed that while the initial three layers behave as a multi-band pass filter, the later layers act as low-pass filters. Specifically, the first three layers' cumulative magnitude frequency response shows peaks in the frequency range of 0-800Hz and 1500-3000HZ. Comparing to the acoustic characteristics of the human voice in American English{~\cite{hillenbrand1995acoustic}}, the first peak (0-800Hz) is specifically suited for capturing the fundamental frequency (F0) and first formant (F1) of the human voice (the average F0 is 195Hz and average F1 is 595Hz) and the second peak (1500-3000HZ) can capture the second (F2) and third (F3) formants of the human voice (the average F2 is 1734Hz and the average F3 is 2826Hz). Therefore, the initial layers of the DeepVOX model learn to capture important speaker-dependent speech characteristics (F0, F1, F2, and F3) from input speech audio and are well-suited for application in a speaker recognition system.
    \end{description}

    \vspace{-0.1cm}
	\section{Applications of DeepVOX}
	\vspace{-0.1cm}

	Speaker recognition systems find applications in several different domains, including telephone banking{~\cite{HSBC}}, E-Commerce{~\cite{epay}} and forensics{~\cite{criminalID}}, and personal virtual assistants{~\cite{alexavoice}} in the form of voice-controlled user interfaces. However, with the increase in the applications of speaker verification technology, its surface area for incoming threats of circumvention or misuse is also increased. For example, the authors in{~\cite{li2019adversarial}} developed an adversarial audio sample that can be played to stop Amazon Alexa from being activated, thus launching a form of denial-of-service attack (DoS attack). Another form of attack called voice spoofing{~\cite{kinnunen2012vulnerability}} can be used to impersonate a target user and fraudulently gain access to sensitive user data. Recently, with the advent of DeepFake technology, it is now possible to synthesize realistic speech audio or create convincing alterations of existing speech audios in the public domain to cause widespread panic and confusion{~\cite{stupp2019fraudsters}}. In such scenarios, a robust speaker recognition system, such as DeepVOX, can help thwart the attempts of unauthorized users to illegitimately access a voice interface through voice spoofing or launching DOS attacks. 

    Towards that end, in this work, we specifically evaluated DeepVOX's generalizability across a wide variety of speech audio, ranging from telephonic speech conversations in the Fisher Speech Corpora to the nearly-unconstrained interview speeches from the VOXCeleb dataset. Furthermore, we also explored a wide variety of speech audio degradations and assessed their impact on the speaker verification performance of DeepVOX-based models. We specifically introduced the experiments with a mismatch in the audio degradations in the train and evaluation sets (Experiments 3 and 5 in Table{~\ref{tab:Experiments_Fisher}}) and mismatch in spoken language in experiments 14 and 15 in Table{~\ref{tab:Experiments_SRE_lang}} to simulate the effect of domain mismatch on the speaker verification performance. While we do notice a performance drop in case of domain mismatch across all the methods and feature combinations tested in this work, the negative impact of domain mismatch is notably reduced across all the scenarios when the DeepVOX replaces traditional speech features (MFCC and LPC). This demonstrates that the DeepVOX in its current form is relatively robust to the adverse effects of mismatch between the training and testing conditions and can be applied to a wide variety of applications discussed above, where domain mismatch is expected. Furthermore, we believe it is possible to tweak the DeepVOX hyperparameters such as the number of DeepVOX filterbanks, the type, length, and stride of the windowing function to adapt the DeepVOX method to specific datasets and audio conditions and further improve its performance.
    
    Additionally, from an implementation perspective, it is important to note that we designed the DeepVOX to offer either an end-to-end learnable or a drop-in replacement for handcrafted filterbanks such as MFCC. For example, on the one hand, we can train DeepVOX end-to-end with any speaker embedding extraction system, as our experiments do with the 1D-Triplet-CNN system. On the other hand, we can replace a fixed MFCC-based feature extraction pipeline with a pre-trained DeepVOX filterbank to asynchronously train a speech embedding method, as shown in our experiments with the xVector and iVector based methods. Also, note that DeepVOX features are generated at the frame level like traditional MFCC features and it should not be confused with a fixed-dimensional speech embedding extractor such as xVector. Therefore, in summary, DeepVOX provides a learnable a time-domain speech filter bank that can either be used to train robust end-to-end speaker recognition systems from scratch or retrofit into existing speaker recognition frameworks. Furthermore, the DeepVOX-based speaker recognition system's robustness to various non-ideal audio conditions, such as background noise, language mismatch, and short audio duration, makes it an essential tool in the arsenal of digital audio forensics to protect and verify the integrity of data before using it.

	\vspace{-0.1cm}
	\section{Conclusion}
	\vspace{-0.1cm}
	The performance of short-term speech feature extraction techniques, such as MFCC, is dependent on the design of handcrafted filterbanks such as the Mel filterbank. While such techniques are easy to use and do not require any training data, they do not adapt well to diverse non-ideal audio conditions. Therefore, it is beneficial to develop feature extraction techniques, such as DeepVOX, that can robust across diverse non-ideal audio conditions, as evident in the experimental results. The frequency analysis of the learned DeepVOX filterbanks indicates that it can extract spectral information from a large frequency range ($0$ to $4000$Hz) and also extract the fundamental frequency (F0) information for representing the speaker in speech audio. It is also important to make note of rare cases such as Experiment $8$ in Table~\ref{tab:Experiments_SRE}, where certain combinations of noise characteristics in the training and testing sets create challenging scenarios for the proposed DeepVOX. Therefore, it is important to continue research in the development of feature extraction algorithms that builds upon and improves the currently proposed algorithm. Furthermore, as discussed in Section~\ref{speech_prep}, the DeepVOX algorithm has a limitation of being trained only $200$ audio frames at a time; hence, it cannot benefit from training on longer audio samples. We plan to extend our DeepVOX model by incorporating methods for automatically learning from audio samples of varying lengths.
	
	\vspace{-0.3cm}
	\section*{Acknowledgment}
	We thank Dr. Shantanu Chakrabartty and Dr. Kenji Aono from Washington University in St. Louis for providing us their audio degradation tool, used in this work. We also thank Dr. Joseph P. Campbell from MIT Lincoln Lab for the useful discussions.

	\ifCLASSOPTIONcaptionsoff
	\newpage
	\fi
	
	\vspace{-0.3cm}
	{\small
		\bibliographystyle{ieee}
		\bibliography{ref}
	}

\end{document}